\documentclass[12pt,a4paper]{article}
\usepackage{amsmath}
\usepackage{amssymb}
\usepackage{amsthm}
\usepackage{float}
\usepackage{amsfonts}
\usepackage{graphicx}
\usepackage[]{cite}
\usepackage{float}
\usepackage{subfigure}
\usepackage{verbatim}
\usepackage[left=2cm,right=2cm,top=3cm,bottom=2.5cm]{geometry}
\usepackage[numbers]{natbib}
\usepackage[utf8]{inputenc}
\usepackage[usenames,dvipsnames,svgnames]{xcolor}
\usepackage[colorlinks=true,
      linkcolor=red,
      urlcolor=gray,
      citecolor=blue]{hyperref}

\def\myalign#1{%
  \def\trule{\noalign{\smallskip\hrule\medskip}}
  \def\nebc{\nearrow\bigcup}
  \def\sebc{\searrow\bigcup}
  \def\pminf{{}_{-\infty}|^{+\infty}}
  \let\Inf\infty
  \def\amp{&} 
  \vbox{\mathsurround0pt\openup1\jot
    \halign{%
      &$\displaystyle##\hfil\tabskip0pt$&\amp##\tabskip1em\crcr
      \noalign{\hrule height1pt\smallskip}#1\noalign{\smallskip\hrule height1pt}\crcr}}}

\begin{document}
\begin{center}
\textbf{Multifluid cosmology in $f(G)$ gravity}
\end{center}
\hfill\\
Albert Munyeshyaka$^{1}$, Abraham Ayirwanda$^{2,3}$,Fidele Twagirayezu $^{2,3}$,Beatrice Murorunkwere $^{2}$ and Joseph Ntahompagaze$^{2}$\\
\hfill\\
%$^{1}$Department of Physics, College of Science and Technology, University of Rwanda, Rwanda\;\;\; \; \hfill\\
$^{1}$Department of Physics, Mbarara University of Science and Technology, Mbarara, Uganda\;\;\; \; \;\hfill\\
$^{2}$Department of Physics, College of Science and Technology, University of Rwanda, Rwanda\;\;\; \; \;\hfill\\ 
$^{3}$ International Center For Theoretical Physics (ICTP)-East African Institute for Fundamental Research, University of Rwanda, Kigali, Rwanda
\\
\hfill\\
Correspondence:munalph@gmail.com\;\;\;\;\;\;\;\;\;\;\;\;\;\;\;\;\;\;\;\;\;\;\;\;\;\;\;\;\;\;\;\;\;\;\;\;\;\;\;\;\;\;\;\;\;\;\;\;\;\;\;\;\;\;\;\;\;\;\;\;\;\;\;\;\;\;\;\;\;\;\;\;\;\;\;
\begin{center}
\textbf{Abstract}
\end{center}
The treatment of $1+3$ covariant perturbation in a multifluid cosmology with the consideration of $f(G)$ gravity, $G$ being the Gauss-Bonnet term, is done in the present paper. We  define a set of covariant and gauge-invariant variables to describe density, velocity and entropy perturbations for both the total matter and component fluids. We then use different techniques such as scalar decomposition, harmonic decomposition, quasi-static aproximation together with the redshift transformation to get  simplified perturbation equations for analysis. We then discuss  number of interesting applications like the case where the universe is filled with a mixture of radiation and Gauss-Bonnet fluids as well as dust with Gauss-Bonnet fluids for both short- and long-wavelength limits.  Considering polynomial $f(G)$ model, we get numerical solutions of energy density perturbations and show that they decay with increase in redshift. This feature shows that under $f(G)$ gravity, specifically under the considered $f(G)$ model,  one expects that the formation of the structure in the late Universe is enhanced.
\\
\hfill\\
\textit{keywords:} $f(G)$ gravity--- Covariant formalism--- Cosmological perturbations.This manuscript was accepted for publication in International Journal of Geometric Methods in Modern Physics.
\section{Introduction}\label{introduction}
The recent observations have revealed that the Universe is experiencing cosmic accelaration \cite{hogg2005cosmic,lu2007obtaining}. The aforementioned cosmic accelaration is sometimes explained by imposing dark energy, a fluid with  a negative pressure. This component may be a vacuum energy giving rise to the standard model of cosmology dubbed $\Lambda$ CDM, or a scalar field \cite{copeland2006dynamics}. \\
On the other side, cosmologists suggested that General Relativity (GR) need to be modified on large scale and at late time \cite{bertschinger2008distinguishing}. This suggestion helps to explain the source of cosmic acceleration giving rise from the dynamics of modified theories of gravity rather than the dark energy as substance   \cite{nojiri2007introduction, riess1998observational}.\\ \\
 Additionally, studying cosmological perturbations help to analyse the dynamics of the Universe on both GR and modified theories of gravity. One can use metric formalism, a gauge-dependent formalism \cite{de2008evolution} to treat perturbations. This formalism brings in some complications when one needs to extract physical information from the perturbation variables \cite{bardeen1980gauge,kodama1984cosmological,bertschinger2000cosmological,dunsby1992cosmological,dunsby1991gauge}.\\
On the other hand, many authors applied the $1+3$ covariant and Gauge-Invariant formalism, which leaves no physical modes in the evolution of perturbations \cite{challinor2000microwave}. Several authors have adopted this framework in GR \cite{dunsby1992cosmological,dunsby1991gauge,ellis1989covariant}, in $f(R)$ \cite{ellis2011inhomogeneity,hawking1966perturbations,ellis1989covariant,abebe2012covariant, abebe2015breaking} ,  in scalar tensor theories \cite{ntahompagaze2017f, ntahompagaze2018study, carloni2006gauge},in $f(T)$ \cite{sami2021covariant} and in $f(G)$ \cite{li2007cosmology}, $R$ is the Ricci scalar, $G$ is the Gauss-Bonnet term and $T$ is the torsion scalar.\\ \\
In the present work, we advocate to use the $1+3$ covariant formalism to treating perturbations in multifluid universe.
The $1+3$ covariant linear perturbations theory have been employed to explain cosmic large scale structure formation. In the recent paper \cite{munyeshyaka2021cosmological}, we  studied cosmological perturbations in $f(G)$ gravity for a two fluid system at linear order using exponential, logarithmic and trigonometric $f(G)$ models. These models have been proposed by \cite{de2009construction,inagaki2020gravitational}. In the current paper, we extend our analysis to linear perturbations for a multifluid system in $f(G)$ theory of gravity.\\
We treat our discussion of a multifluid perturbation as an extension of the previous work with a consideration of different $f(G)$ model. This is necessary since the Universe is composed of different fluids such as relativistic particles, radiation, dust, cold dark matter and many others. The extension to a multifluid cosmology is found interesting also in different  works in modified theories of gravity such as in  $f(R)$ \cite{dunsby1991gauge,dunsby1992covariant,abebe2012covariant}, and in scalar tensor theories of gravity \cite{ntahompagaze2020multifluid}. \\ \\ The $f(G)$ theories of gravity received green light to be able to explain the cosmic acceleration motivated by the epoch for which one can experience the cosmic acceleration  reproduced without the implication of dark energy hypothesis \cite{nojiri2007introduction,li2007cosmology} and
 can also be adequate to study cosmology of early and late time universe \cite{nojiri2006modified}. For instance in the work done by\cite{de2015cosmological}, cosmological inflation in $f(R,G)$ gravity was studied. The consideration of $f(R,G)$ produces the situation where the cosmic dynamic is driven by double inflation scenario due to the consideration of non-linear $R$ and $G$ terms. In the work done in \cite{de2020tracing}, the cosmic history was explored under the same consideration of $f(R,G)$ gravity by taking into account the cosmography where it was clearly shown that double inflation can naturarly be achieved to evantually produce large and very large structure. \\ \\
To study the $1+3$ covariant perturbations in a multifluid universe within $f(G)$ theories of gravity and to know how these perturbations evolve worth attention. In the present work, we develop and consider a polynomial $f(G)$ model to analyse the energy density perturbations in a multifluid system, where we consider non-interacting fluids.   All the considered models in both works are proven to be viable models that are compatible with cosmological observations and they are representative examples of models that could account for the late-time acceleration of the universe without the need for dark energy \cite{li2011large}.\\
It is believed that the dynamical evolution of small energy density perturbations seeded in the early universe led to  large scale structure formation.
The $1+3$ covariant formalism is used for studying cosmological perturbations, developed to analyse the evolution of linear perturbations of Friedmann-Robertson-Walker (FRW) models in different theories of gravity. In \cite{munyeshyaka2021cosmological}, the evolution of scalar perturbations of FRW was developed for a single barotropic fluid using the $1+3$ covariant formalism in $f(G)$ gravity focussing on two-fluid systems. The solutions of the perturbation equations on large scale structure showed that the energy density perturbations decay with increase in redshift, implying more structure formation rate today. However since the Universe is made of a mixture of fluids, a complete treatment of perturbations requires taking all the fluid into account.\\ \\
The aim of this paper is therefore to present a general framework for studying multifluid cosmological perturbations with a complete general equation of state in an $f(G)$ theory of gravity, using the $1+3$ covariant formalism. In this context, we define the gradient variables of Gauss-Bonnet fluid in addition to the gradient variables of the physical standard total matter and component fluids to get a complete set of linear perturbation equations. \\ \\For analysis, we use harmonic decomposition method to get a set of ordinary differential equations which are time dependent.  Using redshift transformation technique, quasistatic approximation, specifically when considering short wavelength limit, together with a polynomial $f(G)$ model, we get a simplified set of perturbation equations for  numerical results. \\ \\ 
The next part of this paper is organised as follows: in Section $2$, we cover the general description
of the $1+3$ covariant approach. In Section $3$,  we present matter description where we  discuss the choice of frame and define
the key variables used in the description of perturbations in the total fluid and the
individual fluid components. Equations for these variables are given in Section $4$. In
Sections $5$, $6$ and $7$, respectively, the scalar , second order equations and the harmonically decomposed forms of the developed 
 equations are considered. Applications to a radiation-dust-Gauss-Bonnet cosmological medium and  the analysis of the short- and long-wavelength modes of the perturbation equations are given in Section $8$, with Section $9$ devoted to discussion and conlusion.\\ \\
The adopted spacetime signature is $(-,+,+,+)$ and unless stated otherwise, we
use $\mu$, $\nu$ . . . $= 0$, $1$, $2$, $3$ and $8\pi G_{ N} = c = 1$, where $G_{ N}$ is the gravitational constant
and $c$ is the speed of light and we consider Friedmann–Robertson–Walker (FRW) spacetime background in this work. The symbols $\bigtriangledown$  represents the usual covariant derivative and 
$\partial$ corresponds to partial differentiation and an over-dot shows differentiation with
respect to proper time.\\
For an arbitrary $f (G)$ gravity the action can be written as
\begin{eqnarray}
 S=\frac{1}{2\kappa^{2}}\int d^{4}x \sqrt{-g}\left( R+f(G)+\mathcal{L}_{m} \right).
 \label{eq01}
\end{eqnarray}
Here  $ \kappa = 8\pi G_{N}$ is a constant, $f(G)$ is a differentiable function of the Gauss-Bonnet term $G$ and $\mathcal{L}_{m}$ is the matter Lagrangian.
The Gauss-Bonnet term is given by~ 
 \begin{equation}
  G=R^{2}-4R_{\mu \nu}R^{\mu\nu}+R_{\mu\nu \sigma \lambda}R^{\mu\nu \sigma \lambda},
 \end{equation}
where  $R$, $R_{\mu \nu}$ and $R_{\mu\nu \sigma \lambda}$ are the Ricci
scalar, Ricci tensor and Riemann tensor respectively. The information about the  content
is contained within the energy-momentum tensor $ T_{\mu \nu} $.  Different authors working on  Gauss-Bonnet gravity as alternative theory to GR 
 \cite{kawai1999evolution,nojiri2011unified,satoh2008circular,satoh2008higher,kawai1998instability,sberna2017nonsingular,odintsov2020rectifying,nojiri2019ghost, zheng2011growth, oikonomou2015singular}.
 By varrying the action (eq. \ref{eq01}) with respect to the metric $g_{\mu\nu}$  and set $\kappa=1$, we get the modified Einsten equations
  \begin{eqnarray}
   && R_{\mu\nu}-\frac{1}{2}g^{\mu\nu}R= T^{m}_{\mu\nu}+\frac{1}{2}g^{\mu\nu}f-2f'RR^{\mu\nu}+4f'R^{\mu}_{\lambda}R^{\nu\lambda}-2f'R^{\mu\nu\sigma\tau}R^{\lambda\sigma\tau}_{b}\nonumber\\
   && \quad  \quad -4f'R^{\mu\lambda\sigma\nu}R_{\lambda\sigma}+2R\bigtriangledown^{\mu}\bigtriangledown_{\nu}f'-2Rg^{\mu\nu}\bigtriangledown^{2}f'-4R^{\nu\lambda}\bigtriangledown_{\lambda}\bigtriangledown^{\mu}f'\nonumber\\
   && \quad \quad-4R^{\mu\lambda}\bigtriangledown_{\lambda}\bigtriangledown^{\nu}f'+4R^{\mu\nu}\bigtriangledown^{2}f' +4g^{\mu\nu}R^{\lambda\sigma}\bigtriangledown_{\lambda}\bigtriangledown_{\sigma}f'-4R^{\mu\lambda\nu\sigma}\bigtriangledown_{\lambda}\bigtriangledown_{\sigma}f'\;,
   \label{eq3}
  \end{eqnarray}
where $f\equiv f(G)$ and $f'=\frac{\partial f}{\partial G}$. $T^{m}_{\mu\nu}$ is the energy momentum tensor of the fluid matter (Photons, baryons, cold dark matter, and light neutrinos).
writing equation eq. \ref{eq3} in more compact form as 
\begin{equation}
 R_{\mu\nu}-\frac{1}{2}g^{\mu\nu}R= T^{m}_{\mu\nu}+T_{\mu\nu}^{G},
 \label{eq4}
\end{equation} where
\begin{eqnarray}
 &&T_{\mu\nu}^{G}=\frac{1}{2}g^{\mu\nu}f-2f'RR^{\mu\nu}+4f'R^{\mu}_{\lambda}R^{\nu\lambda}-2f'R^{\mu\nu\sigma\tau}R^{\lambda\sigma\tau}_{b}-4f'R^{\mu\lambda\sigma\nu}R_{\lambda\sigma}\nonumber \\
 && \quad \quad+2R\bigtriangledown^{\mu}\bigtriangledown_{\nu}f'
 -2Rg^{\mu\nu}\bigtriangledown^{2}f'-4R^{\nu\lambda}\bigtriangledown_{\lambda}\bigtriangledown^{\mu}f'-4R^{\mu\lambda}\bigtriangledown_{\lambda}\bigtriangledown^{\nu}f'\nonumber\\
 &&\quad\quad+4R^{\mu\nu}\bigtriangledown^{2}f'+4g^{\mu\nu}R^{\lambda\sigma}\bigtriangledown_{\lambda}\bigtriangledown_{\sigma}f'-4R^{\mu\lambda\nu\sigma}\bigtriangledown_{\lambda}\bigtriangledown_{\sigma}f'\;.
 \label{eq5}
\end{eqnarray}
For the case $f(G)=G$, equation eq. \ref{eq5} vanishes  ($T_{\mu\nu}^{G}=0)$, hence the Einstein gravity is recovered
\begin{equation}
 R_{\mu\nu}-\frac{1}{2}g^{\mu\nu}R= T^{m}_{\mu\nu}.
\end{equation}
For a spatially flat FRW universe, 
\begin{equation}
 ds^{2}=-dt^{2}+a^{2}dX^{2},
\end{equation}
 where $X=x,y,z$, the equation (0,0 component of eq. \ref{eq3}) corresponding to the Friedmann equation is presented as follows:
\begin{eqnarray}
 &&3H^{2}=\frac{1}{2}\left( Gf'-f-24\dot{G}H^{3}f''\right)+\rho_{m}\;,
 \label{eq8} \\
 && \quad \quad G=24H^{2}(\dot{H}+H^{2})\;,
 \label{eq9}\\
  && \quad\quad R=6(\dot{H}+2H^{2})\;,
  \label{eq10}
\end{eqnarray}
where $H=\frac{\dot{a}}{a}$ is the Hubble parameter.
The Friedmann equations in the Einstein gravity is presented as 
\begin{eqnarray}
&& \rho_{m}=3H^{2}\;,
 \label{eq11}\\
&& \quad \quad p_{m}=-(3H^{2}+2\dot{H})\;,
 \label{eq12}
\end{eqnarray}
it follows that the FRW like equations in $f(G)$ gravity are given as 
\begin{eqnarray}
 &&\rho_{total}=3H^{2}\;,
 \label{eq13}\\
&& \quad\quad p_{total}=-(3H^{2}+2\dot{H})\;,
 \label{eq14}\\
&&\quad \quad \rho_{total}=\rho_{G}+\rho_{m}\;,
 \label{eq15}\\
&&\quad\quad p_{total}=p_{G}+p_{m}\;,
 \label{eq16}
\end{eqnarray}

where \cite{garcia2011energy}
\begin{eqnarray}
&& \rho_{G}=\frac{1}{2} \left(Gf'-f-24\dot{G}H^{3}f''\right)\;,\\
&&\quad\quad p_{G}=\frac{1}{2}\left( f-f'G+\frac{2G\dot{G}}{3H}f''+8H^{2}\ddot{G}f''+8H^{2}\dot{G}^{2}f'''\right).
\end{eqnarray}
For the case $f(G)=G$, $\rho_{G}=0=p_{G}$,  so that eq. \ref{eq15} and eq. \ref{eq16} ressemble to eq. \ref{eq11} and eq. \ref{eq12} respectively. 
We assume that matter can be described by a barotropic perfect fluid such that $p=w\rho$.\\ In the FRW universe, the energy conservation law can be expressed as standard equation as 
\begin{equation}
 \dot{\rho}+3H(\rho+p)=0.
\end{equation}
 The general solution is given as 
\begin{equation}
 \rho=\rho_{0}t^{-3m(1+w)}.
 \label{eq20}
\end{equation}
We assume an exact power law solution for the field equations  to be 
\begin{equation}
 a(t)=a_{0}t^{m},
 \label{eq21}
\end{equation}
where $m$ is a positive constant. Using eq. \ref{eq21} in eq. \ref{eq9} and eq. \ref{eq10}, we have
\begin{equation}
 G=\frac{24m^{3}(m-1)}{t^{4}},
 \label{eq22}
\end{equation}
\begin{equation}
 \dot{G}=-\frac{96m^{3}(m-1)}{t^{5}},
 \label{eq23}
\end{equation}
and
\begin{equation}
 R=\frac{6m(2m-1)}{t^{2}}.
 \label{eq24}
\end{equation}
Using eq. \ref{eq21} through to eq. \ref{eq24} in eq. \ref{eq8}, we get a differential equation for the function $f(G)$ in $G$ space presented as 
\begin{equation}
 \frac{4G^{2}}{m-1}f''+Gf'-f+\rho_{0}\left(\frac{G}{24m^{3}(m-1)}\right)^{\frac{3}{4}(m(1+w))}-\left(\frac{3mG}{8(m-1)}\right)^{\frac{1}{2}}=0.
 \label{eq25}
\end{equation}
The general solutions of eq. \ref{eq25} is given as 
\begin{equation}
 f(G)=C_{1}G+C_{2}G^{-\frac{1}{4}(m-1)}-\frac{1}{2}\left(\sqrt{\frac{6m(m-1)G}{(m+1)^{2}}}+AG^{\frac{3}{4}m(1+w)}\right),
 \label{eq26}
\end{equation}
\begin{equation}
 A=\frac{8\rho_{0}(m-1)\left[13824m^{9}(m-1)^{3}\right]^{-\frac{1}{4}m(1+w)}}{4+m\left[3m(1+w)(w+\frac{4}{3})-18w-19\right]},
 \label{eq27}
\end{equation}
where $C_{1}$ and $C_{2}$ are constants of integration.
 This solution (eq. \ref{eq26}) is in agreement with the ones obtained in \cite{goheer2009coexistence,rastkar2012phantom} for a flat universe. Assuming $C_{1}=1$,$C_{2}=0$, the solution  eq. \ref{eq26} can be represented as  
\begin{equation}
 f(G)=G-\frac{1}{2}\left(\sqrt{\frac{6m(m-1)G}{(m+1)^{2}}}+AG^{\frac{3}{4}m(1+w)}\right).
 \label{eq28}
\end{equation}
For the case $m=1$, $f(G)\sim G$, $G\sim 0$ (eq. \ref{eq22}) and $R\sim t^{-2}$ (eq. \ref{eq24}) so that we recover GR case with a power law solution as presened in eq. \ref{eq21} and energy density $\rho_{m}$ as presented in eq. \ref{eq20}. Referring to eq. \ref{eq21} we can set $m=\frac{2}{3(1+w)}$ for the GR limit. For $m=1$, the equation of state parameter is set to $w=-\frac{1}{3}$, to account for a negative pressure but not yet an accelerating universe. In order to produce  an accelerating universe, we set $m\succ 1$ and $4+m\left[3m(1+w)(w+\frac{4}{3})-18w-19\right]\neq 0$, with $w=[-1,1]$.
In the late times, the $f(G)$ term should become dominant as compared with the matter lagrangian density. For more about the period where Einstein or Gauss-Bonnet term dominates one another, see \cite{cognola2006dark}. This $f(G)$ cosmological model can also give the possibility to produce inflation driven by $G$ thanks to the use of linear $R$ and non-linear $G$, since linear $R$ produce GR while linear $G$ vanishes. 
Power-law  solutions were obtained in \cite{bajardi2020f} while treating pure $f(G)$ Noether cosmology. In \cite{capozziello2014noether} a cosmological dynamical system analysis was  done in power-law $f(R,G)$ model and discussed the possibility to fix the form of $f(R,G)$-lagrangian by existence of of symmetry using particulary the Noether symmetry approach.
We will use the defined $f(G)$ model  for the purpose of using the $1+3$ perturabtion equations in  the analysis sections in order to see whether the  formation of structure is enhanced or not and to explain  the current cosmic acceleration. 
\section{ The $1+3$ covariant perturbations in $f(G)$ gravity}
\subsection{Introduction}
The $1+3$ covariant formalism  leaves no physical modes in the evolution of the fluctuations when it comes to extracting physical information from the perturbation variables. In this formalism, spacetime is split into space and time, where $1+3$ refers to the number of dimensions involved in each slice to investigate the deviation from homogeneity and isotropy of the Universe.
The $4$-velocity field vector $u^{a}$ is defined as 
\begin{equation}
 u^{a}=\frac{dx^{a}}{d \tau},
\end{equation}
where $\tau$ is the proper time such that $u^{a}u_{a}=-1$.
\subsection{Kinematic quantities}
The geometry of the fluid flow lines is determined by the kinematics of $u^{a}$ \cite{gidelew2013beyond,carloni2010conformal} as
\begin{equation}
 \bigtriangledown_{b}u_{a}=\tilde{\bigtriangledown}_{b}u_{a}-\dot{u}_{a}u_{b},
 \label{eq30}
\end{equation}
with
\begin{equation}
 \tilde{\bigtriangledown}_{b}u_{a}=\frac{1}{3}\theta h_{ab} +\sigma_{ab}+\omega_{ab}.
 \label{eq31}
\end{equation}
Substituting eq. \ref{eq30}  into eq. \ref{eq31}, we get an important equation relating our key kinematic quantities
\begin{equation}
 \bigtriangledown_{a}u_{a}=-u_{b}\dot{u}_{a}+\frac{1}{3}\theta h_{ab} +\sigma_{ab}+\omega_{ab}.
\end{equation}
Where $ \dot{u}_{a}$ is the acceleration of fluid flow, $\theta$ is the expansion, $h_{ab}=g_{ab}+u_{a}u_{b}$ is the projection tensor, $g_{ab}$ is the metric, $\sigma_{ab}$ is shear and $\omega_{ab}$ is the vorticity.\\
Another key equations are defined as:\\
The conservation equations:
\begin{eqnarray}
 &&\dot{\rho}-\theta \left(\rho+p\right)=0\;,\\ 
 &&\tilde{\bigtriangledown}_{a} p-\left(\rho+p\right)\dot{u}_{a}=0.
\end{eqnarray}
The propagation equation for the expansion - the Raychaudhuri equation for the FRW background, the equation of state and the Friedmann equation :
\begin{eqnarray}
&& \dot{\theta}-\tilde{\bigtriangledown}_{a}\dot{u}_{a}=-\frac{1}{3}\theta^{2}-\frac{1}{2}\left(\rho+3p\right)\;,\\
&& p= p(\rho, s)\;,\\
&& \theta^{2}+\frac{9K}{a^{2}}-3\rho=0\;,
\end{eqnarray}
form a closed system of equations and completely characterize the kinematics of the
background cosmological model \cite{li2011large,dunsby1992covariant,sahlu2020scalar,ehlers2007ak}.  $K$ stands for the curvature of the universe (for flat universe, $K=0$) and $a$ stand for the cosmological scale factor. 
\section{Matter description}
\subsection{Effective total energy-momentum tensor}
The total energy momentum tensor in a general frame is sourced by 
\begin{equation}
 T_{ab}=\rho u_{a}u_{b}+p h_{ab}+2q_{a}u_{b}+\pi_{ab}=T_{ab}^{m}+T_{ab}^{G},
\end{equation}
where $\rho$, $p$, $q_{a}$
and $\pi_{ab}$ represent the energy density, isotropic pressure, energy flux and the anisotropic pressure respectively. Note that the superscript $G$ is neither a power nor a running index, it shows the contribution from Gauss-Bonnet term. With
\begin{eqnarray}
 &&\rho^{total}=T_{ab}^{total}u^{a}u^{b}=\rho_{m}+\rho_{G}\;,
 \label{eq40}\\
&& p^{total}=\frac{1}{3}T_{ab}^{total}h_{ab}=p_{m}+p_{G}\;,
 \label{eq41}\\
&& q^{total}_{a}=-T_{bc}^{total}h^{b}_{a}u^{c}=q_{a}^{m}+q^{G}_{a}\;,\\
&& \pi^{total}_{ab}=T_{cd}^{total}h_{a}^{c}h^{d}_{c}=\pi^{m}_{ab}+\pi^{G}_{ab}\;,
\end{eqnarray}
where $\rho_{m}$, $p_{m}$, $q^{m}_{a}$ and $\pi^{m}_{ab}$ represent the effective thermodynamic quantities of matter and $\rho_{G}$, $p_{G}$, $q^{G}_{a}$ and $\pi^{G}_{ab}$ represent the thermodynamic quantities of the Gauss-Bonnet fluid contribution.
For multi component fluid, we have 
\begin{equation}
  T_{ab}^{m}=\sum_{i}T^{i}_{ab}\;,\\
\end{equation}
where
\begin{eqnarray}
 &&T_{ab}^{i}=\rho_{i}u_{a}^{i}u^{i}_{b}+p_{i}h^{i}_{ab}+q^{i}_{a}u^{i}_{b}+q^{i}_{b}u^{i}_{a}+\pi^{i}_{ab}\;,\\
&& h^{i}_{ab}=g_{ab}+u_{a}^{i}u_{b}^{i}\;,
\end{eqnarray}
$u^{i}_{a}$ being the normalized $4$-velocity vectorfor the $i^{th}$ component.
Decomposing the matter stress energy momentum tensor with respect to the $4$ velocity $u^{a}$ gives the following thermodynamic quantities
\begin{eqnarray}
&& \rho_{m}=T^{m}_{ab}u^{a}u^{b}=\sum^{N}_{i=1}\rho_{i}\;,\\
&& p_{m}=\frac{1}{3}T^{m}_{ab}h_{ab}=\sum^{N}_{i=1}p_{i}\;,\\
&& q^{m}_{a}=-T^{m}_{bc}h^{b}_{c}u^{c}=\sum^{N}_{i=1}\left(\rho_{i}+p_{i}\right)V^{i}_{a}\;,\\
&& \pi^{m}_{ab}=T^{m}_{cd}h^{c}_{a}h^{d}_{b}=0\;.
\end{eqnarray}
There are two different frame:The particle frame also known as Eckart choice where $u^{a}=u_{N}^{a}$, an observer $u_{N}$ sees no particle drift and the Energy frame $u^{a}=u_{E}^{a}$ also known as the Landau choice, an observer measures no energy flux $q_{a}=q^{E}_{a}=0$ along the flow line. Choosing a relevant frame is important in the covariant formulation of perturbation theories \cite{dunsby1992covariant,abebe2012covariant, dunsby1992cosmological}. In this paper we choose the fluid flow vector $u^{a}$ to coincide with the energy frame $u_{E}^{a}$ so that the exact FRW models will be characterized by a vanishing shear and vorticity of $u^{a}$ and all spatial gradient variables orthogonal to $u^{a}$ of any scalar quantity.

By Choosing the energy frame, we can set $q^{i}_{a}=q^{a}_{Ei}=0$ and for a perfect fluid, a unique hydrodynamic $4$-velocity $u^{a}$ can be defined for the fluid flow so that there is no energy flux and no anisotropic pressure, hence the energy momentum tensor is presented as 
\begin{equation}
T_{ab}=\rho u_{a}u_{b}+p h_{ab},
\end{equation}
where $\rho$ and $p$ are related by the equation of state
\begin{equation}
 p=p(\rho,s),
\end{equation}
$s$ represent the entropy,
 and for the component fluid, we have 
\begin{equation}
 T_{ab}^{i}=\rho_{i}u_{a}^{i}u^{i}_{b}+p_{i}h^{i}_{ab}.
\end{equation}
The velocity of the $i^{th}$ fluid component  relative to the fundamental observer is defined as
\begin{equation}
 V_{i}^{a}=u_{i}^{a}-u^{a}.
\end{equation}
In the background FRW universe, $V_{i}^{a}=0$ and all perfect fluid components have the same $4$-velocity. Using Stewart-Walker lemma \cite{stewart1974perturbations}, all spatial gradients orthogonal  to $u^{a}$ of any scalar quantity vanish so that
\begin{equation}
 \sigma_{ab}=\omega_{ab}=0, \tilde{\bigtriangledown}_{a}X=0,
\end{equation}
it follows 
\begin{equation}
 X_{a}=\tilde{\bigtriangledown}\rho=0, Y_{a}=\tilde{\bigtriangledown} p, Z_{a}=\tilde{\bigtriangledown} \theta=0
\end{equation}
in the background, then $\rho=\rho(t)$, $p=p(t)$ and $\theta=\theta(t)$.
\subsection{Standard inhomogeneity variables for the total matter}
The inhomogeneities of matter are characterized by
\begin{eqnarray}
&&D^{m}_{a}=\frac{a\tilde{\bigtriangledown}_{a}\rho_{m}}{\rho_{m}}\;,
 Z_{a}=a\tilde{\bigtriangledown}_{a}\theta\;,  Y^{m}_{a}=\tilde{\bigtriangledown}_{a}p_{m}\;,\varepsilon_{a}=\frac{a}{p_{m}}(\frac{\partial p}{\partial s})\tilde{\bigtriangledown}_{a}s\;.
 \end{eqnarray}
Where $a=a(t)$ is the usual FRW cosmological scale factor. 
$D^{m}_{a} $
and $Z_{ a}$ define
the comoving fractional density gradient and comoving gradient of the expansion respectively and can in principle be measured observationally. The subscript $a$ presented in $D^{m}_{a} $
and $Z_{ a}$ is not a scale factor nor a running index but an index. 
We define the effective barotropic
equation of state and speed of sound of the total matter fluid, respectively as
\begin{eqnarray}
&& w=\frac{p}{\rho}\;, c_{s}^{2}=\frac{\partial p}{\partial \rho}\;.
\end{eqnarray}
\subsection{Matter inhomogeneity variables for the components}
The variables characterizing inhomogeneities of matter for the $i^{ th}$ - component fluid
are defined as
\begin{eqnarray}
&&D^{i}_{a}=\frac{a\tilde{\bigtriangledown}_{a}\rho_{i}}{\rho_{i}}\;,  Y^{i}_{a}=\tilde{\bigtriangledown}_{a}p_{m}^{i}\;,\varepsilon^{i}_{a}=\frac{a}{p_{i}}(\frac{\partial p^{i}}{\partial s_{i}})\tilde{\bigtriangledown}_{a}s_{i}\;.
 \end{eqnarray}
 We define the relation
 \begin{equation}
  p\varepsilon_{a}=\sum_{i}p_{i}\varepsilon^{i}_{a}+\frac{1}{2}\sum_{i,j}\frac{h_{i}h_{j}}{h}\left(c_{si}^{2}-c_{sj}^{2}\right)S^{ij}_{a},
  \label{eq43}
 \end{equation}
 that contain the dimensionless variable $\varepsilon_{a} $ that quantifies entropy perturbations in the total fluid, $ h_{i}=\rho_{m}+p_{m} $ for the total matter and $ h_{i}=\rho_{i}+p_{i}$ for the component fluid, $ c_{si}^{2} $ and $ c_{sj}^{2} $ denote the speed of sound of the component matter fluid given as $c^{2}_{si}=\frac{\partial p_{i}}{\partial \rho_{i}}$.
\subsection{Gauss-Bonnet fluid variables}
The Gauss-Bonnet fluid variables are defined as 
\begin{eqnarray}
 &&\mathcal{G}_{a}=a\tilde{\nabla} _{a}G\;,
 \mathsf{G}_{a}=a\tilde{\nabla}  _{a}\dot{G}\;.
 \end{eqnarray}
  The above gradient variables  characterize perturabtions due to Gauss-Bonnet parameter $G$ and its momentum $\dot{G}$ and they descibe the inhomogeneities in the Gauss-Bonnet fluid.
\section{Linear evolution equations in the Energy frame}
We derive linear evolution equations for the defined gradient variables in the energy frame of matter fluid.
\subsection{Total fluid equations}
Total fluid equations characterize the temporal fluctuations of inhomogeneities in perfect cosmological fluid with an equation of state parameter evolving as $\dot{w}=(1+w)(w-c^{2}_{s})\theta$, and are given as 
\begin{eqnarray}
 && \dot{D_{a}^{m}}=-(1+w)Z_{a}+\theta c^{2}_{s}D_{a}^{m}+w\theta \varepsilon_{a}\;,\\
&&\dot{Z}_{a}=\left(-\frac{1}{3}\theta^{2}+\frac{3}{2}\left(w-1\right)\rho_{m}-\frac{1}{2} f+\frac{1}{2}Gf'-\frac{2}{3}\theta^{2}\ddot{G}f''+\frac{2}{9}\theta^{3}\dot{G}f''-\frac{3}{2\theta}G \dot{G}f''-\frac{2}{3}\theta^{2}\dot{G}^{2}f''' \right) \left(\frac{c^{2}_{s}}{1+w}D^{m}_{a}\right)\nonumber\\
&&\quad\quad-\frac{c^{2}_{s}}{1+w}\tilde{\bigtriangledown}^{2}_{a}D^{m}_{a}  +\{f''(\theta f'-G+\frac{3\dot{G}}{2\theta}) -\frac{3G\dot{G}f'''}{2\theta}-\frac{3f''\dot{G}}{2\theta}-\frac{2}{3}\theta^{2}\dot{G}^{2}f^{iv}-\frac{2}{3}\theta^{2}\ddot{G}f'''\nonumber\\
&&\quad\quad  +\frac{f'''}{f''}(\frac{9G}{16\theta^{2}}+\frac{\theta^{2}}{6}+\frac{1-3w}{2}\rho_{m}-f+Gf'-\frac{2\theta^{2}\dot{G}^{2}}{3}f''')+\frac{1}{2}f''
 +\frac{2}{9}\theta^{3}\dot{G}f'''\}\mathcal{G}_{a}\nonumber\\
 &&\quad\quad+\left(-\frac{1}{3}\theta^{2}-\frac{1}{2}\left(1+3w\right)\rho_{m}-\frac{1}{2} f+\frac{1}{2}Gf'-\frac{2}{3}\theta^{2}\ddot{G}f''+\frac{2}{9}\theta^{3}\dot{G}f''-\frac{3}{2\theta}G\dot{G}f''-\frac{2}{3}\theta^{2}(\dot{G})^{2}f''' \right)\left(\frac{w}{1+w}\varepsilon_{a}\right)\nonumber\\
 &&\quad\quad-\frac{w}{1+w}\tilde{\bigtriangledown}^{2}_{a}\varepsilon_{a}
  +\{\frac{2}{9}\theta^{3}f''+\frac{3G}{2\theta}f''+\frac{4\theta^{2}\dot{G}}{3}f''' +\frac{2}{9}\theta^{3}f''-\frac{4}{3}\theta^{2}\dot{G}f'''-\frac{3f''G}{2\theta}\}\mathsf{G}_{a}\nonumber\\
  &&\quad\quad+\{\frac{9}{4\theta^{3}}+\frac{1-3w}{\theta}\rho_{m}
  +\frac{G}{2\theta}f'+f''(\frac{2\theta^{2}\dot{G}}{9}-\frac{9G\dot{G}}{2\theta^{2}})-\frac{1}{2\theta}f-\frac{2}{3}\theta +\frac{2}{3}\theta^{2}\dot{G}f''
 +\frac{3G\dot{G}f''}{2\theta^{2}}\nonumber\\
 &&\quad\quad -\frac{4}{3}\theta \dot{G}^{2}f'''-\frac{4}{3}\theta\ddot{G}f'' \}Z_{a}\;,\\
  &&\dot{\mathcal{G}_{a}}=\mathsf{G}_{a}+\frac{c^{2}_{s}}{1+w}\dot{G}D_{a}^{m}-w\dot{G}\varepsilon_{a}\;,\\
&&  \dot{\mathsf{G}_{a}}=\{ \frac{3(1-3w)}{4}\frac{\rho_{m}}{\theta^{2}f''}-\frac{c^{2}_{s}}{(1+w)}\ddot{G}\}D^{m}_{a}+\{-\frac{\theta}{3}-\frac{9}{4}\frac{G}{\theta^{3}}-2\frac{f'''\dot{G}}{f''}\}\mathsf{G}_{a}+\{\frac{1}{\theta^{2}f''}(\frac{27}{32\theta^{2}}+\frac{3}{2}f')\nonumber\\
&&\quad\quad
  +\frac{f'''}{\theta^{2}f''}\left(-\frac{27}{32}\frac{G}{\theta^{2}}-\frac{\theta^{2}}{4}
  -\frac{3(1-3w)}{4}\rho_{m}+\frac{3}{2}f-\frac{3}{2}Gf'+\dot{G}^{2}\theta^{2}f'''\right)-\frac{3}{2\theta}f'+\frac{3}{2\theta^{2}}G\nonumber\\
  &&\quad\quad
  -\frac{9}{4\theta}^{3}\dot{G}-\frac{\dot{G}^{2}f^{iv}}{f''}\}\mathcal{G}_{a}
  +\{\frac{1}{\theta^{2}f''}\left(-\frac{27}{8\theta^{3}}-\frac{3(1-3w)\rho_{m}}{2\theta}+\frac{3}{\theta}f-\frac{3}{\theta}Gf'\right)-\frac{\dot{G}}{3}+\frac{27}{4\theta^{4}}G\dot{G}\}Z_{a}\nonfrenchspacing\\
  &&+\frac{w\ddot{G}}{1+w}\varepsilon_{a}\;.
 \end{eqnarray}
\subsection{Component equations}
For the component matter and velocity fluctuations,
 the equations describing the evolution of the individual fluid component fluctuations  are given as 
 \begin{eqnarray}
 && \dot{D}_{a}^{i}=-(1+w_{i})Z_{a}+\left(w_{i}- c^{2}_{si}\right)\theta D^{i}_{a}+\left(-\frac{1+w_{i}}{1+w}c^{2}_{s}\theta \right)D_{a}+\left(-(1+w_{i})\frac{w}{1+w}\theta \right)\varepsilon_{a}\nonumber\\&&\quad\quad-\theta w_{i}\varepsilon^{i}_{a}-(1+w_{i})a\tilde{\bigtriangledown}_{a}\tilde{\bigtriangledown}^{b}V^{i}_{b}\;,\\
 && \dot{V}^{i}_{a}-\left(3c^{2}_{si}-1\right)\frac{\theta}{3}V^{i}_{a}=\frac{1}{a(1+w)}\left(c^{2}_{s} D_{a}+w \varepsilon_{a}\right)-\frac{1}{a(1+w_{i})}\left(c^{2}_{si}D^{i}_{a}+w_{i}\varepsilon^{i}_{a}\right).
\end{eqnarray}
The equations involving the gradients of the inhomogeneities in the expansion and Gauss-Bonnet variables ($Z_{a}$, $\mathcal{G}_{a}$, $\mathsf{G}_{a}$) remain the same as in the total fluid components since they do not represent the individual components of matter in the fluids.
\subsection{Relative equations}
We define the variables that relate features of different components of the fluid and derive their evolution equations
\begin{eqnarray}
 &&V^{ij}_{a}=V^{i}_{a}-V^{j}_{a}\;,\\
&& S^{ij}_{a}=\frac{\rho_{i}D_{i}}{h_{i}}-\frac{\rho_{j}D^{i}_{a}}{h_{j}}\;,
\end{eqnarray}
with $h=\rho+p$ and $h_{i}=\rho_{i}+p_{i}$.\\
These quantities help us to distinguish adiabatic and isothermal perturbations.
Their evolution equations are given as
\begin{eqnarray}
 &&\dot{S}^{ij}_{a}=-a\tilde{\bigtriangledown}_{a}\tilde{\bigtriangledown}^{b}V^{ij}_{b}\;,\\
&& \dot{V}^{ij}_{a}=\left(c^{2}_{si}-c^{2}_{sj} \right)\theta V^{i}_{a}+\left(3c^{2}_{sj}-1\right)\frac{\theta}{3}V^{ij}_{a}+\left(\frac{c^{2}_{sj}}{a(1+w_{i})}-\frac{c^{2}_{si}}{a(1+w_{i})}\right)D^{i}_{a}\nonumber\\
&&\quad\quad-\frac{w_{i}}{a(1+w_{i})}\varepsilon^{i}_{a}-\frac{c^{2}_{sj}}{a}S^{ij}_{a}+\frac{w_{j}}{a(1+w_{j})}\varepsilon^{j}_{a}\;.
\end{eqnarray}
The quantities we descibe so far are general evolution equations containing both the scalar and vector  parts. In the next section, We present the spherically symetric, scalar density perturbations since it is belived that structure formation on cosmological scales follow spherical clustering.
\section{Scalar equations}
We extract the scalar part of the perturbation vectorial gradients by taking the divergence of the gradiant quantities
\subsection{Scalar gradient variables}
On the basis of the above decomposition scheme, we have\\
\begin{eqnarray} &&\Delta^{m}=a\tilde{\bigtriangledown}^{a}D^{m}_{a},
  Z=a\tilde{\bigtriangledown}^{a}Z_{a},
  \mathcal{G}=a\tilde{\bigtriangledown}^{a} \mathcal{G}_{a},
  G=a\tilde{\bigtriangledown}^{a} G_{a}, 
  \varepsilon=a \tilde{\bigtriangledown}^{a}\varepsilon_{a},\nonumber\\&&\quad\quad
  \Delta_{m}^{i}=a \tilde{\bigtriangledown}^{a}D^{i}_{a},
  V_{ij}=a \tilde{\bigtriangledown}^{a} V^{ij}_{a},
  S_{ij}=a \tilde{\bigtriangledown}^{a} S^{ij}_{a} and
  V_{i}=a \tilde{\bigtriangledown}^{a}V^{i}_{a}.
  \end{eqnarray}
\subsection{Linear evolution equations for Scalar variables}
The scalar gradient variables describing the total fluid evolve as 
  \begin{eqnarray}
   &&\dot{\Delta^{m}}=-(1+w)Z+c^{2}_{s}\theta                                                                                      \Delta^{m}+w\theta \varepsilon\;,\\
 && \dot{Z}=\left(-\frac{1}{3}\theta^{2}-\frac{1}{2}\left(1+3w\right)\rho_{m}-\frac{1}{2} f+\frac{1}{2}Gf'-\frac{2}{3}\theta^{2}\ddot{G}f''+\frac{2}{9}\theta^{3}\dot{G}f''-\frac{3}{2\theta}G \dot{G}f''-\frac{2}{3}\theta^{2}\dot{G}^{2}f''' \right) \left(\frac{c^{2}_{s}}{1+w}\Delta^{m}\right)\nonumber\\
 &&\quad\quad-(1-3w)\rho_{m}\Delta^{m}-\frac{c^{2}_{s}}{1+w}\tilde{\bigtriangledown}^{2}_{a}\Delta^{m}  +\{f''(\theta f'-G+\frac{3\dot{G}}{2\theta}) -\frac{3G\dot{G}f'''}{2\theta}-\frac{3f''\dot{G}}{2\theta}-\frac{2}{3}\theta^{2}\dot{G}^{2}f^{iv}\nonumber\\
 &&\quad\quad-\frac{2}{3}\theta^{2}\ddot{G}f'''
  +\frac{f'''}{f''}(\frac{9G}{16\theta^{2}}+\frac{\theta^{2}}{6}+\frac{1-3w}{2}\rho_{m}-f+Gf'-\frac{2\theta^{2}\dot{G}^{2}}{3}f''')+\frac{1}{2}f''
 +\frac{2}{9}\theta^{3}\dot{G}f'''\}\mathcal{G}\nonumber\\
 &&\quad\quad+\left(-\frac{1}{3}\theta^{2}-\frac{1}{2}\left(1+3w\right)\rho_{m}-\frac{1}{2} f+\frac{1}{2}Gf'-\frac{2}{3}\theta^{2}\ddot{G}f''+\frac{2}{9}\theta^{3}\dot{G}f''-\frac{3}{2\theta}G\dot{G}f''-\frac{2}{3}\theta^{2}(\dot{G})^{2}f''' \right)\left(\frac{w}{1+w}\varepsilon \right)\nonumber\\
 &&\quad\quad-\frac{w}{1+w}\bigtriangledown^{2}\varepsilon
  +\frac{4}{9}\theta^{3}f''\mathsf{G}+\{\frac{9}{4\theta^{3}}+\frac{1-3w}{\theta}\rho_{m}
  +\frac{G}{2\theta}f'+f''(\frac{2\theta^{2}\dot{G}}{9}-\frac{9G\dot{G}}{2\theta^{2}})-\frac{1}{2\theta}f-\frac{2}{3}\theta \nonumber\\
 &&\quad\quad+\frac{2}{3}\theta^{2}\dot{G}f''
 +\frac{3G\dot{G}f''}{2\theta^{2}}
  -\frac{4}{3}\theta \dot{G}^{2}f'''-\frac{4}{3}\theta\ddot{G}f'' \}Z\;,\\
&&\dot{ \mathcal{G}}=\mathsf{G}+\frac{c^{2}_{s}}{1+w}\dot{G}\Delta^{m}-w\dot{G}\varepsilon\;,\\
 && \dot{\mathsf{G}}=\{ \frac{3(1-3w)}{4}\frac{\rho_{m}}{\theta^{2}f''}-\frac{c^{2}_{s}}{(1+w)}\ddot{G}\}\Delta^{m}+\{-\frac{\theta}{3}-\frac{9}{4}\frac{G}{\theta^{3}}-2\frac{f'''\dot{G}}{f''}\}\mathsf{G}+\{\frac{1}{\theta^{2}f''}(\frac{27}{32\theta^{2}}+\frac{3}{2}f')\nonumber\\
 &&\quad\quad
  +\frac{f'''}{\theta^{2}f''}\left(-\frac{27}{32}\frac{G}{\theta^{2}}-\frac{\theta^{2}}{4}
  -\frac{3(1-3w)}{4}\rho_{m}+\frac{3}{2}f-\frac{3}{2}Gf'+\dot{G}^{2}\theta^{2}f'''\right)-\frac{3}{2\theta}f'+\frac{3}{2\theta^{2}}G\nonumber\\&&\quad\quad
  -\frac{9}{4\theta}^{3}\dot{G}-\frac{\dot{G}^{2}f^{iv}}{f''}\}\mathcal{G}
  +\{\frac{1}{\theta^{2}f''}\left(-\frac{27}{8\theta^{3}}-\frac{3(1-3w)\rho_{m}}{2\theta}+\frac{3}{\theta}f-\frac{3}{\theta}Gf'\right)-\frac{\dot{G}}{3}+\frac{27}{4\theta^{4}}G\dot{G}\}Z\nonumber\\
&&  +\ddot{G}\frac{w}{(1+w)}\varepsilon\;.
 \end{eqnarray}
The scalar variables describing components inhomogeneities and interactions in the fluids are given as
\begin{eqnarray}
 &&\dot{\Delta}_{m}^{i}=-(1+w_{i})Z+\left(w_{i}- c^{2}_{si}\right) \theta \Delta_{m}^{i}
 -\left(\frac{1+w_{i}}{1+w}c^{2}_{s}\right)
 \theta \Delta_{m}-(1+w_{i})\frac{w}{1+w}\theta  \varepsilon
 -w_{i} \theta \varepsilon^{i}\nonumber\\&& \quad\quad-a\left(1+w_{i} \right) \bigtriangledown^{2}V_{i}\;,\\
&& \dot{V}_{i}=\left(c^{2}_{si}-\frac{1}{3}\right)\theta V_{i}+\frac{1}{a(1+w)}\left(c^{2}_{s}\Delta_{m} +w \varepsilon \right)-\frac{1}{a(1+w_{i})}\left(c^{2}_{si}\Delta_{m}^{i}+w_{i}\varepsilon^{i}\right)\;,\\
 &&\dot{V}_{ij}=\left(c^{2}_{si}-c^{2}_{sj} \right)\theta V_{i}+\left(c^{2}_{sj}-\frac{1}{3}\right)\theta V_{ij}-\left(\frac{c^{2}_{si}}{a(1+w_{i})}-\frac{c^{2}_{sj}}{a(1+w_{i})}\right)\Delta_{m}^{i}-\frac{w_{i}}{a(1+w_{i})}\varepsilon^{i}\nonumber\\&&\quad\quad-\frac{c^{2}_{sj}}{a}S_{ij}+\frac{w_{j}}{a(1+w_{j})}\varepsilon_{j}\;,\\ 
&&  \dot{S}_{ij}=-a\bigtriangledown^{2}V_{ij}.
 \end{eqnarray}
\section{Second order equations}
All the derived first order equations  can be reduced to a set of linearly independent second order equations for simplicity and making comparison to general relativity easier. By making second derivaive  of the linear evolution equations, we have
\newpage
\begin{eqnarray}
  &&\ddot{\Delta}^{m}=-(-\theta^{2}+
  \frac{9}{4\theta^{2}}-\dot{\theta}+\frac{(w^{2}+\frac{2}{3}w-\frac{1}{3})}{c^{2}_{s}}\rho_{m}+(\frac{1}{2}-\frac{9}{2}w)\rho_{m}- f+Gf'\nonumber\\&&\quad\quad-2\theta^{2}\ddot{G}f''+\frac{8}{9}\theta^{3}\dot{G}f''-2\theta^{2}\dot{G}^{2}f''' 
  +f''(\frac{2\theta^{3}\dot{G}}{9}-\frac{9G\dot{G}}{2\theta})
 ) \left(c^{2}_{s}\Delta^{m}\right)\nonumber\\&&\quad\quad
  -\{\frac{9}{4\theta^{3}}-(\frac{2}{3}+c^{2}_{s})\theta+\frac{1-3w}{\theta}\rho_{m}
  +\frac{G}{2\theta}f'+f''(\frac{2\theta^{2}\dot{G}}{9}-\frac{9G\dot{G}}{2\theta^{2}})-\frac{1}{2\theta}f\nonumber\\
  &&\quad\quad +\frac{2}{3}\theta^{2}\dot{G}f''
 +\frac{3G\dot{G}f''}{2\theta^{2}}
  -\frac{4}{3}\theta \dot{G}^{2}f'''-\frac{4}{3}\theta\ddot{G}f'' \}
  \dot{\Delta}^{m}  
  +c^{2}_{s}\tilde{\bigtriangledown}^{2}_{a}\Delta^{m}\nonumber\\
  &&\quad\quad-(1+w)\{f''\left(\frac{1}{2}+\theta f'-G+\frac{3\dot{G}}{2\theta}\right)-\frac{3G\dot{G}f'''}{2\theta}-\frac{3f''\dot{G}}{2\theta}-\frac{2}{3}\theta^{2}\dot{G}^{2}f^{iv}\nonumber\\
  &&\quad\quad-\frac{2}{3}\theta^{2}\ddot{G}f'''
  +\frac{f'''}{f''}\left(\frac{9G}{16\theta^{2}}+\frac{\theta^{2}}{6}+\frac{1-3w}{2}\rho_{m}-f+Gf'-\frac{2\theta^{2}\dot{G}^{2}}{3}f'''\right)
 +\frac{2}{9}\theta^{3}\dot{G}f'''\}\mathcal{G}
 \nonumber\\&&\quad\quad-\{\frac{9}{4\theta^{2}}-\theta^{2}-\dot{\theta}+(\frac{1}{2}-\frac{9}{2} w)\rho_{m}
  +Gf'+f''\left(\frac{10}{9}\theta^{3}\dot{G}-\frac{9}{2\theta}G\dot{G}-2\theta^{2}\ddot{G}\right)
  \nonumber\\
  &&\quad\quad-2\theta^{2} \dot{G}^{2}f''' \} \left(w\varepsilon \right)+w\theta \dot{\varepsilon}+w\bigtriangledown^{2}\varepsilon
  -(1+w)\frac{4}{9}\theta^{3}f''\mathsf{G}\;,\\
 && \ddot{ \mathcal{G}}=\{ -\frac{c^{2}_{s}\theta \dot{G}}{3(1+w)}+\frac{27c^{2}_{s}}{4(1+w)\theta^{3}}G\dot{G}+\frac{1}{\theta^{2}f''}(\frac{3(1-3w)}{4}\rho_{m}-\frac{27}{8\theta^{2}}\frac{c^{2}_{s}}{1+w}-\frac{3(1-3w)}{2}\frac{c^{2}_{s}}{1+w}\rho_{m}\nonumber\\&&\quad\quad+3\frac{c^{2}_{s}}{1+w}f-3\frac{c^{2}_{s}}{1+w}Gf')\}\Delta^{m}\nonumber\\
  &&\quad\quad
  +\{\frac{1}{\theta^{2}f''}\left(-\frac{27}{8(1+w)\theta^{3}}-\frac{3(1-3w)}{2(1+w)\theta}\rho_{m}+\frac{3}{(1+w)\theta}f-\frac{3}{(1+w)\theta}Gf'\right)
  \nonumber\\
  &&\quad\quad
  \frac{3c^{2}_{s}-1}{3(1+w)}\dot{G}+\frac{27}{4(1+w)\theta^{4}}G\dot{G}\}\dot{\Delta}^{m}+\{-\frac{\theta}{3}-\frac{9}{4}\frac{G}{\theta^{3}}-2\dot{G}\frac{f'''}{f''}\}\mathsf{G}\nonumber\\
  && \quad\quad
  +\{\frac{1}{\theta^{2}f''}\left(-\frac{27}{8\theta^{2}}\frac{w}{1+w}-\frac{3(1-3w)}{2}\frac{w}{1+w}\rho_{m}+\frac{3w}{1+w}f-\frac{3w}{1+w}Gf'\right)\nonumber\\
  &&\quad\quad-\frac{w\theta \dot{G}}{3(1+w)}+\frac{27w}{4(1+w)\theta^{3}}G\dot{G}+\ddot{G}\frac{w}{(1+w)}-w\ddot{G}\}\varepsilon  -w\dot{G}\dot{\varepsilon}
  \nonumber\\
  &&\quad\quad+\{\frac{1}{\theta^{2}f''}(\frac{27}{32\theta^{2}}+\frac{3}{2}f')-\frac{3}{2\theta}f'+\frac{3}{2\theta^{2}}G
  -\frac{9}{4\theta}^{3}\dot{G}-\frac{\dot{G}^{2}f^{iv}}{f''}\nonumber\\
  &&\quad\quad
  +\frac{f'''}{\theta^{2}f''}\left(-\frac{27}{32}\frac{G}{\theta^{2}}-\frac{\theta^{2}}{4}
  -\frac{3(1-3w)}{4}\rho_{m}+\frac{3}{2}f-\frac{3}{2}Gf'+\dot{G}^{2}\theta^{2}f'''\right)\}\mathcal{G}\;,\\
 && \ddot{\Delta}_{m}^{i}=-\frac{(1+w_{i})c^{2}_{s}}{1+w}(-\theta^{2}+\frac{9}{4\theta^{2}}+
 \dot{\theta}+(\frac{1}{2}-\frac{9}{2}w)\rho_{m}-f+Gf'-2\theta^{2}\dot{G}^{2}f'''  -\frac{(1-3w)(1+w)}{c^{2}_{s}}\rho_{m}\nonumber\\
 &&\quad\quad
+f''(\frac{10\theta^{3}\dot{G}}{9}-\frac{9G\dot{G}}{2\theta}-2\theta^{2}\ddot{G})
 )\Delta^{m}\nonumber\\
 &&\quad\quad
  -\frac{(1+w_{i})}{1+w}\{\frac{9}{4\theta^{3}}+c^{2}_{s}
 \theta+\frac{1-3w}{\theta}\rho_{m}
  +\frac{G}{2\theta}f'+f''(\frac{8\theta^{2}\dot{G}}{9}-\frac{6G\dot{G}}{2\theta^{2}}-\frac{4}{3}\theta\ddot{G})-\frac{1}{2\theta}f-\frac{2}{3}\theta 
  -\frac{4}{3}\theta \dot{G}^{2}f'''\}\dot{\Delta}^{m}\nonumber\\
  &&\quad\quad
  -(1+w_{i})\{f''(\theta f'-G+\frac{1}{2}) +f'''(-\frac{3G\dot{G}}{2\theta}-\frac{2}{3}\theta^{2}\ddot{G}+\frac{2}{9}\theta^{3}\dot{G})\nonumber\\
  &&\quad\quad-\frac{2}{3}\theta^{2}\dot{G}^{2}f^{iv}
  +\frac{f'''}{f''}\left(\frac{9G}{16\theta^{2}}+\frac{\theta^{2}}{6}+\frac{1-3w}{2}\rho_{m}-f+Gf'-\frac{2\theta^{2}\dot{G}^{2}}{3}f'''\right)\}\mathcal{G}\nonumber\\
  &&\quad\quad-\frac{(1+w_{i})w}{1+w}(-\theta^{2}+\frac{9}{4\theta^{2}}+\dot{\theta}+\left(\frac{1}{2}-\frac{9}{2} w\right)\rho_{m}- f+Gf'-2\theta^{2}\dot{G}^{2}f'''\nonumber\\
  &&\quad\quad
  +f''(\frac{10\theta^{3}\dot{G}}{9}-\frac{9G\dot{G}}{2\theta}-2\theta^{2}\ddot{G})) \varepsilon
  -(1+w_{i})\frac{4}{9}\theta^{3}f''\mathsf{G}
  +\left(w_{i}- c^{2}_{si}\right) \dot{\theta} \Delta_{m}^{i}+\left(w_{i}- c^{2}_{si}\right) \theta\dot{ \Delta}_{m}^{i}\nonumber\\
  &&\quad\quad
 -(1+w_{i})\frac{w}{1+w}\theta  \dot{\varepsilon}
 -w_{i} \dot{\theta} \varepsilon^{i}-w_{i} \theta \dot{\varepsilon}^{i}-a(1+w_{i})\left(c^{2}_{si}-\frac{2}{3}\right)\theta \bigtriangledown^{2} V_{i}+c^{2}_{si}\bigtriangledown^{2}\Delta_{m}^{i}+w_{i}\bigtriangledown^{2}\varepsilon^{i}.  
 \end{eqnarray}
 The second order evolution equations governing the propagation of entropy perturbations for a general $\varepsilon$ or $S_{ij}$ will be presented for radiation-dust systems.
\section{Harmonic analysis}
The harmonic decomposition technique  is used in such a way that the evolution equations can be converted into ordinary differential equations for each mode \cite{abebe2015breaking,carloni2006gauge,dunsby1992covariant,munyeshyaka2021cosmological,ntahompagaze2020multifluid}. On an almost FRW space-time, we consider the differential equation of the form
\begin{equation}
 \ddot{X}+A_{d}\dot{X}+A_{r}X=A_{s}(Y,\dot{Y}),
 \label{eq67}
\end{equation}
with $A_{d}$, $A_{r}$ and $A_{s}$ represent damping, restoring and source terms respectively. The separation of variable technique for solving Eq. \ref{eq67}  can be applied such that $X(\vec{x})$ and $Y(\vec{x})$ depend on spatial variable $x$ only and $X(t)$ and $Y(t)$ depend on time variable $t$ so that 
\begin{equation}
 X(\vec{x},t)=X(\vec{x}).X(t),
\end{equation}
and
\begin{equation}
 Y(\vec{x},t)=Y(\vec{x}).Y(t).
\end{equation}
To make a summation over a wavenumber $k$, we use the eigenfunctions $Q_{k}$ so that
\begin{eqnarray}
 &&X=\sum_{k}X^{k}(t)Q_{k}(x)\;,\\
&& Y=\sum_{k}Y^{k}(t)Q_{k}(x)\;,
\end{eqnarray}
where $Q_{k}$ are the eigenfunctions of the covariant Laplace-Beltrami operator such that
\begin{equation}
 \tilde{\bigtriangledown}  Q=-\frac{k^{2}}{a^{2}}Q.
\end{equation}
The Laplace-Beltrami operator is covariantly constant, it means $\dot{Q}_{k}(x)=0$ and $k=\frac{2\pi a}{\lambda}$ is the order of the harmonic, and $k$ is wavelength.
Applying the harmonic decomposition scheme, the first order total and component fluid equations can be represented as
\begin{eqnarray}                                                                                                    &&\dot{\Delta_{m}^{k}}=-(1+w)Z^{k}+c^{2}_{s}\theta \Delta_{m}^{k}+w\theta \varepsilon^{k}\;,\\                                                                                                                      &&\dot{Z^{k}}=\left(-\frac{1}{3}\theta^{2}+\frac{3}{2}\left(w-1\right)\rho_{m}-\frac{1}{2} f+\frac{1}{2}Gf'-\frac{2}{3}\theta^{2}\ddot{G}f''+\frac{2}{9}\theta^{3}\dot{G}f''-\frac{3}{2\theta}G \dot{G}f''-\frac{2}{3}\theta^{2}\dot{G}^{2}f''' \right) \left(\frac{c^{2}_{s}}{1+w}\Delta_{m}^{k}\right)\nonumber\\
 &&\quad\quad
+\frac{c^{2}_{s}}{1+w}\frac{k^{2}}{a^{2}}\Delta^{k}  +\{f''(\theta f'-G+\frac{3\dot{G}}{2\theta}) -\frac{3G\dot{G}f'''}{2\theta}-\frac{3f''\dot{G}}{2\theta}-\frac{2}{3}\theta^{2}\dot{G}^{2}f^{iv}-\frac{2}{3}\theta^{2}\ddot{G}f'''\nonumber\\
&&\quad\quad
  +\frac{f'''}{f''}(\frac{9G}{16\theta^{2}}+\frac{\theta^{2}}{6}+\frac{1-3w}{2}\rho_{m}-f+Gf'-\frac{2\theta^{2}\dot{G}^{2}}{3}f''')+\frac{1}{2}f''
 +\frac{2}{9}\theta^{3}\dot{G}f'''\}\mathcal{G}^{k}\nonumber\\
 &&\quad\quad+\left(-\frac{1}{3}\theta^{2}-\frac{1}{2}\left(1+3w\right)\rho_{m}-\frac{1}{2} f+\frac{1}{2}Gf'-\frac{2}{3}\theta^{2}\ddot{G}f''+\frac{2}{9}\theta^{3}\dot{G}f''-\frac{3}{2\theta}G\dot{G}f''-\frac{2}{3}\theta^{2}(\dot{G})^{2}f''' \right)\left(\frac{w}{1+w}\varepsilon^{k} \right)\nonumber\\
 &&\quad\quad-\frac{w}{1+w}\bigtriangledown^{2}\varepsilon
  +\{\frac{4}{9}\theta^{3}f''\}\mathsf{G}^{k}+\{\frac{9}{4\theta^{3}}+\frac{1-3w}{\theta}\rho_{m}
  +\frac{G}{2\theta}f'+f''(\frac{2\theta^{2}\dot{G}}{9}-\frac{9G\dot{G}}{2\theta^{2}})\nonumber\\
  &&\quad\quad-\frac{1}{2\theta}f-\frac{2}{3}\theta +\frac{2}{3}\theta^{2}\dot{G}f''
 +\frac{3G\dot{G}f''}{2\theta^{2}}
  -\frac{4}{3}\theta \dot{G}^{2}f'''-\frac{4}{3}\theta\ddot{G}f'' \}Z^{k}\;,\\
&&\dot{ \mathcal{G}}^{k}=\mathsf{G}^{k}+\frac{c^{2}_{s}}{1+w}\dot{G}\Delta^{k}-w\dot{G}\varepsilon^{k}\;,\\
 &&  \dot{\mathsf{G}}^{k}=\{ \frac{3(1-3w)}{4}\frac{\rho_{m}}{\theta^{2}f''}-\frac{c^{2}_{s}}{(1+w)}\ddot{G}\}\Delta^{k}+\{-\frac{\theta}{3}-\frac{9}{4}\frac{G}{\theta^{3}}-2\frac{f'''\dot{G}}{f''}\}\mathsf{G}^{k}+\{\frac{1}{\theta^{2}f''}(\frac{27}{32\theta^{2}}+\frac{3}{2}f')\nonumber\\
 &&\quad\quad
  +\frac{f'''}{\theta^{2}f''}\left(-\frac{27}{32}\frac{G}{\theta^{2}}-\frac{\theta^{2}}{4}
  -\frac{3(1-3w)}{4}\rho_{m}+\frac{3}{2}f-\frac{3}{2}Gf'+\dot{G}^{2}\theta^{2}f'''\right)-\frac{3}{2\theta}f'+\frac{3}{2\theta^{2}}G\nonumber\\
  &&\quad\quad
  -\frac{9}{4\theta}^{3}\dot{G}-\frac{\dot{G}^{2}f^{iv}}{f''}\}\mathcal{G}^{k}\nonumber\\
  &&\quad\quad
  +\left[\frac{1}{\theta^{2}f''}\left(-\frac{27}{8\theta^{3}}-\frac{3(1-3w)\rho_{m}}{2\theta}+\frac{3}{\theta}f-\frac{3}{\theta}Gf'\right)-\frac{\dot{G}}{3}+\frac{27}{4\theta^{4}}G\dot{G}\right]Z^{k}
  +\ddot{G}\frac{w}{(1+w)}\varepsilon^{k}\;,\\
&& \dot{\Delta}_{k}^{i}=-(1+w_{i})Z^{k}+\left(w_{i}- c^{2}_{si}\right) \theta \Delta^{k}_{i}
 -\left(\frac{1+w_{i}}{1+w}c^{2}_{s}\right)
 \theta \Delta^{k}-(1+w_{i})\frac{w}{1+w}\theta  \varepsilon^{k}\nonumber\\
 &&\quad\quad
 -w_{i} \theta \varepsilon_{i}^{k}+\left(1+w_{i} \right) \frac{k^{2}}{a} V^{k}_{i}\;,\\
&& \dot{V}^{k}_{i}=\left(c^{2}_{si}-\frac{1}{3}\right)\theta V^{k}_{i}+\frac{1}{a(1+w)}\left(c^{2}_{s}\Delta^{k} +w \varepsilon^{k} \right)-\frac{1}{a(1+w_{i})}\left(c^{2}_{si}\Delta^{k}_{i}+w_{i}\varepsilon_{i}^{k}\right)\;,\\
&&\dot{V}^{k}_{ij}=\left(c^{2}_{si}-c^{2}_{sj} \right)\theta V^{k}_{i}+\left(c^{2}_{sj}-\frac{1}{3}\right)\theta V^{k}_{ij}-\left(\frac{c^{2}_{si}}{a(1+w_{i})}-\frac{c^{2}_{sj}}{a(1+w_{i})}\right)\Delta^{k}_{i}\nonfrenchspacing\\
&&\quad\quad-\frac{w_{i}}{a(1+w_{i})}\varepsilon_{i}^{k}-\frac{c^{2}_{sj}}{a}S^{k}_{ij}+\frac{w_{j}}{a(1+w_{j})}\varepsilon^{k}_{j}\;,\\
&&  \dot{S}^{k}_{ij}=\frac{k^{2}}{a^{2}} V^{k}_{ij}\;.
 \end{eqnarray} 
 The harmonically decomposed set of second order equations for total fluid and component fluids are given respectively as
 \begin{eqnarray}
  &&\ddot{\Delta}_{m}^{k}=-(-\theta^{2}+
  \frac{9}{4\theta^{2}}-\dot{\theta}+\frac{(w^{2}+\frac{2}{3}w-\frac{1}{3})}{c^{2}_{s}}\rho_{m}+(\frac{1}{2}-\frac{9}{2}w)\rho_{m}- f+Gf'-2\theta^{2}\ddot{G}f''\nonumber\\
  &&\quad\quad+\frac{8}{9}\theta^{3}\dot{G}f''-2\theta^{2}\dot{G}^{2}f''' 
  +f''(\frac{2\theta^{3}\dot{G}}{9}-\frac{9G\dot{G}}{2\theta})
 ) \left(c^{2}_{s}\Delta_{m}^{k}\right)
  -\{\frac{9}{4\theta^{3}}-(\frac{2}{3}+c^{2}_{s})\theta\nonumber\\
  &&\quad\quad+\frac{1-3w}{\theta}\rho_{m}
  +\frac{G}{2\theta}f'+f''(\frac{2\theta^{2}\dot{G}}{9}-\frac{9G\dot{G}}{2\theta^{2}})-\frac{1}{2\theta}f +\frac{2}{3}\theta^{2}\dot{G}f''
 +\frac{3G\dot{G}f''}{2\theta^{2}}
  -\frac{4}{3}\theta \dot{G}^{2}f'''\nonumber\\
  &&\quad\quad-\frac{4}{3}\theta\ddot{G}f'' \}
  \dot{\Delta}_{m}^{k}  
  -c^{2}_{s}\frac{k^{2}}{a^{2}}\Delta_{m}^{k}-(1+w)\{f''\left(\frac{1}{2}+\theta f'-G+\frac{3\dot{G}}{2\theta}\right)-\frac{3G\dot{G}f'''}{2\theta}-\frac{3f''\dot{G}}{2\theta}\nonumber\\
  &&\quad\quad-\frac{2}{3}\theta^{2}\dot{G}^{2}f^{iv}-\frac{2}{3}\theta^{2}\ddot{G}f'''
  +\frac{f'''}{f''}\left(\frac{9G}{16\theta^{2}}+\frac{\theta^{2}}{6}+\frac{1-3w}{2}\rho_{m}-f+Gf'-\frac{2\theta^{2}\dot{G}^{2}}{3}f'''\right)\nonumber\\
  &&\quad\quad
 +\frac{2}{9}\theta^{3}\dot{G}f'''\}\mathcal{G}^{k}
 -\{\frac{9}{4\theta^{2}}-\theta^{2}-\dot{\theta}+(\frac{1}{2}-\frac{9}{2} w)\rho_{m}
  +Gf'+f''\left(\frac{10}{9}\theta^{3}\dot{G}-\frac{9}{2\theta}G\dot{G}-2\theta^{2}\ddot{G}\right)\nonumber\\
  &&\quad\quad
  -2\theta^{2} \dot{G}^{2}f''' \} \left(w\varepsilon^{k} \right)+w\theta \dot{\varepsilon}^{k}-w\frac{k^{2}}{a^{2}}\varepsilon^{k}
  -(1+w)\frac{4}{9}\theta^{3}f''\mathsf{G}^{k}\;,\\
  &&\ddot{ \mathcal{G}}^{k}=\{ -\frac{c^{2}_{s}\theta \dot{G}}{3(1+w)}+\frac{27c^{2}_{s}}{4(1+w)\theta^{3}}G\dot{G}+\frac{1}{\theta^{2}f''}(\frac{3(1-3w)}{4}\rho_{m}-\frac{27}{8\theta^{2}}\frac{c^{2}_{s}}{1+w}\nonumber\\&&\quad\quad-\frac{3(1-3w)}{2}\frac{c^{2}_{s}}{1+w}\rho_{m}+3\frac{c^{2}_{s}}{1+w}f-3\frac{c^{2}_{s}}{1+w}Gf')\}\Delta^{k}\nonumber\\&&\quad\quad
  +\{\frac{1}{\theta^{2}f''}\left(-\frac{27}{8(1+w)\theta^{3}}-\frac{3(1-3w)}{2(1+w)\theta}\rho_{m}+\frac{3}{(1+w)\theta}f-\frac{3}{(1+w)\theta}Gf'\right)\nonumber\\
  &&\quad\quad\frac{3c^{2}_{s}-1}{3(1+w)}\dot{G}+\frac{27}{4(1+w)\theta^{4}}G\dot{G}\}\dot{\Delta}^{k}+\{-\frac{\theta}{3}-\frac{9}{4}\frac{G}{\theta^{3}}-2\dot{G}\frac{f'''}{f''}\}\mathsf{G}^{k}\nonumber\\
  &&\quad\quad
  +\{\frac{1}{\theta^{2}f''}\left(-\frac{27}{8\theta^{2}}\frac{w}{1+w}-\frac{3(1-3w)}{2}\frac{w}{1+w}\rho_{m}+\frac{3w}{1+w}f-\frac{3w}{1+w}Gf'\right)\nonumber\\&&\quad\quad-\frac{w\theta \dot{G}}{3(1+w)}+\frac{27w}{4(1+w)\theta^{3}}G\dot{G}+\ddot{G}\frac{w}{(1+w)}-w\ddot{G}\}\varepsilon^{k}  -w\dot{G}\dot{\varepsilon}^{k}\nonumber\\
  &&\quad\quad
  +\{\frac{1}{\theta^{2}f''}(\frac{27}{32\theta^{2}}+\frac{3}{2}f')-\frac{3}{2\theta}f'+\frac{3}{2\theta^{2}}G
  -\frac{9}{4\theta}^{3}\dot{G}-\frac{\dot{G}^{2}f^{iv}}{f''}\nonumber\\
  &&\quad\quad
  +\frac{f'''}{\theta^{2}f''}\left(-\frac{27}{32}\frac{G}{\theta^{2}}-\frac{\theta^{2}}{4}
  -\frac{3(1-3w)}{4}\rho_{m}+\frac{3}{2}f-\frac{3}{2}Gf'+\dot{G}^{2}\theta^{2}f'''\right)\}\mathcal{G}^{k}\;,\\
 &&\ddot{\Delta}^{k}_{i}=-\frac{(1+w_{i})c^{2}_{s}}{1+w}(-\theta^{2}+\frac{9}{4\theta^{2}}+
 \dot{\theta}+(\frac{1}{2}-\frac{9}{2}w)\rho_{m}-f+Gf'-2\theta^{2}\dot{G}^{2}f'''\nonumber\\
&&\quad\quad  -\frac{(1-3w)(1+w)}{c^{2}_{s}}\rho_{m}+f''(\frac{10\theta^{3}\dot{G}}{9}-\frac{9G\dot{G}}{2\theta}-2\theta^{2}\ddot{G})
 )\Delta^{k}\nonumber\\
 &&\quad\quad
  -\frac{(1+w_{i})}{1+w}\{\frac{9}{4\theta^{3}}+c^{2}_{s}
 \theta+\frac{1-3w}{\theta}\rho_{m}
  +\frac{G}{2\theta}f'+f''(\frac{8\theta^{2}\dot{G}}{9}-\frac{6G\dot{G}}{2\theta^{2}}-\frac{4}{3}\theta\ddot{G})\nonumber\\
&&\quad\quad-\frac{1}{2\theta}f-\frac{2}{3}\theta 
  -\frac{4}{3}\theta \dot{G}^{2}f'''\}\dot{\Delta}^{k}
  -(1+w_{i})\{f''(\theta f'-G+\frac{1}{2}) +f'''(-\frac{3G\dot{G}}{2\theta}-\frac{2}{3}\theta^{2}\ddot{G}+\frac{2}{9}\theta^{3}\dot{G})\nonumber\\
  &&\quad\quad-\frac{2}{3}\theta^{2}\dot{G}^{2}f^{iv}
  +\frac{f'''}{f''}\left(\frac{9G}{16\theta^{2}}+\frac{\theta^{2}}{6}+\frac{1-3w}{2}\rho_{m}-f+Gf'-\frac{2\theta^{2}\dot{G}^{2}}{3}f'''\right)\}\mathcal{G}^{k}\nonumber\\
  &&\quad\quad-\frac{(1+w_{i})w}{1+w}(-\theta^{2}+\frac{9}{4\theta^{2}}+\dot{\theta}+\left(\frac{1}{2}-\frac{9}{2} w\right)\rho_{m}- f+Gf'-2\theta^{2}\dot{G}^{2}f''' \nonumber\\
  &&\quad\quad
  +f''(\frac{10\theta^{3}\dot{G}}{9}-\frac{9G\dot{G}}{2\theta}-2\theta^{2}\ddot{G})) \varepsilon^{k}
  -(1+w_{i})\{\frac{4}{9}\theta^{3}f''\}\mathsf{G}^{k}
  +\left(w_{i}- c^{2}_{si}\right) \dot{\theta} \Delta^{k}_{i}+\left(w_{i}- c^{2}_{si}\right) \theta\dot{ \Delta}^{k}_{i}\nonumber\\
  &&\quad\quad
 -(1+w_{i})\frac{w}{1+w}\theta  \dot{\varepsilon}^{k}
 -w_{i} \dot{\theta} \varepsilon^{k}_{i}-w_{i} \theta \dot{\varepsilon}_{i}^{k}+(1+w_{i})\left(c^{2}_{si}-\frac{2}{3}\right)\theta \frac{k^{2}}{a} V_{i}-c^{2}_{si}\frac{k^{2}}{a^{2}}\Delta^{k}_{i}-w_{i}\frac{k^{2}}{a^{2}}\varepsilon_{i}^{k}\;.
 \end{eqnarray}
 For a two component fluid, the entropy and velocity perturbation equations are given by
 \begin{eqnarray}
  &&\ddot{S}^{k}_{ij}=\frac{k^{2}}{a^{2}}\dot{V}^{k}_{ij}-\frac{1}{3}\theta \frac{k^{2}}{a^{2}} V^{k}_{ij}\;,\\
 &&\ddot{V}^{k}_{ij}=\left(c^{2}_{si}-c^{2}_{sj} \right)\dot{\theta} V^{k}_{i}+\left(c^{2}_{si}-c^{2}_{sj} \right)\theta \dot{V}^{k}_{i}+\left(c^{2}_{sj}-\frac{1}{3}\right)\dot{\theta} V^{k}_{ij}+\left(c^{2}_{sj}-\frac{1}{3}\right)\theta\dot{ V}^{k}_{ij}\nonumber\\
 &&\quad\quad-\left(\frac{c^{2}_{si}-c^{2}_{sj}}{a(1+w_{i})}\right)\dot{\Delta}^{k}_{i}+\left(\frac{c^{2}_{si}-c^{2}_{sj}}{3a(1+w_{i})}\right)\theta \Delta^{k}_{i}-\frac{w_{i}}{a(1+w_{i})}\dot{\varepsilon}_{i}^{k}+\frac{w_{i}}{3a(1+w_{i})}\theta \varepsilon^{k}_{i}\nonumber\\
 &&\quad\quad-\frac{c^{2}_{sj}}{a}\dot{S}^{k}_{ij}+\frac{c^{2}_{sj}}{3a}\theta S^{k}_{ij}+\frac{w_{j}}{a(1+w_{j})}\dot{\varepsilon}^{k}_{j}-\frac{w_{j}}{3a(1+w_{j})}\theta \varepsilon^{k}_{j}\;.
\end{eqnarray}
\section{Perturbations in a radiation-dust universe}
\subsection{Basics of the radiation-dust mixture}
We need to reduce the derived perturbation equations for a general multifluid system, for an application of the equations for a cosmological medium. We consider a Universe filled with a non interacting radiation and dust together with Gauss-Bonnet fluid, the three form of the fluid we are considering. We also assume a flat homogeneous and isotropic Universe as a background (FRW with $K=0$). The evolution equation for radiation energy density $\rho_{r}$ and dust energy density $\rho_{d}$ as
\begin{equation}
 \dot{\rho}_{r}=-\frac{4}{3}\theta \rho_{r}
 \label{eq86}
\end{equation}
and
\begin{equation}
 \dot{\rho}_{d}=-\theta \rho_{d}
\end{equation}
with $r$ and $d$ represent radiation and dust respectively.
 The equation of state parameter for dust is considered to be $w_{ d} = 0$ and that
of radiation is $w_{ r} =\frac{1}{3}$ . With this in mind, the equation of state parameter of the
total matter fluid  is given as
\begin{equation}
 w=\frac{p_{m}}{\rho_{m}}=\frac{\rho_{r}}{3\left(\rho_{d}+\rho_{r}\right)}.
\end{equation}
The speed of sound in this total matter fluid is given as
\begin{equation}
 c^{2}_{s}=\frac{\dot{p}_{m}}{\dot{\rho}_{m}}=\frac{4\rho_{r}}{3\left(3\rho_{d}+4\rho_{r}\right)}.
\end{equation}
Define another parameter conecting two speeds of sound $c^{2}_{sd}$ and $c^{2}_{sr}$ as
\begin{eqnarray}
 &&c^{2}_{z}=\frac{1}{h}\left(h_{r}c_{sd}^{2}+h_{d}c^{2}_{sr}\right)=\frac{\rho_{d}}{3\rho_{d}+4\rho_{r}}\;,\\
&& \Delta_{m}=\frac{\rho_{d}}{\rho_{d}+\rho_{r}}\Delta_{d}+\frac{\rho_{r}}{\rho_{d}+\rho_{r}}\Delta_{r}\;,\\
&& \dot{\Delta}_{m}=\frac{w\theta \rho_{d}}{\rho_{d}+\rho_{r}}\Delta_{d}-\frac{w\theta \rho_{d}}{\rho_{d}+\rho_{r}}\Delta_{r}+\frac{\rho_{d}}{\rho_{d}+\rho_{r}}\dot{\Delta}_{d}+3w\dot{\Delta}_{r}\;,\\
&& S_{dr}=\frac{\rho_{d}}{h_{d}}\Delta_{d}-\frac{\rho_{r}}{h_{r}}\Delta_{r}=\Delta_{d}-\frac{3}{4}\Delta_{r}\;,\\
&& \dot{S}_{dr}=\dot{\Delta}_{d}-\frac{3}{4}\dot{\Delta}_{r}\;.
 \label{eq94}
\end{eqnarray}
All the above defined parameters Eq. \ref{eq86} through to Eq. \ref{eq94} help during the analysis of dust and radiation doninated universe.
We can confine our discussions to polynomial $f(G)$ model  and look for solutions  in the short wavelength and long wavelength approximations for perturbations deep in the radiation and dust dominated epochs \cite{cognola2006dark,carloni2005cosmological}.
In $f(G)$ gravity, the expressions for the expansion, the effective matter energy density   are given respectively as \cite{dunsby1991gauge, abebe2012covariant}
\begin{eqnarray}
&& \theta=\frac{2m}{(1+w)t}\;,\\
&&{\displaystyle{
 \rho_{m}=\left(\frac{3}{4}\right)^{1-m}\left[\frac{4m^{2}-3m(1+w)}{(1+w)^{2}t^{2}}\right]^{m-1}\frac{4m^{3}-2m(m-1)\left(2m(3w+5)-3(1+w)\right)}{(1+w)^{2}t^{2}}}}\;.
 \label{eq48}
\end{eqnarray}
Start with Eq. \ref{eq43},
\begin{equation}
 p\varepsilon_{a}=\sum_{i} p_{i}\varepsilon^{i}_{a}+\frac{1}{2}\sum_{ij}\frac{h_{i}h_{j}}{h}\left(c^{2}_{si}-c^{2}_{sj}\right)S^{ij}_{a},
\end{equation}
for a perfect fluid, $\varepsilon^{i}_{a}=0$ and consider a mixture of dust-radiation, we have
\begin{equation}
 p\varepsilon_{a}=\frac{h_{d}h_{r}}{h}\left(c^{2}_{sd}-c^{2}_{sr}\right)S^{dr}_{a},
\end{equation}
using $h=\rho+p$, $h_{i}=\rho_{i}+p_{i}$ and $p_{d}=0$, $p_{r}=\frac{1}{3}\rho_{r}$, we have
\begin{equation}
 p\varepsilon_{a}=\frac{4}{3}\left(\frac{\rho_{d}\rho_{r}}{\frac{3\rho_{d}+4\rho_{r}}{3}}\right) \left(c^{2}_{sd}-c^{2}_{sr}\right)S^{dr}_{a},
 \end{equation}
 using $c^{2}_{sd}=0$ and $c^{2}_{sr}=\frac{1}{3}$, we have
 \begin{equation}
 p\varepsilon_{a}=-\frac{4\rho_{d}\rho_{r}}{3(3\rho_{d}+4\rho_{r})} S^{dr}_{a},
 \end{equation}
 using $p=\frac{1}{3}\rho_{r}$, we have
 \begin{equation}
 \varepsilon_{a}=-\frac{4\rho_{d}}{3\rho_{d}+4\rho_{r}} S^{dr}_{a}.
 \label{eq102}
 \end{equation}
 Eq. \ref{eq102} represents the entropy of the system.
 Its scalar equation becomes
 \begin{equation}
 \varepsilon=-\frac{4\rho_{d}}{3\rho_{d}+4\rho_{r}} S_{dr}.
 \end{equation}
 It's harmonic decomposition gives
 \begin{equation}
 \varepsilon^{k}=-\frac{4\rho_{d}}{3\rho_{d}+4\rho_{r}} S^{k}_{dr}.
 \end{equation}
 The evolution equation for the entropy is thus 
 \begin{equation}
 \dot{\varepsilon}=-\frac{16\theta \rho_{d}\rho_{r}}{3(3\rho_{d}+4\rho_{r})^{2}}S_{dr}-\frac{4\rho_{d}}{3\rho_{d}+4\rho_{r}}\dot{S}_{dr}
 \end{equation}
and its second order equation is presented as 
\begin{multline}
  \ddot{\varepsilon}=\frac{8\theta\rho_{r}}{3(3\rho_{d}+4\rho_{r})}\dot{\varepsilon}+\left(\frac{4\rho_{r}\dot{\theta}}{3(3\rho_{d}+4\rho_{r})}-\frac{4\theta^{2}\rho_{d}\rho_{r}}{3(3\rho_{d}+4\rho_{r})^{2}}\left(1+4\frac{\rho_{r}}{\rho_{d}}\right)\right)\varepsilon-\frac{4\rho_{d}}{3\rho_{d}+4\rho_{r}}\ddot{S}_{dr}.~~~~~~~~~~~~~~~~~~~~~~~~~~~~~~~~~~~~~~~~
  \label{eq127}
 \end{multline}
 \subsection{Total fluid equations}
 We use the defined relations Eq. \ref{eq86} through to Eq. \ref{eq94} and applying the general total fluid second order equations
to the radiation-dust mixture, to have
\begin{eqnarray}
  &&\ddot{\Delta}_{m}^{k}=(\theta^{2}-
  \frac{9}{4\theta^{2}}+\dot{\theta}-\frac{(w^{2}+\frac{2}{3}w-\frac{1}{3})}{c^{2}_{s}}\rho_{m}-(\frac{1}{2}-\frac{9}{2}w)\rho_{m}+ f-Gf'+2\theta^{2}\dot{G}^{2}f'''\nonumber\\ 
  &&\quad\quad+f''(\frac{2\theta^{3}\dot{G}}{9}+\frac{9G\dot{G}}{2\theta}-2\theta^{2}\ddot{G})
 -\frac{k^{2}}{a^{2}}) \left(c^{2}_{s}\Delta_{m}^{k}\right)
  -\{\frac{9}{4\theta^{3}}-(\frac{2}{3}+c^{2}_{s})\theta+\frac{1-3w}{\theta}\rho_{m}
  +\frac{G}{2\theta}f'\nonumber\\&&\quad\quad+f''(\frac{8\theta^{2}\dot{G}}{9}-\frac{6G\dot{G}}{2\theta^{2}}-\frac{4}{3}\theta \ddot{G})-\frac{1}{2\theta}f 
  -\frac{4}{3}\theta \dot{G}^{2}f''' \}
  \dot{\Delta}_{m}^{k}+\frac{4wc^{2}_{z}}{3}\{\frac{9}{4\theta^{2}}-\theta^{2}-\dot{\theta}+(\frac{1}{2}-\frac{9}{2} w)\rho_{m}
  +Gf'\nonumber\\&&\quad\quad+f''\left(\frac{10}{9}\theta^{3}\dot{G}-\frac{9}{2\theta}G\dot{G}-2\theta^{2}\ddot{G}\right)
  -2\theta^{2} \dot{G}^{2}f''' -3\theta^{2}c^{2}_{s}+\frac{k^{2}}{a^{2}}\} S^{k}_{dr} -4w\theta c^{2}_{z}\dot{S}^{k}_{dr}\nonumber\\
&&\quad\quad  
 -(1+w)\{f''\left(\frac{1}{2}+\theta f'-G\right)-(\frac{3G\dot{G}}{2\theta}+\frac{2\theta^{2}\ddot{G}}{3}-\frac{2\theta^{3}\dot{G}}{9})f'''-\frac{2}{3}\theta^{2}\dot{G}^{2}f^{iv}\nonumber\\
&&\quad\quad
  +\frac{f'''}{f''}\left(\frac{9G}{16\theta^{2}}+\frac{\theta^{2}}{6}+\frac{1-3w}{2}\rho_{m}-f+Gf'-\frac{2\theta^{2}\dot{G}^{2}}{3}f'''\right) \}\mathcal{G}^{k}
  -(1+w)\frac{4}{9}\theta^{3}f''\dot{ \mathcal{G}}^{k}\;,
 \end{eqnarray}
 \begin{eqnarray}
  &&\ddot{ \mathcal{G}}=\{ \frac{9c^{2}_{s}}{(1+w)\theta^{3}}G\dot{G}+\frac{1}{\theta^{2}f''}(\frac{3(1-3w)}{4}\rho_{m}-\frac{27}{8\theta^{2}}\frac{c^{2}_{s}}{1+w}-\frac{3(1-3w)}{2}\frac{c^{2}_{s}}{1+w}\rho_{m}\nonumber \\
  &&\quad\quad+3\frac{c^{2}_{s}}{1+w}f-3\frac{c^{2}_{s}}{1+w}Gf')+\frac{2c^{2}_{s}\dot{G}^{2}f'''}{(1+w)f''}\}\Delta^{m}\nonumber \\
  &&\quad\quad
  +\{\frac{1}{\theta^{2}f''}\left(-\frac{27}{8(1+w)\theta^{3}}-\frac{3(1-3w)}{2(1+w)\theta}\rho_{m}+\frac{3}{(1+w)\theta}f-\frac{3}{(1+w)\theta}Gf'\right)\frac{3c^{2}_{s}-1}{3(1+w)}\dot{G}\nonumber\\
  &&\quad\quad+\frac{27}{4(1+w)\theta^{4}}G\dot{G}\}\dot{\Delta}^{m}+\{-\frac{\theta}{3}-\frac{9}{4}\frac{G}{\theta^{3}}-2\dot{G}\frac{f'''}{f''}\}\dot{ \mathcal{G}}^{k}\nonumber\\
  && \quad\quad
  -4c^{2}_{z}\{\frac{1}{\theta^{2}f''}\left(-\frac{27}{8\theta^{2}}\frac{w}{1+w}-\frac{3(1-3w)}{2}\frac{w}{1+w}\rho_{m}+\frac{3w}{1+w}f-\frac{3w}{1+w}Gf'\right)\nonumber\\
  &&\quad\quad-\frac{w\theta \dot{G}}{3(1+w)}+\frac{27w}{4(1+w)\theta^{3}}G\dot{G}+\ddot{G}\frac{w}{(1+w)}-w\ddot{G}-w\theta\dot{G}c^{2}_{s}\} S^{k}_{dr} -4w\dot{G}c^{2}_{z}\dot{S}^{k}_{dr}\nonumber\\
  &&\quad\quad
  +\{\frac{1}{\theta^{2}f''}(\frac{27}{32\theta^{2}}+\frac{3}{2}f')-\frac{3}{2\theta}f'+\frac{3}{2\theta^{2}}G-\frac{9}{4\theta}^{3}\dot{G}-\frac{\dot{G}^{2}f^{iv}}{f''}\nonumber\\
  &&\quad\quad
  +\frac{f'''}{\theta^{2}f''}\left(-\frac{27}{32}\frac{G}{\theta^{2}}-\frac{\theta^{2}}{4}
  -\frac{3(1-3w)}{4}\rho_{m}+\frac{3}{2}f-\frac{3}{2}Gf'+\dot{G}^{2}\theta^{2}f'''\right)\}\mathcal{G}\;,\\
 &&\ddot{S}^{k}_{ij}=(c^{2}_{si}-\frac{2}{3})\theta \dot{S}^{k}_{ij}-c^{2}_{sj}\frac{k^{2}}{a^{2}} S^{k}_{ij}-\left(\frac{c^{2}_{si}}{1+w_{i}}-\frac{c^{2}_{sj}}{1+w_{i}}\right)\frac{k^{2}}{a^{2}}\Delta^{k}_{m}\;,
\end{eqnarray}
where $\Delta_{m}$ and $S_{dr}$  are given by $\Delta_{m}=\frac{\rho_{d}\Delta_{d}+\rho_{r}\Delta_{r}}{\rho_{d}+\rho_{r}}$, $S_{dr}=\Delta_{d}-\frac{3}{4}\Delta_{r}$.

\subsection{Component equations}
The energy density perturbations equation for the  component fluids is presented as
\begin{eqnarray}
 &&\ddot{\Delta}^{k}_{i}=-\frac{(1+w_{i})c^{2}_{s}}{1+w}(-\theta^{2}+\frac{9}{4\theta^{2}}+
 \dot{\theta}+(\frac{1}{2}-\frac{9}{2}w)\rho_{m}-f+Gf'-2\theta^{2}\dot{G}^{2}f''' \nonumber\\
&&\quad\quad -\frac{(1-3w)(1+w)}{c^{2}_{s}}\rho_{m}
+f''(\frac{10\theta^{3}\dot{G}}{9}-\frac{9G\dot{G}}{2\theta}-2\theta^{2}\ddot{G})
 )\Delta_{m}^{k}\nonumber\\
&&\quad\quad
  -\frac{(1+w_{i})}{1+w}\{\frac{9}{4\theta^{3}}+c^{2}_{s}
 \theta+\frac{1-3w}{\theta}\rho_{m}
  +\frac{G}{2\theta}f'+f''(\frac{8\theta^{2}\dot{G}}{9}\nonumber\\
  &&\quad\quad-\frac{6G\dot{G}}{2\theta^{2}}-\frac{4}{3}\theta\ddot{G})-\frac{1}{2\theta}f-\frac{2}{3}\theta 
  -\frac{4}{3}\theta \dot{G}^{2}f'''\}\dot{\Delta}_{m}^{k}\nonumber\\
&&\quad\quad
  -(1+w_{i})\{f''(\theta f'-G+\frac{1}{2}) +f'''(-\frac{3G\dot{G}}{2\theta}-\frac{2}{3}\theta^{2}\ddot{G}+\frac{2}{9}\theta^{3}\dot{G})-\frac{2}{3}\theta^{2}\dot{G}^{2}f^{iv}
  \nonumber\\
&&\quad\quad+\frac{f'''}{f''}\left(\frac{9G}{16\theta^{2}}+\frac{\theta^{2}}{6}+\frac{1-3w}{2}\rho_{m}-f+Gf'-\frac{2\theta^{2}\dot{G}^{2}}{3}f'''\right)\}\mathcal{G}^{k}\nonumber\\
  &&\quad\quad-\frac{(1+w_{i})w}{1+w}(-\theta^{2}+\frac{9}{4\theta^{2}}+\dot{\theta}+\left(\frac{1}{2}-\frac{9}{2} w\right)\rho_{m}- f+Gf'-2\theta^{2}\dot{G}^{2}f''' \nonumber\\
&&\quad\quad
  +f''(\frac{10\theta^{3}\dot{G}}{9}-\frac{9G\dot{G}}{2\theta}-2\theta^{2}\ddot{G})) \varepsilon^{k}
  -(1+w_{i})\{\frac{4}{9}\theta^{3}f''\}\mathsf{G}^{k}\nonumber\\
&&\quad\quad
  +\left(w_{i}- c^{2}_{si}\right) \dot{\theta} \Delta^{k}_{i}+\left(w_{i}- c^{2}_{si}\right) \theta\dot{ \Delta}^{k}_{i}
 -(1+w_{i})\frac{w}{1+w}\theta  \dot{\varepsilon}^{k}\nonumber\\
 &&\quad\quad
 -w_{i} \dot{\theta} \varepsilon^{k}_{i}-w_{i} \theta \dot{\varepsilon}_{i}^{k}+(1+w_{i})\left(c^{2}_{si}-\frac{2}{3}\right)\theta \frac{k^{2}}{a} V_{i}-c^{2}_{si}\frac{k^{2}}{a^{2}}\Delta^{k}_{i}-w_{i}\frac{k^{2}}{a^{2}}\varepsilon_{i}^{k}. 
 \end{eqnarray}
 By considering the fluid to be perfect ($\varepsilon_{i}=0$) and setting the direction of the unit velocity vector  $V_{i}$ to be in the same directtion as that of total fluid, implies vanishing relative velocity $V_{i}=0$,and setting $i$ as radiation and $c^{2}_{si}=\frac{1}{3}$, we have
 \begin{multline}
 \ddot{\Delta}^{k}_{r}=\{\frac{w\theta}{1+w}\left(-3c^{2}_{s}+4\ddot{G}f''+4\dot{G}^{2}f'''+\frac{2}{3}\right)-\frac{4w\theta^{2}}{3(1+w)}\left(2\dot{G}f''+c^{2}_{z}\right)-\frac{27w}{4(1+w)\theta^{3}}\\-\frac{3w}{(1+w)\theta}\left((1-3w)\rho_{d}+(1-3w)\rho_{r}+\frac{Gf'}{2}-\frac{f}{2}\right)+\frac{9wG\dot{G}f''}{(1+w)\theta^{2}}\}\dot{\Delta}^{k}_{r}\\+\{\frac{8\theta^{2}}{3(1+w)}\left(\frac{\dot{G}\rho_{d}f''}{3(\rho_{d}+\rho_{r})}+2c^{2}_{z}\right)+\frac{\theta \rho_{d}}{(1+w)(\rho_{d}+\rho_{r})}\left(-c^{2}_{s}+\frac{4}{3}\ddot{G}f''+\frac{2}{3}+\frac{4}{3}\dot{G}^{2}f'''\right)\\-\frac{9\rho_{d}}{4(1+w)\theta^{3}(\rho_{d}+\rho_{r})}+\frac{3G\dot{G}\rho_{d}f''}{(1+w)\theta^{2}(\rho_{d}+\rho_{r})}+\frac{\rho_{d}}{(1+w)(\rho_{d}+\rho_{r})}\left((3w-1)-\frac{Gf'}{2}+\frac{f}{2}\right)\}\dot{\Delta}_{d}^{k}\\ +\{\theta^{2}\left(\frac{c^{2}_{s}}{1+w}\frac{\rho_{d}}{\rho_{d}+\rho_{r}}(\frac{4}{3}-w+\frac{2w}{3c^{2}_{s}})-\frac{16wc^{2}_{z}}{3(1+w)}(1-c^{2}_{s})\right)\\+\theta^{2}\dot{G}^{2}f'''\left(\frac{4\rho_{d}}{3(1+w)(\rho_{d}+\rho_{r})}(2c^{2}_{s}+w)-\frac{
 32wc^{2}_{z}}{3(1+w)}\right)+\theta^{2}\ddot{G}f''\left(\frac{4\rho_{d}}{3(1+w)(\rho_{d}+\rho_{r})}(2c^{2}_{s}+w)\right)\\+\frac{32wc^{2}_{z}}{3(1+w)}\theta^{2}\dot{G}f''-\frac{1}{\theta^{2}}\left(\frac{3\rho_{d}}{(1+w)(\rho_{d}+\rho_{r})}\left(c^{2}_{s}+\frac{3}{4}w\right)+\frac{12wc^{2}_{z}}{1+w}\right)\\-\frac{8\dot{G}\theta^{3}f''}{9(1+w)}\left(\frac{\rho_{d}}{\rho_{d}+\rho_{r}}\left(c^{2}_{s}+w\right)-\frac{20c^{2}_{z}}{3}\right)+\frac{4\dot{\theta}}{3(1+w)}\left(-\frac{c^{2}_{s}\rho_{d}}{\rho_{d}+\rho_{r}}+4wc^{2}_{z}\right)\\ +\frac{3\dot{G}f''}{(1+w)\theta}\left(\frac{\rho_{d}}{\rho_{d}+\rho_{r}}\left(2c^{2}_{s}+w\right)-8wc^{2}_{z}\right)+\frac{f}{1+w}\left(\frac{\rho_{d}}{\rho_{d}+\rho_{r}}\left(\frac{4}{3}c^{2}_{s}+\frac{1}{2}w\right)-\frac{16wc^{2}_{z}}{3}\right)\\ +\frac{Gf'}{1+w}\left(-\frac{\rho_{d}}{\rho_{d}+\rho_{r}}\left(\frac{4}{3}c^{2}_{s}+\frac{1}{2}w\right)+\frac{16wc^{2}_{z}}{3}\right)+\frac{4(1-3w)\rho_{d}}{3(\rho_{d}+\rho_{r})}+\frac{(\frac{1}{3}-3w)(8wc^{2}_{z}\rho_{d})}{1+w}\\ +\frac{\rho^{2}_{d}}{(1+w)(\rho_{d}+\rho_{r})}\left(2c^{2}_{s}\left(-\frac{1}{3}+3w\right)-(1-3w)w\right)\\+\frac{\rho_{d}}{\rho_{d}+\rho_{r}}\left(\frac{2c^{2}_{s}\rho_{d}}{1+w}\left(-\frac{1}{3}+3w\right)-\frac{(1-3w)w\rho_{d}}{1+w}+\frac{4(1-3w)}{3}\right)\\+\frac{w\rho_{d}}{1+w}\left(c^{2}_{s}(6w-\frac{2}{3})+(w-\frac{1}{3})(w+4)\right)+\frac{8(\frac{1}{3}-3w)wc^{2}_{z}\rho_{r}}{1+w} \}\Delta^{k}_{d}
 \\+\{\frac{4w\theta^{2}\ddot{G}f''}{1+w}\left(-\frac{\rho_{d}}{3(\rho_{d}+\rho_{r})}+2c^{2}_{s}+2c^{2}_{z}\right)+\frac{8w\theta^{3}\dot{G}f''}{3(1+w)}\left(\frac{\rho_{d}}{3(\rho_{d}+\rho_{r})}-c^{2}_{s}-\frac{5}{3}c^{2}_{z}\right)\\ +\frac{4w\theta^{2}\dot{G}^{2}f'''}{1+w}\left(-\frac{\rho_{d}}{3(\rho_{d}+\rho_{r})}+2c^{2}_{s}+2c^{2}_{z}\right)+\frac{18wG\dot{G}c^{2}_{z}f''}{1+w}\\\frac{wGf'}{1+w}\left(\frac{\rho_{d}}{2(\rho_{d}+\rho_{r})}-4(c^{2}_{s}+c^{2}_{z})\right)+\frac{3wG\dot{G}}{(1+w)\theta}\left(-\frac{\rho_{d}f''}{\rho_{d}+\rho_{r}}+9c^{2}_{s}f'\right)\\+\frac{wf}{1+w}\left(-\frac{\rho_{d}}{2(\rho_{d}+\rho_{r})}+4(c^{2}_{s}+c^{2}_{z})\right)+\frac{w\theta^{2}\rho_{d}}{(1+w)(\rho_{d}+\rho_{r})}\left(c^{2}_{s}-\frac{2}{3}\right)+\frac{4w\theta^{2}}{1+w}\left(c^{2}_{s}+c^{2}_{z}(1-c^{2}_{s})\right)\\-\frac{4w\dot{\theta}}{1+w}\left(c^{2}_{s}-c^{2}_{z}\right)+\frac{9w}{(1+w)\theta^{2}}\left(\frac{\rho_{d}}{4(\rho_{d}+\rho_{r})}-3c^{2}_{s}-c^{2}_{z}\right)+\frac{(1-3w)w\rho^{2}_{d}}{(1+w)(\rho_{d}+\rho_{r})}\\+\frac{w\rho_{d}}{1+w}\left((1-3w)w-2c^{2}_{s}(1-9w)-2c^{2}_{z}(1-9w)\right)+4w(1-3w)\rho_{d}\\ \frac{w\rho_{r}}{1+w}\left(-2c^{2}_{s}(1-9w)-2c^{2}_{z}\right)+4w(1-3w)\rho_{r}+\frac{18w^{2}c^{2}_{z}}{1+w}-\frac{k^{2}}{a^{2}}\}\Delta_{r}^{k}\\
  -\frac{4}{3}\{f''(\theta f'-G+\frac{1}{2}) +f'''(-\frac{3G\dot{G}}{2\theta}-\frac{2}{3}\theta^{2}\ddot{G}+\frac{2}{9}\theta^{3}\dot{G})\\-\frac{2}{3}\theta^{2}\dot{G}^{2}f^{iv}
  +\frac{f'''}{f''}\left(\frac{9G}{16\theta^{2}}+\frac{\theta^{2}}{6}+\frac{1-3w}{2}\rho_{m}-f+Gf'-\frac{2\theta^{2}\dot{G}^{2}}{3}f'''\right)\}\mathcal{G}^{k}
  -\frac{16}{27}\theta^{3}f''\dot{ \mathcal{G}}^{k}.~~~~~~~~~~~~~~~~~~~~~~~~~~~~~~~~~~~~~~~~~~~~~~~~~~~~~~~~~~~~~
 \end{multline}
 Similary, the dust component is presented as
 \begin{multline}
 \ddot{\Delta}^{k}_{d}=\{\frac{c^{2}_{s}\rho_{d}\theta^{2}}{(1+w)(\rho_{d}+\rho_{r})}\left(1+2\dot{G}f'''+2\ddot{G}f''-w\right)+\frac{wc^{2}_{z}}{1+w}\left(\left(2-18w\right)\left(\rho_{d}+\rho_{r}\right)-4\left(f-Gf'\right)\right)\\+\frac{w\rho_{d}\theta^{2}}{(1+w)(\rho_{d}+\rho_{r})}\left(\frac{4\ddot{G}f''}{3}+\frac{2}{3}+\frac{4\dot{G}^{2}f'''}{3}\right)-\frac{4wc^{2}_{z}\theta^{2}}{1+w}\left(1+2\dot{G}^{2}f'''-c^{2}_{s}+4w\ddot{G}f''\right)\\ +\frac{\dot{\theta}}{1+w}\left(4wc^{2}_{z}-\frac{c^{2}_{s}\rho_{d}}{\rho_{d}+\rho_{r}}\right)+\frac{3G\dot{G}}{\theta}\left(-6wc^{2}_{z}f''+\frac{\rho_{d}f''}{(1+w)(\rho_{d}+\rho_{r})}\left(\frac{3c^{2}_{s}}{2}+w\right)\right)\\+\frac{1}{(1+w)\theta^{2}}\left(wc^{2}_{z}-\frac{9\rho_{d}}{4(\rho_{d}+\rho_{r})}\left(c^{2}_{s}+w\right)\right)+\frac{2\dot{G}\theta^{3}f''}{3(1+w)}\left(20wc^{2}_{z}-\frac{\rho_{d}}{\rho_{d}+\rho_{r}}\left(c^{2}_{s}+\frac{4w}{3}\right)\right)\\+\frac{\rho_{d}}{(1+w)(\rho_{d}+\rho_{r})}\left(c^{2}_{s}\left(f-\left(\frac{1}{2}-\frac{9w}{2}\right)\rho_{d}-Gf'\right)+\frac{w}{2}\left(-Gf'+f \right)\right)\\ +(1-3w)(1+w)\frac{\rho_{d}}{\rho_{d}+\rho_{r}}\left(\rho_{d}\left(1-\frac{w}{1+w}\right)+\rho_{r}\left(1-\frac{w}{1+w}\right)\right)\}\Delta^{k}_{d}\\ \{\frac{c^{2}_{s}\rho_{r}\theta^{2}}{(1+w)(\rho_{d}+\rho_{r})}\left(1+2\dot{G}^{2}f'''+2\ddot{G}f''\right)+\frac{w\rho_{d}\theta^{2}}{(1+w)(\rho_{d}+\rho_{r})}\left(c^{2}_{s}-4\ddot{G}f''-\frac{2}{3}-\frac{4\dot{G}^{2}f'''}{3}\right)\\ +\frac{w \theta^{2}}{1+w}\left(-6c^{2}_{z}\dot{G}^{2}f'''-6c^{2}_{z}\ddot{G}f''-3c^{2}_{z}c^{2}_{s}\right)+\frac{2\theta^{3}f''}{3(1+w)}\left(\frac{c^{2}_{s}\dot{G}\rho_{r}}{\rho_{d}+\rho_{r}}+\frac{15wc^{2}_{z}\dot{G}}{3}+\frac{4w\dot{G}\rho_{d}}{3(\rho_{d}+\rho_{r})}\right)\\ +\frac{3G\dot{G}f''}{(1+w)\theta}\left(\frac{3c^{2}_{s}\rho_{r}}{2(\rho_{d}+\rho_{r})}-\frac{9wc^{2}_{z}}{2}-\frac{w\rho_{d}}{\rho_{d}+\rho_{r}}\right)+\frac{\dot{\theta}}{1+w}\left(-\frac{c^{2}_{s}\rho_{r}}{\rho_{d}+\rho_{r}}+3wc^{2}_{z}\right)\\+\frac{9}{4(1+w)\theta^{2}}\left(-\frac{c^{2}_{s}\rho_{r}}{\rho_{d}+\rho_{r}}+3wc^{2}_{z}+\frac{rho_{d}}{\rho_{d}+\rho_{r}}\right)+\frac{3wc^{2}_{z}}{1+w}\left(\frac{3}{2}-\frac{9w}{2}-f+Gf'\right)\\+\frac{w\rho_{d}}{(1+w)(\rho_{d}+\rho_{r})}\left((1-3w)\rho_{d}+\frac{Gf'}{2(1+w)}-\frac{f}{2}-\frac{(\frac{1}{2}-\frac{9w}{2})c^{2}_{s}\rho_{r}}{w}+(1-3w)\rho_{r}\right)\\+\frac{\rho_{r}}{(1+w)(\rho_{d}+\rho_{r})}\left(-c^{2}_{s}\left((\frac{1}{2}-\frac{9w}{2})\rho_{r}+f+Gf'\right)+(1-3w)(1+w)\rho_{r}+(1-3w)(1+w)\rho_{d}\right)\}\Delta^{k}_{r} \\ \{-\frac{8G\rho_{d}\theta^{2}f''}{9(1+w)(\rho_{d}+\rho_{r})}+\theta \left(\frac{\rho_{d}}{(1+w)(\rho_{d}+\rho_{r})}\left(-c^{2}_{s}+\frac{4\ddot{G}f''}{3}+\frac{2}{3}+\frac{4\dot{G}^{2}f'''}{3}\right)+\frac{4wc^{2}_{z}}{1+w}\right)\\+\frac{\rho_{d}}{\theta(1+w)(\rho_{d}+\rho_{r})}\left(-(1-3w)\rho_{d}-(1-3w)\rho_{r}-\frac{Gf'}{2}+\frac{f}{2}\right)\\-\frac{9\rho_{d}}{4(1+w)\theta^{3}(\rho_{d}+\rho_{r})}+\frac{3G\dot{G}f'' \rho_{d}}{(1+w)\theta^{2}(\rho_{d}+\rho_{r})}\}\dot{\Delta}^{k}_{d}\\ \{-\frac{27w}{4(1+w)\theta^{3}}-\frac{8w\dot{G}\theta^{2}f''}{3(1+w)}+\frac{9wG\dot{G}f''}{(1+w)\theta^{2}}+\frac{w\theta}{1+w}\left(-3(c^{2}_{s}+c^{2}_{z})+4\ddot{G}f''+2+4\dot{G}^{2}f'''\right)\\+\frac{w}{(1+w)\theta}\left(-3(1-3w)(\rho_{d}+\rho_{r})-\frac{3Gf''}{2}+\frac{3f}{2}\right)\}\dot{\Delta}^{k}_{r}\\
  -\{f''(\theta f'-G+\frac{1}{2}) +f'''(-\frac{3G\dot{G}}{2\theta}-\frac{2}{3}\theta^{2}\ddot{G}+\frac{2}{9}\theta^{3}\dot{G})\\-\frac{2}{3}\theta^{2}\dot{G}^{2}f^{iv}
  +\frac{f'''}{f''}\left(\frac{9G}{16\theta^{2}}+\frac{\theta^{2}}{6}+\frac{1-3w}{2}\rho_{m}-f+Gf'-\frac{2\theta^{2}\dot{G}^{2}}{3}f'''\right)\}\mathcal{G}^{k}
  -\frac{4}{9}\theta^{3}f''\dot{ \mathcal{G}}^{k}.~~~~~~~~~~~~~~~~~~~~~~~~~~~~~~~~~~~~~~~~~~~~~~~~~~~~~
 \end{multline}
\subsection{Short wavelength solutions}
In this section, we study the evolution of the short wavelength modes, it means large
values of the wave number $k$ for a radiation and dust mixture. The general results will then be considered for polynomial $f(G)$ model
and we will use quasi-static approximation for the matter perturbations for both radiation and dust dominated epochs. In that approximation,
widely used in the literature \cite{silvestri2009approaches,dunsby1992covariant,ntahompagaze2017f}, all the time derivative terms for the Gauss-Bonnet term and its momentum are discarded, and only those including energy density perturbations $\Delta_{m}$ are kept \cite{ntahompagaze2020multifluid}.
\subsubsection{Perturbations in the radiation-dominated epoch}
Let us now look at the case where the characteristic size of the fluid inhomogeneities
is much less than the Jeans length for the radiation fluid but is still larger than the
mean free path of the photon, i.e., $ \lambda \ll \lambda_{H} \ll \lambda_{ J}$ .
We can neglect the interaction between the component fluids
and  assume that the radiation energy density can be taken as homogeneous, meaning
$\Delta_{ r} \approx 0$. Our equations become
\begin{eqnarray}
 &&\dot{\Delta}_{d}^{k}=-Z^{k}- c^{2}_{sd} \theta \Delta_{d}^{k}
 -c^{2}_{s}
 \theta \Delta_{m}^{k}
 + \frac{k^{2}}{a} V^{k}_{d}\;,\\
 && \dot{Z^{k}}=\left(-\frac{1}{4}\theta^{2}-\frac{3}{4}\rho_{r}-\frac{3}{8} f+\frac{3}{4}Gf'-\frac{1}{2}\theta^{2}\ddot{G}f''+\frac{1}{3}\theta^{3}\dot{G}f''-\frac{9}{8\theta}G \dot{G}f''-\frac{1}{2}\theta^{2}\dot{G}^{2}f''' \right) \left(c^{2}_{s}\Delta_{m}^{k}\right)\nonumber\\
 &&\quad\quad+\frac{3c^{2}_{s}}{4}\frac{k^{2}}{a^{2}}\Delta_{m}^{k}  +\{f''(\theta f'-G) -\frac{3G\dot{G}f'''}{2\theta}-\frac{2}{3}\theta^{2}\dot{G}^{2}f^{iv}-\frac{2}{3}\theta^{2}\ddot{G}f'''\nonumber\\
 &&\quad\quad
  +\frac{f'''}{f''}(\frac{9G}{16\theta^{2}}+\frac{\theta^{2}}{6}-f+Gf'-\frac{2\theta^{2}\dot{G}^{2}}{3}f''')+\frac{1}{2}f''
 +\frac{2}{9}\theta^{3}\dot{G}f'''\}\mathcal{G}^{k}\nonumber\\
 &&\quad\quad+\left(-\frac{1}{12}\theta^{2}-\frac{1}{4}\rho_{r}-\frac{1}{8} f+\frac{1}{8}Gf'-\frac{1}{6}\theta^{2}\ddot{G}f''+\frac{1}{18}\theta^{3}\dot{G}f''-\frac{3}{8\theta}G\dot{G}f''-\frac{1}{6}\theta^{2}\dot{G}^{2}f''' \right)\left(\varepsilon^{k} \right)-\frac{1}{4}\frac{k^{2}}{a^{2}}\varepsilon^{k}\nonumber\\
 &&\quad\quad
  +\{\frac{4}{9}\theta^{3}f''\}\mathsf{G}^{k}+\{\frac{9}{4\theta^{3}}
  +\frac{G}{2\theta}f'+f''(\frac{8\theta^{2}\dot{G}}{9}-\frac{6G\dot{G}}{2\theta^{2}}-\frac{4}{3}\theta\ddot{G})-\frac{1}{2\theta}f-\frac{2}{3}\theta 
  -\frac{4}{3}\theta \dot{G}^{2}f''' \}Z^{k}\;,\\
&&\dot{ \mathcal{G}}^{k}=\mathsf{G}^{k}+\frac{c^{2}_{s}}{1+w}\dot{G}\Delta_{m}^{k}-w\dot{G}\varepsilon^{k}\;,\\
 && \dot{\mathsf{G}}^{k}=\{-\frac{c^{2}_{s}}{(1+w)}\ddot{G}\}\Delta^{k}+\{-\frac{\theta}{3}-\frac{9}{4}\frac{G}{\theta^{3}}-2\frac{f'''\dot{G}}{f''}\}\mathsf{G}^{k}+\{\frac{1}{\theta^{2}f''}(\frac{27}{32\theta^{2}}+\frac{3}{2}f')\nonumber\\
 &&\quad\quad
  +\frac{f'''}{\theta^{2}f''}\left(-\frac{27}{32}\frac{G}{\theta^{2}}-\frac{\theta^{2}}{4}
+\frac{3}{2}f-\frac{3}{2}Gf'+\dot{G}^{2}\theta^{2}f'''\right)-\frac{3}{2\theta}f'+\frac{3}{2\theta^{2}}G
  -\frac{9}{4\theta}^{3}\dot{G}-\frac{\dot{G}^{2}f^{iv}}{f''}\}\mathcal{G}^{k}\nonumber\\
  &&\quad\quad
  +\{\frac{1}{\theta^{2}f''}\left(-\frac{27}{8\theta^{3}}+\frac{3}{\theta}f-\frac{3}{\theta}Gf'\right)-\frac{\dot{G}}{3}+\frac{27}{4\theta^{4}}G\dot{G}\}Z^{k}
  +\ddot{G}\frac{w}{(1+w)}\varepsilon^{k}\;,\\
&& \dot{V}^{k}_{d}=\left(c^{2}_{sd}-\frac{1}{3}\right)\theta V^{k}_{d}+\frac{1}{a(1+w)}\left(c^{2}_{s}\Delta^{k} +w \varepsilon^{k} \right)-\frac{1}{a}\left(c^{2}_{sd}\Delta^{k}_{d}\right)\;,\\
&&\dot{V}^{k}_{dr}=\left(c^{2}_{sd}-c^{2}_{sr} \right)\theta V^{k}_{d}+\left(c^{2}_{sr}-\frac{1}{3}\right)\theta V^{k}_{dr}-\left(\frac{c^{2}_{sd}}{a}-\frac{c^{2}_{sr}}{a}\right)\Delta^{k}_{d}-\frac{c^{2}_{sr}}{a} S^{k}_{dr}+\frac{w_{r}}{a(1+w_{r})}\varepsilon^{k}_{r}\;,\\
  &&\dot{S}^{k}_{dr}=\frac{k^{2}}{a^{2}} V^{k}_{dr}\;.
 \end{eqnarray}
 The matter energy density and entropy 
 $S_{dr}$  are given by $\Delta_{m}=\frac{\rho_{d}\Delta_{d}}{\rho_{d}+\rho_{r}}$ and $S_{dr}=\Delta_{d}$.
 For  
$\Delta_{r}\ll \Delta_{d}$
and using $c^{2}_{sd}=0$ and 
from the homogeneity of radiation energy density as a background,
 we have 
 \begin{eqnarray}
  &&c^{2}_{s}\rho_{m}\Delta^{k}_{m}+p\varepsilon^{k}=\frac{1}{3}\rho_{r}\Delta^{k}_{r}\approx 0\;,\\
 &&\dot{\Delta}_{d}^{k}=-Z^{k}
 + \frac{k^{2}}{a} V^{k}_{d}\;,\\
&&\dot{ \mathcal{G}}^{k}=\mathsf{G}^{k}.
 \end{eqnarray}
 \begin{eqnarray}
  &&\dot{\mathsf{G}}^{k}=\{-\frac{c^{2}_{s}}{(1+w)}\ddot{G}\}(\frac{\rho_{d}}{\rho_{d}+\rho_{r}})\Delta_{d}^{k}+\{-\frac{\theta}{3}-\frac{9}{4}\frac{G}{\theta^{3}}-2\frac{f'''\dot{G}}{f''}\}\dot{\mathcal{G}}^{k}+\{\frac{1}{\theta^{2}f''}(\frac{27}{32\theta^{2}}+\frac{3}{2}f')\nonumber\\
  &&\quad\quad
  +\frac{f'''}{\theta^{2}f''}\left(-\frac{27}{32}\frac{G}{\theta^{2}}-\frac{\theta^{2}}{4}
+\frac{3}{2}f-\frac{3}{2}Gf'+\dot{G}^{2}\theta^{2}f'''\right)-\frac{3}{2\theta}f'+\frac{3}{2\theta^{2}}G\nonumber\\
&&\quad\quad
  -\frac{9}{4\theta^{3}}\dot{G}-\frac{\dot{G}^{2}f^{iv}}{f''}\}\mathcal{G}^{k}
  +\{\frac{1}{\theta^{2}f''}\left(-\frac{27}{8\theta^{3}}+\frac{3}{\theta}f-\frac{3}{\theta}Gf'\right)-\frac{\dot{G}}{3}+\frac{27}{4\theta^{4}}G\dot{G}\}Z^{k}\nonumber\\
  &&\quad\quad
  -\ddot{G}\frac{\rho_{d}}{3\rho_{d}+4\rho_{r}}\Delta^{k}_{d}\;,\\
 &&\dot{V}^{k}_{d}=-\frac{1}{3}\theta V^{k}_{d}\;,\\
&&\dot{V}^{k}_{dr}=-\frac{1}{3}\theta V^{k}_{d}\;,\\
&&  \dot{S}^{k}_{dr}=\frac{k^{2}}{a^{2}} V^{k}_{dr}.
 \end{eqnarray}
At this stage, we can set the
direction of the unit velocity vector $u_{a}^{d}$ of the dust to be in the same direction as
that of the total matter fluid. This implies that we have a vanishing relative velocity
$V_{a}^{d}$. Thus we have
\begin{eqnarray}
 &&V_{d}^{k} = 0\;,\\
&& \dot{\Delta}_{d}^{k}=-Z^{k}\;,\\
&&  \dot{Z^{k}}=\left(-\frac{1}{4}\theta^{2}-\frac{3}{4}\rho_{r}-\frac{3}{8} f+\frac{3}{4}Gf'-\frac{1}{2}\theta^{2}\ddot{G}f''+\frac{1}{3}\theta^{3}\dot{G}f''-\frac{9}{8\theta}G \dot{G}f''-\frac{1}{2}\theta^{2}\dot{G}^{2}f''' \right) \left(\frac{c^{2}_{s}\rho_{d}}{\rho_{d}+\rho_{r}}\Delta_{d}^{k}\right)\nonumber\\
&&\quad\quad+\frac{3c^{2}_{s}}{4}\frac{k^{2}\rho_{d}}{a^{2}\rho_{d}+\rho_{r}}\Delta_{d}^{k}  +\{f''(\theta f'-G) -\frac{3G\dot{G}f'''}{2\theta}-\frac{2}{3}\theta^{2}\dot{G}^{2}f^{iv}-\frac{2}{3}\theta^{2}\ddot{G}f'''\nonumber\\
&&\quad\quad
  +\frac{f'''}{f''}(\frac{9G}{16\theta^{2}}+\frac{\theta^{2}}{6}-f+Gf'-\frac{2\theta^{2}\dot{G}^{2}}{3}f''')+\frac{1}{2}f''
 +\frac{2}{9}\theta^{3}\dot{G}f'''\}\mathcal{G}^{k}\nonumber\\
 &&\quad\quad+\left(-\frac{1}{12}\theta^{2}-\frac{1}{4}\rho_{r}-\frac{1}{8} f+\frac{1}{8}Gf'-\frac{1}{6}\theta^{2}\ddot{G}f''+\frac{1}{18}\theta^{3}\dot{G}f''-\frac{3}{8\theta}G\dot{G}f''-\frac{1}{6}\theta^{2}\dot{G}^{2}f''' \right)\left(-\frac{4\rho_{d}}{3\rho_{d}+4\rho_{r}}\Delta_{d}^{k} \right)\nonumber\\
 &&\quad\quad+\frac{1}{3}\frac{k^{2}\rho_{d}}{a^{2}(3\rho_{d}+4\rho_{r})}\Delta_{d}^{k}
  +\{\frac{4}{9}\theta^{3}f''\}\dot{\mathcal{G}}^{k}-\frac{c^{2}_{s}\dot{G}\theta^{3}f''\rho_{d}}{3(\rho_{d}+\rho_{r})}\Delta_{d}^{k}-\{\frac{9}{4\theta^{3}}
  +\frac{G}{2\theta}f'+f''(\frac{8\theta^{2}\dot{G}}{9}-\frac{6G\dot{G}}{2\theta^{2}}-\frac{4}{3}\theta\ddot{G})\nonumber\\
  &&\quad\quad-\frac{1}{2\theta}f-\frac{2}{3}\theta 
  -\frac{4}{3}\theta \dot{G}^{2}f''' \}\dot{\Delta}_{d}^{k}\;,
 \end{eqnarray}
arranging terms and simplifying to have
\begin{eqnarray}
  &&\dot{Z}^{k}=-\{\frac{9}{4\theta^{3}}
  +\frac{G}{2\theta}f'+f''(\frac{8\theta^{2}\dot{G}}{9}-\frac{6G\dot{G}}{2\theta^{2}}-\frac{4}{3}\theta\ddot{G})-\frac{1}{2\theta}f-\frac{2}{3}\theta 
  -\frac{4}{3}\theta \dot{G}^{2}f''' \}\dot{\Delta}_{d}^{k}\nonumber\\
  &&\quad\quad+ \{(-\frac{1}{4}\theta^{2}-\frac{3}{4}\rho_{r}-\frac{3}{8} f+\frac{3}{4}Gf'-\frac{1}{2}\theta^{2}\ddot{G}f''+\frac{1}{3}\theta^{3}\dot{G}f''-\frac{9}{8\theta}G \dot{G}f''-\frac{1}{2}\theta^{2}\dot{G}^{2}f''' \nonumber\\
  &&\quad\quad+\frac{3k^{2}}{4a^{2}}-\frac{\theta^{3}\dot{G}f''}{3}) \left(\frac{c^{2}_{s}\rho_{d}}{\rho_{d}+\rho_{r}}\right)\nonumber\\
  &&\quad\quad+\left(\frac{1}{3}\theta^{2}+\rho_{r}+\frac{1}{2} f-\frac{1}{2}Gf'+\frac{2}{3}\theta^{2}\ddot{G}f''-\frac{2}{9}\theta^{3}\dot{G}f''+\frac{3}{2\theta}G\dot{G}f''+\frac{2}{3}\theta^{2}\dot{G}^{2}f''' \right)\left(\frac{\rho_{d}}{3\rho_{d}+4\rho_{r}} \right)\}\Delta_{d}^{k}\nonumber\\
  &&\quad\quad+\{f''(\theta f'-G) -\frac{3G\dot{G}f'''}{2\theta}-\frac{2}{3}\theta^{2}\dot{G}^{2}f^{iv}-\frac{2}{3}\theta^{2}\ddot{G}f'''+\frac{1}{2}f''
 +\frac{2}{9}\theta^{3}\dot{G}f'''\nonumber\\
  &&\quad\quad
  +\frac{f'''}{f''}\left(\frac{9G}{16\theta^{2}}+\frac{\theta^{2}}{6}-f+Gf'-\frac{2\theta^{2}\dot{G}^{2}}{3}f'''\right)\}\mathcal{G}^{k}
  +\frac{4}{9}\theta^{3}f''\dot{\mathcal{G}}^{k}\;,
 \end{eqnarray}

so that the second-order equation in $\Delta_{d}^{k}$ becomes
\begin{multline}
 \ddot{\Delta}_{d}^{k}=\{\frac{G}{2\theta}f'+f''(\frac{9}{4\theta^{3}}
  +\frac{8\theta^{2}\dot{G}}{9}-\frac{6G\dot{G}}{2\theta^{2}}-\frac{4}{3}\theta\ddot{G})-\frac{1}{2\theta}f-\frac{2}{3}\theta 
  -\frac{4}{3}\theta \dot{G}^{2}f''' \}\dot{\Delta}_{d}^{k}\\ -\{(-\frac{1}{4}\theta^{2}-\frac{3}{4}\rho_{r}-\frac{3}{8} f+\frac{3}{4}Gf'-\frac{1}{2}\theta^{2}\ddot{G}f''+\frac{1}{3}\theta^{3}\dot{G}f''-\frac{9}{8\theta}G \dot{G}f''-\frac{1}{2}\theta^{2}\dot{G}^{2}f''' \\+\frac{3k^{2}}{4a^{2}}-\frac{\theta^{3}\dot{G}f''}{3}) \left(\frac{c^{2}_{s}\rho_{d}}{\rho_{d}+\rho_{r}}\right)\\-\left(\frac{1}{3}\theta^{2}+\rho_{r}+\frac{1}{2} f-\frac{1}{2}Gf'+\frac{2}{3}\theta^{2}\ddot{G}f''-\frac{2}{9}\theta^{3}\dot{G}f''+\frac{3}{2\theta}G\dot{G}f''+\frac{2}{3}\theta^{2}\dot{G}^{2}f''' \right)\left(\frac{\rho_{d}}{3\rho_{d}+4\rho_{r}} \right)\}\Delta_{d}^{k}\\ -\{f''(\theta f'-G) -\frac{3G\dot{G}f'''}{2\theta}-\frac{2}{3}\theta^{2}\dot{G}^{2}f^{iv}-\frac{2}{3}\theta^{2}\ddot{G}f'''+\frac{1}{2}f''
 +\frac{2}{9}\theta^{3}\dot{G}f'''\\
  +\frac{f'''}{f''}\left(\frac{9G}{16\theta^{2}}+\frac{\theta^{2}}{6}-f+Gf'-\frac{2\theta^{2}\dot{G}^{2}}{3}f'''\right)\}\mathcal{G}^{k}
  -\{\frac{4}{9}\theta^{3}f''\}\dot{\mathcal{G}}^{k}.~~~~~~~~~~~~~~~~~~~~~~~~~~~~~~~~~~~~~~~~~~~~~~~~~~~~~~~~~~~~~~~~~~~~~~~~~~~~~~~~
\end{multline}
Since we are dealing with the epoch where radiation dominates over dust, we also assume that the energy density of radiation is much larger
than that of dust, that is $\rho_{d} \ll \rho_{r}$ , so that the leading equation becomes
\begin{eqnarray}
 &&\ddot{\Delta}_{d}^{k}=\{\frac{G}{2\theta}f'+f''(\frac{9}{4\theta^{3}}
  +\frac{8\theta^{2}\dot{G}}{9}-\frac{6G\dot{G}}{2\theta^{2}}-\frac{4}{3}\theta\ddot{G})-\frac{1}{2\theta}f+\frac{2}{3}\theta 
  -\frac{4}{3}\theta \dot{G}^{2}f''' \}\dot{\Delta}_{d}^{k}\nonumber\\
  &&\quad\quad-\{(-\frac{1}{4}\theta^{2}-\frac{3}{4}\rho_{r}-\frac{3}{8} f+\frac{3}{4}Gf'-\frac{1}{2}\theta^{2}\ddot{G}f''+\frac{1}{3}\theta^{3}\dot{G}f''-\frac{9}{8\theta}G \dot{G}f''-\frac{1}{2}\theta^{2}\dot{G}^{2}f''' \nonumber\\
  &&\quad\quad+\frac{3k^{2}}{4a^{2}}-\frac{\theta^{3}\dot{G}f''}{3}) \left(\frac{c^{2}_{s}\rho_{d}}{\rho_{r}}\right)\nonumber\\
  &&\quad\quad-\left(\frac{1}{3}\theta^{2}+\rho_{r}+\frac{1}{2} f-\frac{1}{2}Gf'+\frac{2}{3}\theta^{2}\ddot{G}f''-\frac{2}{9}\theta^{3}\dot{G}f''+\frac{3}{2\theta}G\dot{G}f''+\frac{2}{3}\theta^{2}\dot{G}^{2}f''' \right)\left(\frac{\rho_{d}}{\rho_{r}} \right)\}\Delta_{d}^{k}\nonumber\\ 
  &&\quad\quad-\{f''(\theta f'-G) -\frac{3G\dot{G}f'''}{2\theta}-\frac{2}{3}\theta^{2}\dot{G}^{2}f^{iv}-\frac{2}{3}\theta^{2}\ddot{G}f'''+\frac{1}{2}f''
 +\frac{2}{9}\theta^{3}\dot{G}f'''\nonumber\\
 &&\quad\quad
  +\frac{f'''}{f''}\left(\frac{9G}{16\theta^{2}}+\frac{\theta^{2}}{6}-f+Gf'-\frac{2\theta^{2}\dot{G}^{2}}{3}f'''\right)\}\mathcal{G}^{k}
  -\frac{4}{9}\theta^{3}f''\dot{\mathcal{G}}^{k}.
\end{eqnarray}
We approximate once more that the product of the dust energy density perturbation
$\Delta^{k}_{d} $ and dust energy density $\rho_{d}$ are small enough so that we neglect $\frac{\rho_{d}\Delta^{k}_{d}}{\rho_{r}} $ over
the other terms. Thus, we have
\begin{multline}
 \ddot{\Delta}_{d}^{k}=\{\frac{9}{4\theta^{3}}f''
  +\frac{G}{2\theta}f'+f''(\frac{8\theta^{2}\dot{G}}{9}-\frac{6G\dot{G}}{2\theta^{2}}-\frac{4}{3}\theta\ddot{G})-\frac{1}{2\theta}f+\frac{2}{3}\theta 
  -\frac{4}{3}\theta \dot{G}^{2}f''' \}\dot{\Delta}_{d}^{k}+\frac{1}{4}\rho_{d}\Delta_{d}^{k}\\ -\{f''(\theta f'-G) -\frac{3G\dot{G}f'''}{2\theta}-\frac{2}{3}\theta^{2}\dot{G}^{2}f^{iv}-\frac{2}{3}\theta^{2}\ddot{G}f'''+\frac{1}{2}f''
 +\frac{2}{9}\theta^{3}\dot{G}f'''\\
  +\frac{f'''}{f''}\left(\frac{9G}{16\theta^{2}}+\frac{\theta^{2}}{6}-f+Gf'-\frac{2\theta^{2}\dot{G}^{2}}{3}f'''\right)\}\mathcal{G}^{k}
  -\frac{4}{9}\theta^{3}f''\dot{\mathcal{G}}^{k}.~~~~~~~~~~~~~~~~~~~~~~~~~~~~~~~~~~~~~~~~~~~~~~~~~~~~~~~~~~~~~~~~~~~~~~~~~~~~~~~~~~~~~~~
\end{multline}
 The evolution equation for the momentum of Gauss-Bonnet parameter is presented as
\begin{multline}
  \dot{\mathsf{G}}^{k}=\{-\frac{\theta}{3}-\frac{9}{4}\frac{G}{\theta^{3}}-2\frac{f'''\dot{G}}{f''}\}\dot{\mathcal{G}}^{k}+\{\frac{1}{\theta^{2}f''}(\frac{27}{32\theta^{2}}+\frac{3}{2}f')\\
  +\frac{f'''}{\theta^{2}f''}\left(-\frac{27}{32}\frac{G}{\theta^{2}}-\frac{\theta^{2}}{4}
+\frac{3}{2}f-\frac{3}{2}Gf'+\dot{G}^{2}\theta^{2}f'''\right)-\frac{3}{2\theta}f'+\frac{3}{2\theta^{2}}G\\
  -\frac{9}{4\theta^{3}}\dot{G}-\frac{\dot{G}^{2}f^{iv}}{f''}\}\mathcal{G}^{k}
  +\{\frac{1}{\theta^{2}f''}\left(-\frac{27}{8\theta^{3}}+\frac{3}{\theta}f-\frac{3}{\theta}Gf'\right)-\frac{\dot{G}}{3}+\frac{27}{4\theta^{4}}G\dot{G}\}\dot{\Delta}^{k}_{d}.~~~~~~~~~~~~~~~~~~~~~~~~~~~~~~~~~~~~~~~~~~~~~~~~~~~~~~~~~~~~~~~~~~~~~~~~~~~~~~
 \end{multline}
 Using \begin{equation*}
\dot{ \mathcal{G}}^{k}=\mathsf{G}^{k},
 \end{equation*}
we have
\begin{equation}
\ddot{ \mathcal{G}}^{k}=\dot{\mathsf{G}}^{k},
 \end{equation}
 so that 
 \begin{multline}
\ddot{ \mathcal{G}}^{k}=
\{-\frac{\theta}{3}-\frac{9}{4}\frac{G}{\theta^{3}}-2\frac{f'''\dot{G}}{f''}\}\dot{\mathcal{G}}^{k}+\{\frac{1}{\theta^{2}f''}(\frac{27}{32\theta^{2}}+\frac{3}{2}f')\\
  +\frac{f'''}{\theta^{2}f''}\left(-\frac{27}{32}\frac{G}{\theta^{2}}-\frac{\theta^{2}}{4}
+\frac{3}{2}f-\frac{3}{2}Gf'+\dot{G}^{2}\theta^{2}f'''\right)-\frac{3}{2\theta}Gf'+\frac{3}{2\theta}G\\
  -\frac{9}{4\theta^{3}}f''\dot{G}-\frac{\dot{G}^{2}f^{iv}}{f''}\}\mathcal{G}^{k}
  +\{\frac{1}{\theta^{2}f''}\left(-\frac{27}{8\theta^{3}}+\frac{3}{\theta}f-\frac{3}{\theta}Gf'\right)-\frac{\dot{G}}{3}+\frac{27}{4\theta^{4}}G\dot{G}\}\dot{\Delta}^{k}_{d}.~~~~~~~~~~~~~~~~~~~~~~~~~~~~~~~~~~~~~~~~~~~~~~~~~~~~~~~~~~~~~~~~~~~~~~~~~~~~
  \end{multline}
  Using quasi-static approximation, we have
  \begin{multline}
 \ddot{\Delta}_{d}^{k}=\{
  \frac{G}{2\theta}f'+f''(\frac{9}{4\theta^{3}}+\frac{8\theta^{2}\dot{G}}{9}-\frac{6G\dot{G}}{2\theta^{2}}-\frac{4}{3}\theta\ddot{G})-\frac{1}{2\theta}f+\frac{2}{3}\theta 
  -\frac{4}{3}\theta \dot{G}^{2}f''' \}\dot{\Delta}_{d}^{k}+\frac{1}{4}\rho_{d}\Delta_{d}^{k}\\ -\{f''(\theta f'-G) -\frac{3G\dot{G}f'''}{2\theta}-\frac{2}{3}\theta^{2}\dot{G}^{2}f^{iv}-\frac{2}{3}\theta^{2}\ddot{G}f'''+\frac{1}{2}f''
 +\frac{2}{9}\theta^{3}\dot{G}f'''\\
  +\frac{f'''}{f''}\left(\frac{9G}{16\theta^{2}}+\frac{\theta^{2}}{6}-f+Gf'-\frac{2\theta^{2}\dot{G}^{2}}{3}f'''\right)\}\mathcal{G}^{k}.~~~~~~~~~~~~~~~~~~~~~~~~~~~~~~~~~~~~~~~~~~~~~~~~~~~~~~~~~~~~~~~~~~~~~~~~~~~~~~~~~~~~~~~~~~~~~~~~~~~~~~~~~~~~~
  \label{eq136}
\end{multline}
For the case $f(G)=G$, we have, \cite{abebe2012covariant}
\begin{equation}
 \ddot{\Delta}_{d}^{k}-\frac{2}{3}\theta \dot{\Delta}_{d}^{k}-\frac{1}{4}\rho_{d}\Delta_{d}^{k}=0,
 \label{eq160}
\end{equation}
and
\begin{equation}
 \mathcal{G}=0.
\end{equation}

\begin{figure}[ht!]
\begin{center}
 \subfigure[ Numerical solutions of Eq. \ref{eq162} using polynomial $f(G)$ model (eq.\ref{eq28}) ]{%
 \label{fig: first }
 \includegraphics[width=80mm]{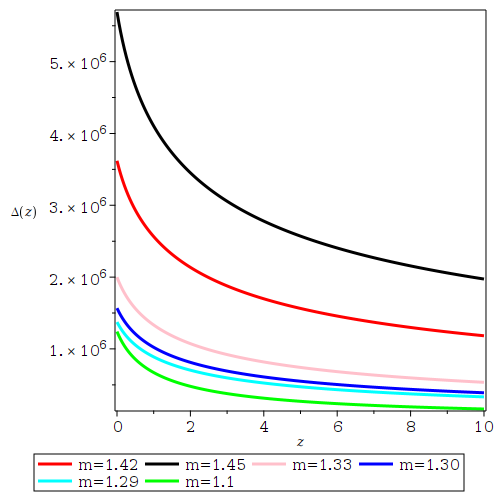}
 }%
 \subfigure[Numerical solutions of Eq. \ref{eq164} for the case $f(G)=G$]{%
 \label{fig: second}
 \includegraphics[width=60mm, height=80mm]{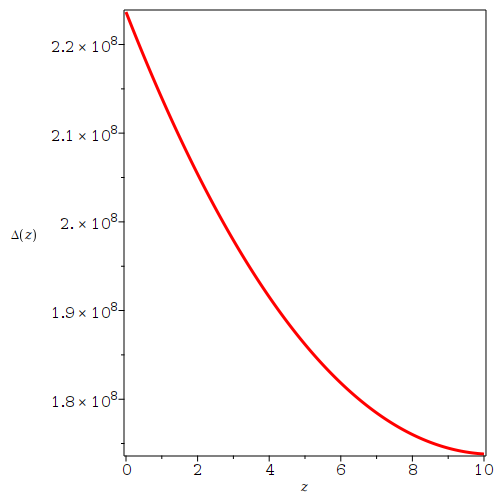}}%
\end{center}
 \caption{%
Plot of energy density perturbations $\Delta(z)$ versus redshift $z$ of  Eq. \ref{eq162} and Eq. \ref{eq164} for short wavelength modes in a radiation dominated Universe using different values of $m$. }%
\label{Fig(1)}
\end{figure}

These equations are  similar to the ones obtained in \cite{abebe2012covariant, ntahompagaze2020multifluid} for the GR limits perturbation equations.
The equations Eq. \ref{eq136} and Eq. \ref{eq160} are time dependent and need to be transformed into redshift space in order to be able to get numerical solutions and  to make comparison with cosmological observation. We use 
\begin{equation}
 a=\frac{1}{1+z},
 \label{eq162}
\end{equation}
to get
the redshift transformed equations and presented  as
\begin{multline}
(1+z)^{2}H^{2} \Delta''_{d}=-(1+z)H\{\frac{G}{2\theta}f'+f''(\frac{9}{4\theta^{3}}
  +\frac{8\theta^{2}G'}{9}-\frac{6GG'}{2\theta^{2}}-\frac{4}{3}\theta G'')-\frac{1}{2\theta}f-\frac{2}{3}\theta \\
  -\frac{4}{3}\theta G'^{2}f''' +(1+z)H'\}\Delta'_{d}+\frac{1}{4}\Delta_{d} -\{f''(\theta f'-G) -\frac{3GG'f'''}{2\theta}-\frac{2}{3}\theta^{2}G'^{2}f^{iv}-\frac{2}{3}\theta^{2}G''f'''\\+\frac{1}{2}f''
 +\frac{2}{9}\theta^{3}G'f'''
  +\frac{f'''}{f''}\left(\frac{9G}{16\theta^{2}}+\frac{\theta^{2}}{6}-f+Gf'-\frac{2\theta^{2}G'^{2}}{3}f'''\right)\}\mathcal{G},~~~~~~~~~~~~~~~~~~~~~~~~~~~~~~~~~~~~~~~~~~~~~~~~~~~~~~~~~~~~~~~~~~~~~~~~~~~
  \label{eq162}
\end{multline}
\begin{equation}
 \Delta''_{d}-\frac{1}{1+z}\Delta'_{d}-\frac{3}{4(1+z)^{2}}\Delta_{d}=0,
 \label{eq164}
\end{equation}
where prime stands for derivative with respect to redshift and 
\begin{equation}
 \theta =\frac{2m}{1+w}(1+z)^{\frac{3(1+w)}{2m}},
 \label{eq165}
\end{equation}
\begin{equation}
 H=\frac{2m}{3(1+w)}(1+z)^{\frac{3(1+w)}{2}},
\end{equation}
and
\begin{equation}
 G=\frac{128}{27}(\frac{m}{1+w})^{4}(1+z)^{\frac{6(1+w)}{m}}(1-(1+z)^{\frac{3(1+z)}{2m}}).
 \label{eq167}
\end{equation}
Equation eq. \ref{eq164} solved analytically and the solution is presented as
\begin{equation}
 \Delta(z)=C_{1}(1+z)^{\alpha_{1}}+C_{2}(1+z)^{\alpha_{2}},
\end{equation}
where $\alpha_{1}=1+\frac{\sqrt{7}}{2}$ and $\alpha_{2}=1-\frac{\sqrt{7}}{2}$. The integration constants $C_{1}$ and $C_{2}$ can be determined by imposing the initial conditions at $z=0$. At $z=0$, we have
\begin{equation}
 \Delta(z=0)=C_{1}+C_{2}.
\end{equation}
\begin{equation}
 \Delta'(z)=C_{1}\alpha_{1}(1+z)^{\alpha_{1}-1}+C_{2}\alpha_{2}(1+z)^{\alpha_{2}-1},
\end{equation}
\begin{equation}
 \Delta'(z=0)=C_{1}\alpha_{1}+C_{2}\alpha_{2},
\end{equation}
Solving simultaneously, we have
\begin{equation}
 C_{1}=\frac{\alpha_{2}\Delta(z=0)-\Delta'(z=0)}{\alpha_{2}-\alpha_{1}},
\end{equation}
and
\begin{equation}
 C_{2}=\frac{\Delta'(z=0)-\alpha_{1}\Delta(z=0)}{\alpha_{2}-\alpha_{1}}.
\end{equation}
For dust dominated Universe, $w=0$  and for radiation dominated Universe, $w=\frac{1}{3}$. Equation eq. \ref{eq162} gives Bessel hypergeometrical solutions, we prefer to present them numerically.
We consider   $f(G)$ model: polynomial (eq. \ref{eq28}), to find numerical solutions for the perturabtion equations. Numerical solutions  of Eq. \ref{eq162} and Eq. \ref{eq164} are presented in Figure \ref{Fig(1)},  the considered $f(G)$  model and for the case $f(G)=G$.

\subsubsection{Perturbations in a dust-dominated epoch}
The energy density of radiation is much smaller than energy density of dust in
this dust dominated epoch, that is, 
$\rho_{ r}\ll \rho_{d} $.
We also assume that the perturbations due to radiation energy
density are small enough compared to the perturbations generated from dust, that is
$\Delta_{r} \ll \Delta_{d}$ .
We keep the
same assumption as we did in radiation dominated epoch of assuming
$\Delta_{r}\approx 0$ .
With the above assumption, the evolution equations governing this system are
\begin{equation}
 \dot{\Delta}_{d}^{k}=-Z^{k}
 + \frac{k^{2}}{a} V^{k}_{d}
\end{equation}
Using the assumption $c^{2}_ {sd} = c_{s}^{2} = 0$ and $w = w_{ d} = 0$, we have
\begin{multline}
  \dot{Z^{k}}=-\rho_{d}\Delta_{d}^{k}+\{f''(\theta f'-G+\frac{3\dot{G}}{2\theta}) -\frac{3G\dot{G}f'''}{2\theta}-\frac{3f''\dot{G}}{2\theta}-\frac{2}{3}\theta^{2}\dot{G}^{2}f^{iv}-\frac{2}{3}\theta^{2}\ddot{G}f'''\\
  +\frac{f'''}{f''}(\frac{9G}{16\theta^{2}}+\frac{\theta^{2}}{6}+\frac{1}{2}\rho_{d}-f+Gf'-\frac{2\theta^{2}\dot{G}^{2}}{3}f''')+\frac{1}{2}f''
 +\frac{2}{9}\theta^{3}\dot{G}f'''\}\mathcal{G}^{k}\\
  +\{\frac{4}{9}\theta^{3}f''\}\mathsf{G}^{k}+\{\frac{9}{4\theta^{3}}+\frac{1}{\theta}\rho_{d}
  +\frac{G}{2\theta}f'+f''(\frac{2\theta^{2}\dot{G}}{9}-\frac{9G\dot{G}}{2\theta^{2}})-\frac{1}{2\theta}f-\frac{2}{3}\theta +\frac{2}{3}\theta^{2}\dot{G}f''
 +\frac{3G\dot{G}f''}{2\theta^{2}}\\
  -\frac{4}{3}\theta \dot{G}^{2}f'''-\frac{4}{3}\theta\ddot{G}f'' \}Z^{k},~~~~~~~~~~~~~~~~~~~~~~~~~~~~~~~~~~~~~~~~~~~~~~~~~~~~~~~~~~~~~~~~~~~~~~~~~~~~~~~~~~~~~~~~~~~~~~~~~~~~~~~~~~~~~~~~~~~~~~~~~~~~~~~~~~~~~~~~~~~
 \end{multline}
 \begin{equation}
\dot{ \mathcal{G}}^{k}=\mathsf{G}^{k},
 \end{equation}
 \begin{multline}
  \dot{\mathsf{G}}^{k}= \frac{3}{4}\frac{\rho_{d}}{\theta^{2}f''}\Delta_{d}^{k}+\{-\frac{\theta}{3}-\frac{9}{4}\frac{G}{\theta^{3}}-2\frac{f'''\dot{G}}{f''}\}\mathsf{G}^{k}+\{\frac{1}{\theta^{2}f''}(\frac{27}{32\theta^{2}}+\frac{3}{2}f')\\
  +\frac{f'''}{\theta^{2}f''}\left(-\frac{27}{32}\frac{G}{\theta^{2}}-\frac{\theta^{2}}{4}
  -\frac{3}{4}\rho_{d}+\frac{3}{2}f-\frac{3}{2}Gf'+\dot{G}^{2}\theta^{2}f'''\right)-\frac{3}{2\theta}f'+\frac{3}{2\theta^{2}}G\\
  -\frac{9}{4\theta}^{3}\dot{G}-\frac{\dot{G}^{2}f^{iv}}{f''}\}\mathcal{G}^{k}
  +\{\frac{1}{\theta^{2}f''}\left(-\frac{27}{8\theta^{3}}-\frac{3\rho_{d}}{2\theta}+\frac{3}{\theta}f-\frac{3}{\theta}Gf'\right)-\frac{\dot{G}}{3}+\frac{27}{4\theta^{4}}G\dot{G}\}Z^{k},~~~~~~~~~~~~~~~~~~~~~~~~~~~~~~~~~~~~~~~~~~~~~~~~~~~~~~~~~~~~~~
 \end{multline}
\begin{equation}
 \dot{V}^{k}_{d}=-\frac{1}{3}\theta V^{k}_{d},
\end{equation}
\begin{equation}
\dot{V}^{k}_{dr}=-\frac{1}{3} \theta V^{k}_{d},
\end{equation}
and
\begin{equation}
  \dot{S}^{k}_{dr}=\frac{k^{2}}{a^{2}} V^{k}_{dr}.
 \end{equation}
 The second order evolution equation in energy density and Gauss-Bonnet parameter perturbations are given by
\begin{multline}
 \ddot{\Delta}_{d}^{k}=\rho_{d}\Delta_{d}^{k}-\{f''(\theta f'-G+\frac{3\dot{G}}{2\theta}) -\frac{3G\dot{G}f'''}{2\theta}-\frac{3f''\dot{G}}{2\theta}-\frac{2}{3}\theta^{2}\dot{G}^{2}f^{iv}-\frac{2}{3}\theta^{2}\ddot{G}f'''\\
  +\frac{f'''}{f''}(\frac{9G}{16\theta^{2}}+\frac{\theta^{2}}{6}+\frac{1}{2}\rho_{d}-f+Gf'-\frac{2\theta^{2}\dot{G}^{2}}{3}f''')+\frac{1}{2}f''
 +\frac{2}{9}\theta^{3}\dot{G}f'''\}\mathcal{G}^{k}\\
  +\{\frac{4}{9}\theta^{3}f''\}\mathsf{G}^{k}+\{\frac{9}{4\theta^{3}}+\frac{1}{\theta}\rho_{d}
  +\frac{G}{2\theta}f'+f''(\frac{2\theta^{2}\dot{G}}{9}-\frac{9G\dot{G}}{2\theta^{2}})-\frac{1}{2\theta}f-\frac{2}{3}\theta +\frac{2}{3}\theta^{2}\dot{G}f''
 +\frac{3G\dot{G}f''}{2\theta^{2}}\\
  -\frac{4}{3}\theta \dot{G}^{2}f'''-\frac{4}{3}\theta\ddot{G}f'' \}\dot{\Delta}^{k}_{d},~~~~~~~~~~~~~~~~~~~~~~~~~~~~~~~~~~~~~~~~~~~~~~~~~~~~~~~~~~~~~~~~~~~~~~~~~~~~~~~~~~~~~~~~~~~~~~~~~~~~~~~~~~~~~~~~~~~~~~~~~~~~~
\end{multline}
and
 \begin{multline}
\ddot{ \mathcal{G}}^{k}= \frac{3}{4}\frac{\rho_{d}}{\theta^{2}f''}\Delta_{d}^{k}+\{-\frac{\theta}{3}-\frac{9}{4}\frac{G}{\theta^{3}}-2\frac{f'''\dot{G}}{f''}\}\dot{\mathcal{G}}^{k}+\{\frac{1}{\theta^{2}f''}(\frac{27}{32\theta^{2}}+\frac{3}{2}f')\\
  +\frac{f'''}{\theta^{2}f''}\left(-\frac{27}{32}\frac{G}{\theta^{2}}-\frac{\theta^{2}}{4}
  -\frac{3}{4}\rho_{d}+\frac{3}{2}f-\frac{3}{2}Gf'+\dot{G}^{2}\theta^{2}f'''\right)-\frac{3}{2\theta}f'+\frac{3}{2\theta^{2}}G\\
  -\frac{9}{4\theta^{3}}\dot{G}-\frac{\dot{G}^{2}f^{iv}}{f''}\}\mathcal{G}^{k}
  -\{\frac{1}{\theta^{2}f''}\left(-\frac{27}{8\theta^{3}}-\frac{3\rho_{d}}{2\theta}+\frac{3}{\theta}f-\frac{3}{\theta}Gf'\right)-\frac{\dot{G}}{3}+\frac{27}{4\theta^{4}}G\dot{G}\}\dot{\Delta}_{d}^{k}.~~~~~~~~~~~~~~~~~~~~~~~~~~~~~~~~~~~~~~~~~~~~~~~~~~~~~~~
 \end{multline}

\begin{figure}[ht!]
\begin{center}
 \subfigure[Numerical solutions of Eq. \ref{eq186} using polynomial $f(G)$ model (eq. \ref{eq28})]{%
 \label{fig: first }
 \includegraphics[width=60mm]{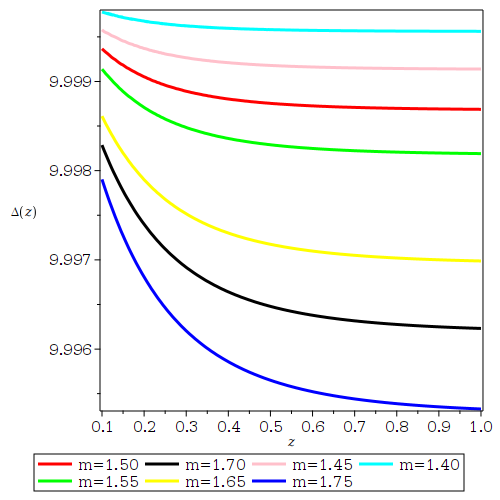}
 }%
 \subfigure[Numerical solutions of Eq. \ref{eq187} for the case $f(G)=G$]{%
 \label{fig: second}
 \includegraphics[width=60mm, height=60mm]{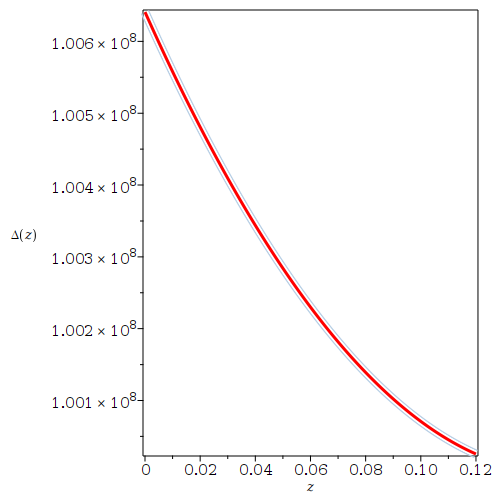}}%
\end{center}
 \caption{%
Plot of energy density perturbations $\Delta(z)$ versus redshift $z$ of Eq. \ref{eq186} and Eq. \ref{eq187} 
for  short wavelength modes in a dust dominated Universe using different values of $m$.}%
\label{Fig(2)}
\end{figure}

By applying the quasi-static approximation, it meas $\dot{\mathcal{G}}$ and $\ddot{\mathcal{G}}$ is set to be equal to zero, we have
\begin{multline}
 \ddot{\Delta}_{d}^{k}=
\{\frac{1}{\theta}\rho_{d}
  +\frac{G}{2\theta}f'+f''(\frac{9}{4\theta^{3}}f''+\frac{8\theta^{2}\dot{G}}{9}-\frac{3G\dot{G}}{\theta^{2}}-\frac{4}{3}\theta\ddot{G})-\frac{1}{2\theta}f-\frac{2}{3}\theta
  -\frac{4}{3}\theta \dot{G}^{2}f''' \}\dot{\Delta}^{k}_{d}\\+\rho_{d}\Delta_{d}^{k}-\{f''(\theta f'-G) -\frac{3G\dot{G}f'''}{2\theta}-\frac{2}{3}\theta^{2}\dot{G}^{2}f^{iv}-\frac{2}{3}\theta^{2}\ddot{G}f'''\\
  +\frac{f'''}{f''}(\frac{9G}{16\theta^{2}}+\frac{\theta^{2}}{6}+\frac{1}{2}\rho_{d}-f+Gf'-\frac{2\theta^{2}\dot{G}^{2}}{3}f''')+\frac{1}{2}f''
 +\frac{2}{9}\theta^{3}\dot{G}f'''\}\mathcal{G}^{k}.~~~~~~~~~~~~~~~~~~~~~~~~~~~~~~~~~~~~~~~~~~~~~~~~~~~~~~~~~~~~~~~~~~~~~~~~~~~~~~~~~~~~~
 \label{eq185}
\end{multline}
For the case $f(G)=G$, we have \cite{munyeshyaka2021cosmological, ntahompagaze2020multifluid, abebe2012covariant}
\begin{equation}
 \ddot{\Delta}_{d}^{k}+(-\frac{\rho_{d}}{\theta}+\frac{2}{3}\theta )\dot{\Delta}_{d}^{k}-\rho_{d}\Delta_{d}^{k}=0.
 \label{eq186}
\end{equation}
Using the same transformation scheme as in a  the radiation dominated Universe (Eq. \ref{eq162} and Eq. \ref{eq165} through to Eq. \ref{eq167}), Eq. \ref{eq185} and Eq. \ref{eq186} transformed as 
\begin{multline}
 (1+z)^{2} H^{2} \Delta''_{d}=-(1+z)H
\{\frac{9}{4\theta^{3}}+\frac{1}{\theta}\rho_{d}
  +\frac{G}{2\theta}f'+f''(\frac{8\theta^{2}G'}{9}-\frac{3GG'}{\theta^{2}}-\frac{4}{3}\theta G'')-\frac{1}{2\theta}f-\frac{2}{3}\theta\\
  -\frac{4}{3}\theta G'^{2}f''' +(1+z)H'\}\Delta'_{d}+\rho_{d}\Delta_{d}-\{f''(\theta f'-G) -\frac{3G\dot{G}f'''}{2\theta}-\frac{2}{3}\theta^{2}G'^{2}f^{iv}-\frac{2}{3}\theta^{2}G''f'''\\
  +\frac{f'''}{f''}(\frac{9G}{16\theta^{2}}+\frac{\theta^{2}}{6}+\frac{1}{2}\rho_{d}-f+Gf'-\frac{2\theta^{2}G'^{2}}{3}f''')+\frac{1}{2}f''
 +\frac{2}{9}\theta^{3}G'f'''\}\mathcal{G},~~~~~~~~~~~~~~~~~~~~~~~~~~~~~~~~~~~~~~~~~~~~~~~~~~~~~~~~~~~~~~~~~~~~~~~~~~~~
 \label{eq186}
\end{multline}
and
\begin{equation}
 \Delta''_{d}-\frac{2}{1+z}\Delta'_{d}-\frac{3}{(1+z)^{2}}\Delta_{d}=0.
 \label{eq187}
\end{equation}
This equation solved analytically and admit the solution
\begin{equation}
 \Delta(z)=C_{3}(1+z)^{\beta_{1}}+C_{4}(1+z)^{\beta_{2}},
\end{equation}
with $\beta_{1}=\frac{3}{2}+\frac{\sqrt{21}}{2}$ and $\beta_{2}=\frac{3}{2}-\frac{\sqrt{21}}{2}$.
The integration constants $C_{3}$ and $C_{4}$ can be determined by imposing the initial conditions at $z=0$. At $z=0$, we have
\begin{equation}
 \Delta(z=0)=C_{1}+C_{2}.
\end{equation}
\begin{equation}
 \Delta'(z)=C_{3}\beta_{1}(1+z)^{\beta_{1}-1}+C_{4}\beta_{2}(1+z)^{\beta_{2}-1},
\end{equation}
\begin{equation}
 \Delta'(z=0)=C_{3}\beta_{1}+C_{4}\beta_{2},
\end{equation}
Solving simultaneously, we have
\begin{equation}
 C_{3}=\frac{\beta_{2}\Delta(z=0)-\Delta'(z=0)}{\beta_{2}-\beta_{1}},
\end{equation}
and
\begin{equation}
 C_{4}=\frac{\Delta'(z=0)-\beta_{1}\Delta(z=0)}{\beta_{2}-\beta_{1}}.
\end{equation}
Numerical solutions of Eq. \ref{eq186} and Eq. \ref{eq187} are presented in Figure  \ref{Fig(2)} for  polynomial model  and for the case $f(G)=G$.

\subsection{Long wavelength solutions}
In this section we analyze the evolution of energy density and Gauss-Bonnet parameter perturbations in the long wavelength limit. In this limit the wavenumber $k$ is so small that
$
\lambda = \frac{2 \pi a}{
k} \gg \lambda_{ H}$ , it means, $\frac
{k^{2}}{a^{ 2} H^{ 2}} \ll 1$. All Laplacian terms can therefore be neglected. We focus our interests on radiation-dominated epoch where the interaction between component fluids is neglected.
\subsubsection{Perturbations in the radiation-dominated epoch}
In this context of radiation dominated over dust component, we assume the homogeneity
of radiation energy density with flat universe K = 0. These assumption results in writing
$\Delta_{ r}\approx 0$. Therefore our leading equations become 
\begin{equation}
 \dot{\Delta}_{d}^{k}=-Z^{k}
 -\frac{1}{3}
 \theta \Delta_{m}^{k},
\end{equation}
\begin{multline}
  \dot{Z^{k}}=\left(-\frac{1}{4}\theta^{2}-\frac{3}{4}\rho_{r}-\frac{3}{8} f+\frac{3}{4}Gf'-\frac{1}{2}\theta^{2}\ddot{G}f''+\frac{1}{3}\theta^{3}\dot{G}f''-\frac{9}{8\theta}G \dot{G}f''-\frac{1}{2}\theta^{2}\dot{G}^{2}f''' \right) \frac{1}{3}\Delta_{m}^{k}\\  +\{f''(\theta f'-G) -\frac{3G\dot{G}f'''}{2\theta}-\frac{2}{3}\theta^{2}\dot{G}^{2}f^{iv}-\frac{2}{3}\theta^{2}\ddot{G}f'''\\
  +\frac{f'''}{f''}(\frac{9G}{16\theta^{2}}+\frac{\theta^{2}}{6}-f+Gf'-\frac{2\theta^{2}\dot{G}^{2}}{3}f''')+\frac{1}{2}f''
 +\frac{2}{9}\theta^{3}\dot{G}f'''\}\mathcal{G}^{k}\\+\left(-\frac{1}{12}\theta^{2}-\frac{1}{4}\rho_{r}-\frac{1}{8} f+\frac{1}{8}Gf'-\frac{1}{6}\theta^{2}\ddot{G}f''+\frac{1}{18}\theta^{3}\dot{G}f''-\frac{3}{8\theta}G\dot{G}f''-\frac{1}{6}\theta^{2}\dot{G}^{2}f''' \right)\left(\varepsilon^{k} \right)\\
  +\{\frac{4}{9}\theta^{3}f''\}\mathsf{G}^{k}+\{\frac{9}{4\theta^{3}}
  +\frac{G}{2\theta}f'+f''(\frac{8\theta^{2}\dot{G}}{9}-\frac{6G\dot{G}}{2\theta^{2}}-\frac{4}{3}\theta\ddot{G})-\frac{1}{2\theta}f-\frac{2}{3}\theta 
  -\frac{4}{3}\theta \dot{G}^{2}f''' \}Z^{k},~~~~~~~~~~~~~~~~~~~~~~~~~~~~~~~~~~~~~~~~~~~~~~~~~~~~
 \end{multline}
 \begin{equation}
\dot{ \mathcal{G}}^{k}=\mathsf{G}^{k}+\frac{1}{4}\dot{G}\Delta_{m}^{k}-\frac{1}{3}\dot{G}\varepsilon^{k},
 \end{equation}
 \begin{multline}
  \dot{\mathsf{G}}^{k}=\{-\frac{1}{4}\ddot{G}\}\Delta_{m}^{k}+\{-\frac{\theta}{3}-\frac{9}{4}\frac{G}{\theta^{3}}-2\frac{f'''\dot{G}}{f''}\}\mathsf{G}^{k}+\{\frac{1}{\theta^{2}f''}(\frac{27}{32\theta^{2}}+\frac{3}{2}f')\\
  +\frac{f'''}{\theta^{2}f''}\left(-\frac{27}{32}\frac{G}{\theta^{2}}-\frac{\theta^{2}}{4}
+\frac{3}{2}f-\frac{3}{2}Gf'+\dot{G}^{2}\theta^{2}f'''\right)-\frac{3}{2\theta}f'+\frac{3}{2\theta^{2}}G\\
  -\frac{9}{4\theta}^{3}\dot{G}-\frac{\dot{G}^{2}f^{iv}}{f''}\}\mathcal{G}^{k}
  +\{\frac{1}{\theta^{2}f''}\left(-\frac{27}{8\theta^{3}}+\frac{3}{\theta}f-\frac{3}{\theta}Gf'\right)-\frac{\dot{G}}{3}+\frac{27}{4\theta^{4}}G\dot{G}\}Z^{k}
  +\frac{1}{4}\ddot{G}\varepsilon^{k},~~~~~~~~~~~~~~~~~~~~~~~~~~~~~~~~~~~~~~~~~~~~~~~~~~~~~~
 \end{multline}
\begin{equation}
 \dot{V}^{k}_{d}=-\frac{1}{3}\theta V^{k}_{d}+\frac{1}{a(1+w)}\left(c^{2}_{s}\Delta^{k} +w \varepsilon^{k} \right),
\end{equation}
\begin{multline}
\dot{V}^{k}_{dr}=-\frac{1}{3}\theta V^{k}_{d}-\left(\frac{c^{2}_{sd}}{a}-\frac{c^{2}_{sr}}{a}\right)\Delta^{k}_{d}-\frac{c^{2}_{sr}}{a} S^{k}_{dr}+\frac{w_{r}}{a(1+w_{r})}\varepsilon^{k}_{r},~~~~~~~~~~~~~~~~~~~~~~~~~~~~~~~~~~~~~~~~~~~~~~~~~~~~~~~~~~~~~~~~~~~~~~~~~~~~~~~~~~~~~~~~~~~~~
\end{multline}
for
a radiation-dust mixture, the equation for the evolution of entropy perturbations is given by
\begin{equation}
  \dot{S}^{k}_{dr}=0.
 \end{equation}
We use $w=c_{sr}=\frac{1}{3}$ and $c_{sd}=0$.
 Knowing that $\Delta^{k}_{m}=\frac{\rho_{d}}{\rho_{d}+\rho_{r}}\Delta^{k}_{d}+\frac{\rho_{r}}{\rho_{d}+\rho_{r}}\Delta^{d}_{r}$
and for $\Delta^{k}_{r}\approx 0$, we have\\
$\Delta^{k}_{m}=\frac{\rho_{d}}{\rho_{d}+\rho_{r}}\Delta^{k}_{d}$,
$S^{k}_{dr}=\Delta^{k}_{d}$,
$\dot{S}^{k}_{dr}=\dot{\Delta}^{k}_{d}$ and 
$\varepsilon^{k}=-\frac{3\rho_{d}}{3\rho_{d}+4\rho_{r}}S^{k}_{dr}$ so that
the leading equations become
\begin{eqnarray}
 &&\dot{\Delta}_{d}^{k}=-Z^{k}
 -\frac{\theta \rho_{d}}{3(\rho_{d}+\rho_{r})}\Delta^{k}_{d}\;,\\
&&  \dot{Z^{k}}=\left(-\frac{1}{4}\theta^{2}-\frac{3}{4}\rho_{r}-\frac{3}{8} f+\frac{3}{4}Gf'-\frac{1}{2}\theta^{2}\ddot{G}f''+\frac{1}{3}\theta^{3}\dot{G}f''-\frac{9}{8\theta}G \dot{G}f''-\frac{1}{2}\theta^{2}\dot{G}^{2}f''' \right)\frac{\rho_{d}}{3(\rho_{d}+\rho_{r})}\Delta^{k}_{d}\nonumber\\
&&\quad\quad+\{f''(\theta f'-G) -\frac{3G\dot{G}f'''}{2\theta}-\frac{2}{3}\theta^{2}\dot{G}^{2}f^{iv}-\frac{2}{3}\theta^{2}\ddot{G}f'''
  +\frac{f'''}{f''}(\frac{9G}{16\theta^{2}}+\frac{\theta^{2}}{6}-f+Gf'\nonumber\\
 &&\quad\quad-\frac{2\theta^{2}\dot{G}^{2}}{3}f''')+\frac{1}{2}f''
 +\frac{2}{9}\theta^{3}\dot{G}f'''\}\mathcal{G}^{k}\nonumber\\
 &&\quad\quad+\left(-\frac{1}{12}\theta^{2}-\frac{1}{4}\rho_{r}-\frac{1}{8} f+\frac{1}{8}Gf'-\frac{1}{6}\theta^{2}\ddot{G}f''+\frac{1}{18}\theta^{3}\dot{G}f''-\frac{3}{8\theta}G\dot{G}f''-\frac{1}{6}\theta^{2}\dot{G}^{2}f''' \right)\left(-\frac{3\rho_{d}}{3\rho_{d}+4\rho_{r}}\Delta^{k}_{d}\right)\nonumber\\
 &&\quad\quad
  +\{\frac{4}{9}\theta^{3}f''\}\mathsf{G}^{k}+\{\frac{9}{4\theta^{3}}
  +\frac{G}{2\theta}f'+f''(\frac{8\theta^{2}\dot{G}}{9}-\frac{6G\dot{G}}{2\theta^{2}}-\frac{4}{3}\theta\ddot{G})-\frac{1}{2\theta}f-\frac{2}{3}\theta 
  -\frac{4}{3}\theta \dot{G}^{2}f''' \}Z^{k}\;,\\
&&\dot{ \mathcal{G}}^{k}=\mathsf{G}^{k}+\frac{1}{4}\dot{G}\frac{\rho_{d}}{\rho_{d}+\rho_{r}}\Delta_{d}^{k}+\dot{G}\frac{\rho_{d}}{3\rho_{d}+4\rho_{r}}\Delta_{d}^{k}\;,\\
&&  \dot{\mathsf{G}}^{k}=\{-\frac{1}{4}\ddot{G}\}\frac{\rho_{d}}{\rho_{d}+\rho_{r}}\Delta_{d}^{k}+\{-\frac{\theta}{3}-\frac{9}{4}\frac{G}{\theta^{3}}-2\frac{f'''\dot{G}}{f''}\}\mathsf{G}^{k}+\{\frac{1}{\theta^{2}f''}(\frac{27}{32\theta^{2}}+\frac{3}{2}f')\nonumber\\
&&\quad\quad
  +\frac{f'''}{\theta^{2}f''}\left(-\frac{27}{32}\frac{G}{\theta^{2}}-\frac{\theta^{2}}{4}
+\frac{3}{2}f-\frac{3}{2}Gf'+\dot{G}^{2}\theta^{2}f'''\right)-\frac{3}{2\theta}f'+\frac{3}{2\theta^{2}}G\nonumber\\
&&\quad\quad
  -\frac{9}{4\theta}^{3}\dot{G}-\frac{\dot{G}^{2}f^{iv}}{f''}\}\mathcal{G}^{k}
  +\{\frac{1}{\theta^{2}f''}\left(-\frac{27}{8\theta^{3}}+\frac{3}{\theta}f-\frac{3}{\theta}Gf'\right)-\frac{\dot{G}}{3}+\frac{27}{4\theta^{4}}G\dot{G}\}Z^{k}
  +\frac{1}{4}\ddot{G}\frac{3\rho_{d}}{3\rho_{d}+4\rho_{r}}\Delta^{k}_{d}\;,\\
 &&\dot{V}^{k}_{d}=-\frac{1}{3}\theta V^{k}_{d}\;,\\
&&\dot{V}^{k}_{dr}=-\frac{1}{3}\theta V^{k}_{d}\;,\\
 && \dot{S}^{k}_{dr}=0.
 \end{eqnarray}
Assume that the energy density of radiation is much larger than that
of dust, that is
$\rho_{d}  \ll \rho_{r}$ ,
and that the product of the dust energy density perturbation $\Delta_{d}^{k}$ and
dust energy density $\rho_{d}$ are small enough so that we neglect ${\rho_{d} \Delta_{d}^{k}}{\rho_{ r}}$, we have
\begin{equation}
 \dot{\Delta}_{d}^{k}=-Z^{k},
 \end{equation}
\begin{multline}
  \dot{Z^{k}}=\{f''(\theta f'-G) -\frac{3G\dot{G}f'''}{2\theta}-\frac{2}{3}\theta^{2}\dot{G}^{2}f^{iv}-\frac{2}{3}\theta^{2}\ddot{G}f'''
  +\frac{f'''}{f''}(\frac{9G}{16\theta^{2}}+\frac{\theta^{2}}{6}-f+Gf'-\frac{2\theta^{2}\dot{G}^{2}}{3}f''')\\+\frac{1}{2}f''
 +\frac{2}{9}\theta^{3}\dot{G}f'''\}\mathcal{G}^{k}
  +\{\frac{4}{9}\theta^{3}f''\}\mathsf{G}^{k}+\{\frac{9}{4\theta^{3}}
  +\frac{G}{2\theta}f'+f''(\frac{8\theta^{2}\dot{G}}{9}-\frac{6G\dot{G}}{2\theta^{2}}-\frac{4}{3}\theta\ddot{G})-\frac{1}{2\theta}f-\frac{2}{3}\theta 
  -\frac{4}{3}\theta \dot{G}^{2}f''' \}Z^{k},
 \end{multline}
 \begin{equation}
\dot{ \mathcal{G}}^{k}=\mathsf{G}^{k},
 \end{equation}
 \begin{multline}
  \dot{\mathsf{G}}^{k}=\{-\frac{\theta}{3}-\frac{9}{4}\frac{G}{\theta^{3}}-2\frac{f'''\dot{G}}{f''}\}\mathsf{G}^{k}+\{\frac{1}{\theta^{2}f''}(\frac{27}{32\theta^{2}}+\frac{3}{2}f')\\
  +\frac{f'''}{\theta^{2}f''}\left(-\frac{27}{32}\frac{G}{\theta^{2}}-\frac{\theta^{2}}{4}
+\frac{3}{2}f-\frac{3}{2}Gf'+\dot{G}^{2}\theta^{2}f'''\right)-\frac{3}{2\theta}f'+\frac{3}{2\theta^{2}}G\\
  -\frac{9}{4\theta}^{3}\dot{G}-\frac{\dot{G}^{2}f^{iv}}{f''}\}\mathcal{G}^{k}
  +\{\frac{1}{\theta^{2}f''}\left(-\frac{27}{8\theta^{3}}+\frac{3}{\theta}f-\frac{3}{\theta}Gf'\right)-\frac{\dot{G}}{3}+\frac{27}{4\theta^{4}}G\dot{G}\}Z^{k},~~~~~~~~~~~~~~~~~~~~~~~~~~~~~~~~~~~~~~~~~~~~~~~~~~~~~~~~~~~~~~~~~~~~~~~~~~~~
 \end{multline}
\begin{equation}
 \dot{V}^{k}_{d}=-\frac{1}{3}\theta V^{k}_{d},
\end{equation}
\begin{equation}
\dot{V}^{k}_{dr}=-\frac{1}{3}\theta V^{k}_{d},
\end{equation}
\begin{equation}
  \dot{S}^{k}_{dr}=0.
 \end{equation}

The second order equation is given as 
\begin{equation}
 \ddot{\Delta}_{d}^{k}=-\dot{Z}^{k}
 \end{equation}
 whih is
 \begin{multline}
 \ddot{\Delta}_{d}^{k}=-
  \{f''(\theta f'-G) -\frac{3G\dot{G}f'''}{2\theta}-\frac{2}{3}\theta^{2}\dot{G}^{2}f^{iv}-\frac{2}{3}\theta^{2}\ddot{G}f'''
  +\frac{f'''}{f''}(\frac{9G}{16\theta^{2}}+\frac{\theta^{2}}{6}-f+Gf'-\frac{2\theta^{2}\dot{G}^{2}}{3}f''')\\+\frac{1}{2}f''
 +\frac{2}{9}\theta^{3}\dot{G}f'''\}\mathcal{G}^{k}
  -\{\frac{4}{9}\theta^{3}f''\}\mathsf{G}^{k}-\{\frac{9}{4\theta^{3}}
  +\frac{G}{2\theta}f'+f''(\frac{8\theta^{2}\dot{G}}{9}-\frac{6G\dot{G}}{2\theta^{2}}-\frac{4}{3}\theta\ddot{G})-\frac{1}{2\theta}f-\frac{2}{3}\theta 
  -\frac{4}{3}\theta \dot{G}^{2}f''' \}Z^{k}.
 \end{multline}
Using $\dot{\Delta}_{d}^{k}=-Z^{k}$ and $\dot{ \mathcal{G}}^{k}=\mathsf{G}^{k}$, we have
\begin{multline}
 \ddot{\Delta}_{d}^{k}=\{\frac{G}{2\theta}f'+f''(\frac{9}{4\theta^{3}}
  +\frac{8\theta^{2}\dot{G}}{9}-\frac{6G\dot{G}}{2\theta^{2}}-\frac{4}{3}\theta\ddot{G})-\frac{1}{2\theta}f-\frac{2}{3}\theta 
  -\frac{4}{3}\theta \dot{G}^{2}f''' \}\dot{\Delta}_{d}^{k}\\ -\frac{4}{9}\theta^{3}f''\dot{\mathcal{G}}^{k}-
  \{f''(\theta f'-G) -\frac{3G\dot{G}f'''}{2\theta}-\frac{2}{3}\theta^{2}\dot{G}^{2}f^{iv}-\frac{2}{3}\theta^{2}\ddot{G}f'''
  +\frac{f'''}{f''}(\frac{9G}{16\theta^{2}}\\+\frac{\theta^{2}}{6}-f+Gf'-\frac{2\theta^{2}\dot{G}^{2}}{3}f''')+\frac{1}{2}f''
 +\frac{2}{9}\theta^{3}\dot{G}f'''\}\mathcal{G}^{k}.~~~~~~~~~~~~~~~~~~~~~~~~~~~~~~~~~~~~~~~~~~~~~~~~~~~~~~~~~~~~~~~~~~~~~~~~~~~~~~~~~~~~~~~~~~~~~~~~~~~~~~~~~~~~~~~~~~~~~~~
 \label{eq218}
 \end{multline} 
 For the case $f(G)=G$, we have\cite{munyeshyaka2021cosmological,abebe2012covariant}
\begin{equation}
 \ddot{\Delta}_{d}^{k}+\frac{2}{3}\theta \dot{\Delta}_{d}^{k}=0.
 \label{eq219}
\end{equation}
 Transforming Eq. \ref{eq218} into redshift, we get   
 \begin{eqnarray}
 &&(1+z)^{2}H^{2}\Delta''_{d}=-(1+z)H\{\frac{9}{4\theta^{3}}
  +\frac{G}{2\theta}f'+f''(\frac{8\theta^{2}G'}{9}-\frac{6G G'}{2\theta^{2}}-\frac{4}{3}\theta G'')\nonumber\\
  &&\quad\quad-\frac{1}{2\theta}f-\frac{2}{3}\theta 
  -\frac{4}{3}\theta G'^{2}f''' +(1+z)H'\}\Delta'_{d}+\frac{4}{9}(1+z)H\theta^{3}f''\mathcal{G}'\nonumber\\
  &&\quad\quad-
  \{f''(\theta f'-G) -\frac{3GG'f'''}{2\theta}-\frac{2}{3}\theta^{2} G'^{2}f^{iv}-\frac{2}{3}\theta^{2}G'f'''\nonumber\\
  &&\quad\quad
  +\frac{f'''}{f''}(\frac{9G}{16\theta^{2}}+\frac{\theta^{2}}{6}-f+Gf'-\frac{2\theta^{2}G'^{2}}{3}f''')+\frac{1}{2}f''
 +\frac{2}{9}\theta^{3}G'f'''\}\mathcal{G}.
 \label{eq219}
 \end{eqnarray} 
 From
 \begin{equation}
\ddot{ \mathcal{G}}^{k}=\dot{\mathsf{G}}^{k},
 \end{equation}
 which is 
 \begin{multline}
\ddot{ \mathcal{G}}^{k}=\{-\frac{\theta}{3}-\frac{9}{4}\frac{G}{\theta^{3}}-2\frac{f'''\dot{G}}{f''}\}\mathsf{G}^{k}+\{\frac{1}{\theta^{2}f''}(\frac{27}{32\theta^{2}}+\frac{3}{2}f')\\
  +\frac{f'''}{\theta^{2}f''}\left(-\frac{27}{32}\frac{G}{\theta^{2}}-\frac{\theta^{2}}{4}
+\frac{3}{2}f-\frac{3}{2}Gf'+\dot{G}^{2}\theta^{2}f'''\right)-\frac{3}{2\theta}f'+\frac{3}{2\theta^{2}}G\\
  -\frac{9}{4\theta}^{3}\dot{G}-\frac{\dot{G}^{2}f^{iv}}{f''}\}\mathcal{G}^{k}
  +\{\frac{1}{\theta^{2}f''}\left(-\frac{27}{8\theta^{3}}+\frac{3}{\theta}f-\frac{3}{\theta}Gf'\right)-\frac{\dot{G}}{3}+\frac{27}{4\theta^{4}}G\dot{G}\}Z^{k}.~~~~~~~~~~~~~~~~~~~~~~~~~~~~~~~~~~~~~~~~~~~~~~~~~~~~~~~~~~~~~~~~~~
 \end{multline}
 Using $\dot{\Delta}_{d}^{k}=-Z^{k}$, $\dot{ \mathcal{G}}^{k}=\mathsf{G}^{k}$ and $\dot{S}^{k}_{dr}=\dot{\Delta}^{k}_{d}=0$, we have
\begin{multline}
\ddot{ \mathcal{G}}^{k}=\{-\frac{\theta}{3}-\frac{9}{4}\frac{G}{\theta^{3}}-2\frac{f'''\dot{G}}{f''}\}\dot{\mathcal{G}}^{k}+\{\frac{1}{\theta^{2}f''}(\frac{27}{32\theta^{2}}+\frac{3}{2}f')
  -\frac{9}{4\theta}^{3}\dot{G}-\frac{\dot{G}^{2}f^{iv}}{f''}\\
  +\frac{f'''}{\theta^{2}f''}\left(-\frac{27}{32}\frac{G}{\theta^{2}}-\frac{\theta^{2}}{4}
+\frac{3}{2}f-\frac{3}{2}Gf'+\dot{G}^{2}\theta^{2}f'''\right)-\frac{3}{2\theta}f'+\frac{3}{2\theta^{2}}G\}\mathcal{G}^{k}.~~~~~~~~~~~~~~~~~~~~~~~~~~~~~~~~~~~~~~~~~~~~~~~~~~~~~~~~~~~~~~~~~~~~~~~~~~~~~~~~~~
\label{eq222}
 \end{multline}
 The redshift transformation of Eq. \ref{eq219} and Eq.  \ref{eq222} are presented as
 \begin{multline}
(1+z)^{2}H^{2} \mathcal{G}''=-(1+z)H\{-\frac{\theta}{3}-\frac{9}{4}\frac{G}{\theta^{3}}-2\frac{f'''G'}{f''}+(1+z)H'\}\mathcal{G}'+\{\frac{1}{\theta^{2}f''}(\frac{27}{32\theta^{2}}+\frac{3}{2}f')
  \\-\frac{9}{4\theta^{3}}G'-\frac{G'^{2}f^{iv}}{f''}
  +\frac{f'''}{\theta^{2}f''}\left(-\frac{27}{32}\frac{G}{\theta^{2}}-\frac{\theta^{2}}{4}
+\frac{3}{2}f-\frac{3}{2}Gf'+G'^{2}\theta^{2}f'''\right)-\frac{3}{2\theta}f'+\frac{3}{2\theta^{2}}G\}\mathcal{G},~~~~~~~~~~~~~~~~~~~~~~~~~~~~~~~~~~~~~~~~~~~~~~~~~~~~~~~~~~~~~~~~~~~~~~~~~~~~~~~~~~~~
\label{eq223}
 \end{multline}
 and \cite{abebe2012covariant}
 \begin{equation}
   \Delta''_{d}-\frac{3}{1+z}\Delta'_{d}=0.
  \label{eq224}
 \end{equation}
 Analytical solution is given as 
 \begin{equation}
  \Delta(z)=C_{5}+C_{6}(1+z)^{4}.
 \end{equation}
 The integration constants $C_{5}$ and $C_{6}$ can be determined by imposing the initial conditions at $z=0$. At $z=0$, we have
\begin{equation}
 \Delta(z=0)=C_{5}+C_{6}.
\end{equation}
\begin{equation}
 \Delta'(z)=4C_{6}(1+z)^{3},
\end{equation}
\begin{equation}
 \Delta'(z=0)=4C_{6}.
\end{equation}
Solving simultaneously, we have
\begin{equation}
 C_{5}=\frac{4\Delta(z=0)-\Delta'(z=0)}{4},
\end{equation}
and
\begin{equation}
 C_{6}=\frac{\Delta'(z=0)}{4}.
\end{equation}

\begin{figure}[ht!]
\begin{center}
 \subfigure[Numerical solutions of Eq. \ref{eq219} and Eq. \ref{eq223} using polynomial $f(G)$ model (eq. \ref{eq28})]{%
 \label{fig: first }
 \includegraphics[width=60mm]{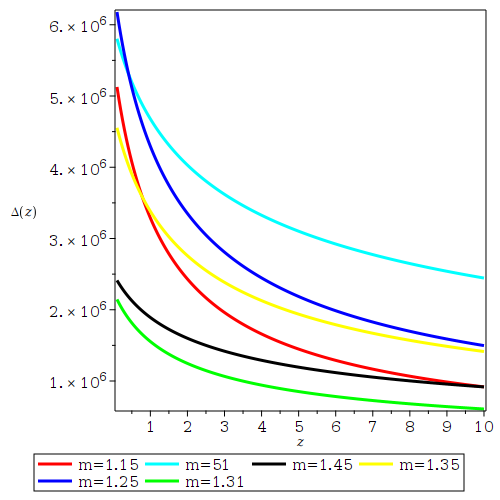}
 }%
 \subfigure[Numerical solutions of Eq. \ref{eq224} for the case $f(G)=G$]{%
 \label{fig: second}
 \includegraphics[width=60mm, height=60mm]{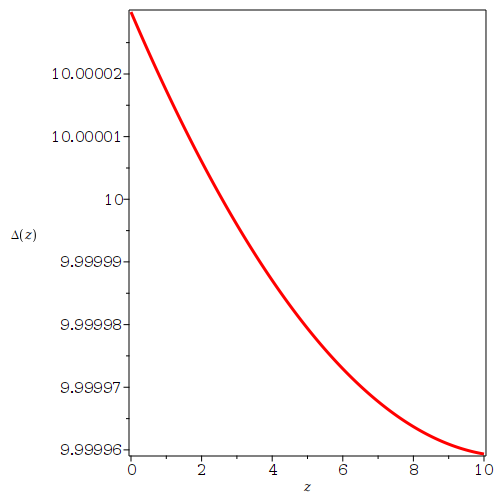}}%
\end{center}
 \caption{%
Plot of energy density perturbations $\Delta(z)$ versus redshift $z$ of Eq. \ref{eq219}, Eq. \ref{eq223} and Eq. \ref{eq224} for long wavelength modes in a radiation dominated Universe using different values of $m$.}%
\label{Fig(3)}
\end{figure}

The numerical solutions of Eq. \ref{eq219}, Eq. \ref{eq223}  and Eq.  \ref{eq224} are presented in Figure \ref{Fig(3)} for polynomial $f(G)$ model and for the case $f(G)=G$. 

\subsubsection{Perturbations in a dust-dominated epoch}
The energy density of radiation is much smaller than energy density of dust in
this dust dominated epoch. that is, 
$\rho_{ r}\ll \rho_{d} $.

 We also assume that the perturbations due to radiation energy
density are small enough compared to the perturbations generated from dust, that is
$\Delta_{r} \ll \Delta_{d}$ .
Proceeding in a similar fashion for the dust dominated, long wavelength regime,
the second order evolution equations  are given as
\begin{eqnarray}
 &&\ddot{\Delta}_{d}^{k}=\rho_{d}\Delta_{d}^{k}-\{f''(\theta f'-G+\frac{3\dot{G}}{2\theta}) -\frac{3G\dot{G}f'''}{2\theta}-\frac{3f''\dot{G}}{2\theta}-\frac{2}{3}\theta^{2}\dot{G}^{2}f^{iv}-\frac{2}{3}\theta^{2}\ddot{G}f'''\nonumber\\
 &&\quad\quad
  +\frac{f'''}{f''}(\frac{9G}{16\theta^{2}}+\frac{\theta^{2}}{6}+\frac{1}{2}\rho_{d}-f+Gf'-\frac{2\theta^{2}\dot{G}^{2}}{3}f''')+\frac{1}{2}f''
 +\frac{2}{9}\theta^{3}\dot{G}f'''\}\mathcal{G}^{k}\nonumber\\
 &&\quad\quad
  -\frac{4}{9}\theta^{3}f''\dot{\mathcal{G}}^{k}\;,
  \label{eq231}\\
&&\ddot{ \mathcal{G}}^{k}=\frac{3}{4}\frac{\rho_{d}}{\theta^{2}f''}\Delta_{d}^{k}+\{-\frac{\theta}{3}-\frac{9}{4}\frac{G}{\theta^{3}}-2\frac{f'''\dot{G}}{f''}\}\dot{\mathcal{G}}^{k}+\{\frac{1}{\theta^{2}f''}(\frac{27}{32\theta^{2}}+\frac{3}{2}f')\nonumber\\
&&\quad\quad
  +\frac{f'''}{\theta^{2}f''}\left(-\frac{27}{32}\frac{G}{\theta^{2}}-\frac{\theta^{2}}{4}
  -\frac{3}{4}\rho_{d}+\frac{3}{2}f-\frac{3}{2}Gf'+\dot{G}^{2}\theta^{2}f'''\right)-\frac{3}{2\theta}f'+\frac{3}{2\theta^{2}}G
  \nonumber\\
  &&\quad\quad-\frac{9}{4\theta^{3}}\dot{G}-\frac{\dot{G}^{2}f^{iv}}{f''}\}\mathcal{G}^{k}\;.
  \label{eq232}
 \end{eqnarray}
 The transformed redshift of Eq. \ref{eq231} and Eq. \ref{eq232} are given as 
\begin{eqnarray}
 &&(1+z)^{2}H^{2}\Delta''_{d}=\rho_{d}\Delta_{d}-\{f''(\theta f'-G+\frac{3G'}{2\theta}) -\frac{3GG'f'''}{2\theta}-\frac{3f''G'}{2\theta}-\frac{2}{3}\theta^{2}G'^{2}f^{iv}\nonumber\\
 &&\quad\quad-\frac{2}{3}\theta^{2}G''f'''
  +\frac{f'''}{f''}(\frac{9G}{16\theta^{2}}+\frac{\theta^{2}}{6}+\frac{1}{2}\rho_{d}-f+Gf'-\frac{2\theta^{2}G'^{2}}{3}f''')+\frac{1}{2}f''
 +\frac{2}{9}\theta^{3}G'f'''\}\mathcal{G}\nonumber\\
 &&\quad\quad
  +\frac{4}{9}(1+z)H\theta^{3}f''\mathcal{G}'\;,
  \label{eq233}\\
&&(1+z)^{2}H^{2} \mathcal{G}''=\frac{3}{4}\frac{\rho_{d}}{\theta^{2}f''}\Delta_{d}-(1+z)H\{-\frac{\theta}{3}-\frac{9}{4}\frac{G}{\theta^{3}}-2\frac{f'''G'}{f''}+(1+z)H'\}\mathcal{G}'\nonumber\\
&&\quad\quad+\{\frac{1}{\theta^{2}f''}(\frac{27}{32\theta^{2}}+\frac{3}{2}f')
  +\frac{f'''}{\theta^{2}f''}\left(-\frac{27}{32}\frac{G}{\theta^{2}}-\frac{\theta^{2}}{4}
  -\frac{3}{4}\rho_{d}+\frac{3}{2}f-\frac{3}{2}Gf'+G'^{2}\theta^{2}f'''\right)\nonumber\\
  &&\quad\quad-\frac{3}{2\theta}f'+\frac{3}{2\theta^{2}}G
  -\frac{9}{4\theta^{3}}G'-\frac{G'^{2}f^{iv}}{f''}\}\mathcal{G}\;,
  \label{eq234}\\
 &&  \Delta''_{d}-\frac{2}{1+z}\Delta'_{d}-\frac{3}{(1+z)^{2}}\Delta_{d}=0\;,
  \label{eq235}
 \end{eqnarray}
 which admit the solution
 \begin{equation}
 \Delta(z)=C_{7}(1+z)^{\sigma_{1}}+C_{8}(1+z)^{\sigma_{2}},
\end{equation}
with $\sigma_{1}=\frac{3}{2}+\frac{\sqrt{21}}{2}$ and $\sigma_{2}=\frac{3}{2}-\frac{\sqrt{21}}{2}$.
The integration constants $C_{7}$ and $C_{8}$ can be determined by imposing the initial conditions at $z=0$. At $z=0$, we have
\begin{equation}
 \Delta(z=0)=C_{7}+C_{8}.
\end{equation}
\begin{equation}
 \Delta'(z)=C_{7}\sigma_{1}(1+z)^{\sigma_{1}-1}+C_{8}\sigma_{2}(1+z)^{\sigma_{2}-1},
\end{equation}
\begin{equation}
 \Delta'(z=0)=C_{7}\sigma_{1}+C_{8}\sigma_{2},
\end{equation}
Solving simultaneously, we have
\begin{equation}
 C_{7}=\frac{\sigma_{2}\Delta(z=0)-\Delta'(z=0)}{\sigma_{2}-\sigma_{1}},
\end{equation}
and
\begin{equation}
 C_{8}=\frac{\Delta'(z=0)-\sigma_{1}\Delta(z=0)}{\sigma_{2}-\sigma_{1}}.
\end{equation}

\begin{figure}[ht!]
\begin{center}
 \subfigure[Numerical solutions of Eq. \ref{eq233}  and Eq. \ref{eq234} using  polynomial  $f(G)$ model (eq.  \ref{eq28})]{%
 \label{fig: first }
 \includegraphics[width=60mm]{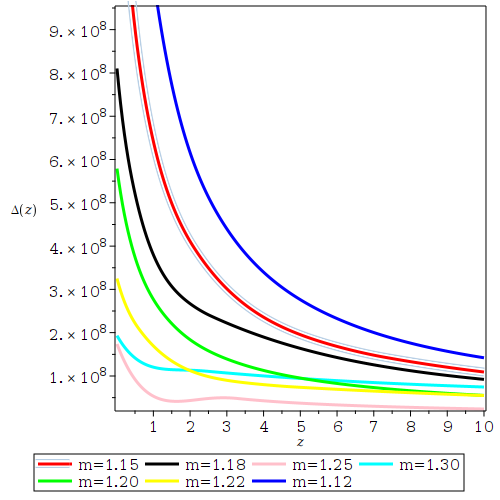}
 }%
 \subfigure[Numerical solutions of Eq. \ref{eq235} for the case $f(G)=G$]{%
 \label{fig: second}
 \includegraphics[width=60mm, height=60mm]{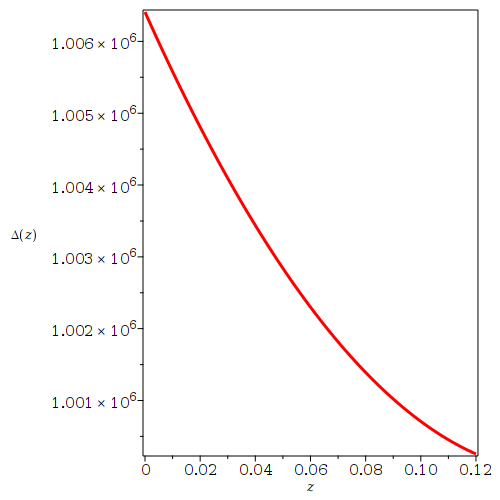}}%
\end{center}
 \caption{%
Plot of energy density perturbations $\Delta(z)$ versus redshift $z$ of  Eq. \ref{eq233}, Eq. \ref{eq234} and Eq. \ref{eq235} for long wavelength modes in a dust dominated Universe using different values of $m$.}%
\label{Fig(4)}
\end{figure}

The numerical solutions of Eq. \ref{eq233}, Eq. \ref{eq234} and Eq. \ref{eq235}  are presented in the Figure \ref{Fig(4)} for polynomial (Eq. \ref{eq28}) $f(G)$ model  and for the case $f(G)=G$.

It can be seen that Eq. \ref{eq223} differs from Eq. \ref{eq162}, Eq. \ref{eq186}, Eq. \ref{eq219}, Eq. \ref{eq233} and Eq. \ref{eq234} since the Gauss-Bonnet fluid perturbations decouple with the matter energy density.
\newpage
 \section{Discussion and Conclusion }
 \subsection{Discussion}
We have developed a theory of cosmological density perturbations in a multifluid cosmological medium using $1+3$ covariant formalism with the consideration of $f(G)$ theory of gravity. We defined vector gradient variables with respect to a FRW background which characterize the time evolution of density and velocity perturbations. We derived a complete set of linear evolution equations for both the total fluid and its components. Using different techniques namely: scalar decomposition, harmonic decomposition and redshift transformation, we analysed the evolution of energy density perturbations in both short and long wavelength modes for radiation-Gauss-Bonnet and dust-Gauss-Bonnet fluid systems. We then considered the case where $f(G)=G$ and  polynomial $f(G)$ model  to get numerical results. The numerical results in  short-wavelength modes for both radiation-Gauss-Bonnet and dust-Gauss-Bonnet fluids are presented in Figure. \ref{Fig(1)}(a) and Figure .\ref{Fig(2)} (a) respectively and Figure. \ref{Fig(1)}(b) and Figure .\ref{Fig(2)} (b) for the case $f(G)=G$. The ones in long-wavelength modes are presented in Figure. \ref{Fig(3)}(a) through to Figure .\ref{Fig(4)}(a) for radiation-Gauss-Bonnet  and dust-Gauss-Bonnet systems and Figure. \ref{Fig(3)}(b) through to Figure .\ref{Fig(4)}(b) for the case $f(G)=G$.
From the plots, for the case $f(G)=G$, we depicted the contribution of radiation and dust component of the Universe for the perturbations of matter energy density where  the energy density perturbations decay with redshift for all Figures: Fig. \ref{Fig(1)}(b), through to    Fig. \ref{Fig(4)}(b). For the   considered polynomial  model, we observed that the energy density perturabtions decay with increase in redshift for all the Figures: Fig. \ref{Fig(1)}(a), through to  Fig. \ref{Fig(4)}(a). We assumed the initial conditions as $\Delta(z_{in})=10^{-6}$ and $\Delta'(z_{in})=0$ \cite{abebe2013large} for each mode $k$ to deal with the growth of matter fluctuations.  The evolution of matter perturbations is scale invariant at all scale in the presence of Gauss-Bonnet term and the growth rate of matter energy density perturbations can be compatible with observations even in the consideration of the contribution from the Gauss-Bonnet interaction. Some of the specific highlights of this work include:
\begin{itemize}
 \item The equations obtained in this paper can be applied in different situations of cosmological interest because they are completely general in terms of fluid properties and interactions and they give a covariant and gauge invariant description of properties of perturbed FRW Universe in $f(G)$ theory of gravity.
 \item We investigated the evolution of the linearly perturbed equations. The  equations Eq. \ref{eq162} and Eq. \ref{eq186} of short-wavelength modes and Eq. \ref{eq219}, Eq. \ref{eq223}, Eq.  \ref{eq233} and Eq. \ref{eq234} of long wavelength modes for the linear matter and Gauss-Bonnet fluid inhomogeneities and Eq. \ref{eq127} for entropy perturbations could be highlighted as one of the main results of this paper since they enabled us to find several results for the evolution of the perturbations in matter-Gauss-bonnet gravity.
 \item We considered only scalar perturbations rather than vector and tensor perturabtions studied in \cite{de2010cosmological}. In both dust and radiation dominated epochs, we have seen that energy density perturbations are $k$ independent and the long and short wavelength modes do not depend on the speed of sound, a result differs in the ones obtained in \cite{de2010cosmological}.
 \item We have found that the velocity perturbations of the perfect fluid propagate with the speed of sound $c^{2}_{s}=\frac{\dot{p}}{\dot{\rho}}$ in agreement with \cite{de2010cosmological}.
 \item The entropy perturbations of the perfect fluid completely decouple from energy density and velocity perturbations for a perfect barotropic equation of state, a feature in agreement with the ones obtained in \cite{de2010cosmological}.
 \item We have shown the ranges of $m$ for which the energy density perturbations decay with redshift  for the polynomial $f(G)$ model.  We considered $1\leq m \leq 1.80$ for both short-and long- wavelength modes. The consideration of different values of $m$ was made basing on the work done in  \cite{ananda2009structure,abebe2013large}.
  The current results show that even at the level of perturbations, the Gauss-Bonnet fluid offers an alternative for the large scale structure formation. This is because we can notice the decay of the energy density perturbations with redshift, which implies that there is an increase in   structure formation rate. This result is in agreement with the astrophysical and cosmological observations and $\Lambda CDM$ model \cite{ananda2009structure,abebe2013large}.
  \item The derived $f(G)$ model shows that the non-linear $G$ can drive the inflation in the early epoch and  describe the late time cosmic acceleration, a result similar to the one obtained in for the considered $f(R,G)$ models, with both $R$ and $G$ being non-linear. This result is in agreement with the recent results by Plank \cite{ade2016planck,aghanim2020planck} and BICEP2 \cite{collaboration2015detection} collaborations.
\end{itemize}
 \subsection{Conclusion}
 In this work, we have presented a detailed analysis of cosmological perturbations in $f(G)$ gravity where the Universe is described by  multi-component fluids with a general equation of state parameter. We explored the  numerical solutions of cosmological perturbation equations in a multifluid cosmological medium in modified Gauss-Bonnet theory of gravity using maple software. We applied the $1+3$ covariant formalism to define gradient variables  and we obtained the linear evolution equations of the matter energy density and Gauss-Bonnet perturbations for both total matter and the component fluid relative to the energy frame. We applied scalar and harmonic decomposition methods to analyse the scalar perturbations of  energy densities involved namely energy density contribution from matter and Gauss-Bonnet fluids. In order to further analyse the perturbation equations, we applied redshift transformation method together with quasi-static approximation to get a set of simplified equations and to be able to make comparison with  observations. We considered different systems such as radiation-Gauss-Bonnet and dust-Gaus-Bonnet fluids in both short- and long-wavelength modes.  We used polynomial $f(G)$ model and considered the case where $f(G)=G$ to get numerical results of the perturbation equations. In conclusion, since all the numerical results presented in Fig.  \ref{Fig(1)}(a) through to Fig. \ref{Fig(4)}(a) and  Fig.  \ref{Fig(1)}(b) through to Fig. \ref{Fig(4)}(b) show the decay of energy density perturbations with an increase in redshift, the formation of  structures is enhanced in $f(G)$ gravity specifically for the model under consideration which is one of the characteristic of cosmic acceleration.  
 
%%%%%%%%%%%%%%%%%%%%%%%%
%%%%%%%%%%%%%%%%%%%%%%%%
\section*{Acknowledgments}
AM, BM and JN acknowledge financial support from International  Science Program (ISP) to the University of Rwanda through Rwanda Astrophysics Space and Climate Science Research Group (RASCSRG), Grant number: RWA 01. FT and AA aknowledge University of Rwanda, College of Science and Technology for the research facilities.
We thank the anonymous reviewers for their careful reading of this manuscript  and their insightful comments and suggestions.
  %%%%%%%%%%%%%%%%%%%here you need to change the path of bib file to the correct one
%\bibliographystyle{ieeetr}
%\bibliography{references}
 
%\section*{Appendix}
 \noindent
{\color{blue} \rule{\linewidth}{1mm} }
\end{document}